\documentclass[twocolumn]{aastex62}
\usepackage{graphicx}
\usepackage{url}

\begin{document}


\title{The Infrared Database of Extragalactic Observables from {\it Spitzer}. -- \\
II. The Database \& The Diagnostic Power of Crystalline Silicate
Features in Galaxy Spectra}

%% Use \author, \affil, plus the \and command to format author and affiliation 
%% information.  If done correctly the peer review system will be able to
%% automatically put the author and affiliation information from the manuscript
%% and save the corresponding author the trouble of entering it by hand.
%%
%% The \affil should be used to document primary affiliations and the
%% \altaffil should be used for secondary affiliations, titles, or email.

\correspondingauthor{Henrik Spoon}
\email{h.spoon@cornell.edu , spoon@astro.cornell.edu}

\author[0000-0002-8712-369X]{H.W.W. Spoon}
\affil{Cornell Center for Astrophysics and Planetary Science (CCAPS),
Department of Astronomy, Cornell University, Ithaca, NY 14853, USA}

\author{A. Hern\'an-Caballero}
\affiliation{Centro de Estudios de F\'isica del Cosmos de Arag\'on (CEFCA), Plaza San Juan, 1, E-44001 Teruel, Spain}

\author{D. Rupke}
\affiliation{Department of Physics, Rhodes College, Memphis, TN 38112, USA}

\author{L.B.F.M. Waters}
\affiliation{Institute for Mathematics, Astrophysics \& Particle
  Physics, Department of Astrophysics, Radboud University, 
P.O. Box 9010, MC 62, 6500 GL Nijmegen, The Netherlands}
\affiliation{SRON Netherlands Institute for Space Research, 
Niels Bohrweg 4, 2333 CA Leiden, The Netherlands}

\author{V. Lebouteiller}
\affiliation{AIM, CEA, CNRS, Universit\'e Paris-Saclay, Universit\'e Paris Diderot, Sorbonne Paris Cit\'e, F-91191 Gif-sur-Yvette, France}

\author{A.G.G.M. Tielens}
\affiliation{Leiden Observatory, Leiden University, P.O. Box 9513, 
2300 RA Leiden, The Netherlands}

\author{T. Loredo}
\affil{Cornell Center for Astrophysics and Planetary Science (CCAPS),
Department of Astronomy, Cornell University, Ithaca, NY 14853, USA}

\author{Y. Su}
\affil{Cornell Center for Astrophysics and Planetary Science (CCAPS),
Department of Astronomy, Cornell University, Ithaca, NY 14853, USA}

\author{V. Viola}
\affiliation{Department of Physics, Rhodes College, Memphis, TN 38112, USA}

\begin{abstract}
We present the Infrared Database of Extragalactic Observables 
from Spitzer (IDEOS), a homogeneous, publicly available, database 
of 77 fitted mid-infrared observables in the 5.4--36\,$\mu$m range, 
comprising measurements for 3335 galaxies observed in the 
low-resolution staring mode of 
the Infrared Spectrometer onboard the Spitzer Space Telescope.
Among the included observables are PAH fluxes and their equivalent 
widths, the strength of the 9.8\,$\mu$m silicate feature, 
emission line fluxes, solid-state features, rest frame continuum fluxes, 
synthetic photometry, and a mid-infrared spectral classification.
The IDEOS spectra were selected from the Cornell Atlas of Spitzer-IRS
Sources.
To our surprise we have detected at a $>$95\% confidence level
crystalline silicates in the spectra of 786
IDEOS galaxies. The detections range from single band detections 
to detections of all fitted crystalline bands (16, 19, 23, 28 and 
33\,$\mu$m). We find the strength of the crystalline silicate bands
to correlate with the amorphous silicate strength, 
and the change from an emission to an absorption feature to occur
at higher obscuration as the wavelength of the crystalline silicate
band is longer. These observed characteristics are consistent with
an origin for the amorphous and crystalline silicate features in 
a centrally heated dust geometry, either an edge-on disk or 
a cocoon. We find the 23 and 33\,$\mu$m crystalline silicate 
bands to be well-suited to classify the obscuration level of galactic
nuclei, even in the presence of strong circumnuclear star formation.
Based on our detection statistics, we conclude that crystalline 
silicates are a common component of the interstellar medium  
of galactic nuclei.
\end{abstract}

\section{Introduction} \label{sec:intro}

The successful mission of the Spitzer Space Telescope
\citep[2003-2020;][]{werner04} has bestowed us with a rich scientific 
legacy comprising many thousands of publications.
In many cases the data that these papers are based on were extracted
by researchers with extensive experience in Spitzer data 
processing.

The individual calibrated frames obtained by the Spitzer instruments,
or basic calibrated data (BCD) in Spitzer speak, are available 
from the Spitzer Heritage Archive\footnote{https://sha.ipac.caltech.edu/applications/Spitzer/SHA/}
(SHA). This is also where the combined frames, extracted spectra, and
celestial maps, the so-called post-BCD products, can be searched and 
downloaded from.

To fill the gap between the bulk-processed products offered by the SHA 
and the level of processing that an instrument specialist can attain, 
the members of the Infrared Spectrograph \citep[IRS;][]{houck04} team 
at Cornell University created CASSIS, the Cornell
Atlas of Spitzer/Infrared Spectrograph Sources\footnote{The Cornell
Atlas of Spitzer/Infrared Spectrograph Sources has since
been renamed and moved to a privately-owned website https://cassis.sirtf.com}\citep{lebouteiller11}.
For each IRS staring-mode observation CASSIS offers the choice of 
several background subtraction options and several spectral extraction 
methods, all of which can be visually compared to select and download 
the best spectrum for the source at hand.

In 2012 we took the next logical step and started working on the
Infrared Database of Extragalactic Observables from Spitzer (IDEOS).
IDEOS builds on the publication-quality IRS spectra in the
CASSIS repository by providing a homogeneous set of measurements of 
spectral features and continuum flux densities resulting from fitting 
and decomposing the IRS low-resolution spectra (R=60--120) of a subset 
of 3335 extragalactic sources.
The selection of the IDEOS spectra and the preparation for their 
analysis are described in \cite{hernan16} (Paper {\sc i}; see 
also Sect.\,\ref{sec:ideosspectra}). In the present paper 
(Paper {\sc ii}) we describe our fitting methods and present our 
observables. Technical details on the IRS modules and their
characteristics can be found in \cite{houck04}.

A large sample of homogeneously measured observables, like IDEOS,
allows for the search for subtle trends that would remain elusive
in smaller or heterogeneously measured data sets.
In Sect.\,\ref{sec:crystsil} we present an analysis of a number of
weak absorption and emission features resulting from the presence of
crystalline silicates in the interstellar medium (ISM) of IDEOS galaxies.

Observations with the Infrared Space Observatory (ISO) in the 
1990s have shown crystalline silicates to be a common component of 
interstellar dust in circumstellar environments 
\citep[][and references there in]{molster05}.
In contrast, ISO observations of the diffuse ISM, as probed by the line
of sight to Sgr\,A$^*$, have set a clear upper limit of 1\% to the 
crystallinity of the diffuse ISM in our galaxy \citep{kemper04,kemper05}. 
The presence of crystalline silicates in extragalactic environments
only became known in 2006 through their detection in the strongly 
silicate-absorbed Spitzer-IRS spectra of twelve deeply enshrouded 
Ultraluminous Infrared Galaxies \citep[ULIRGs;][]{spoon06}.

In recent years, X-ray Si K-band absorption edge studies have 
revealed the presence of a fraction of crystalline silicates in the
ISM of about 0.1, well exceeding the infrared upper limits 
\citep{zeegers19,rogantini19}. The difference has been attributed to the
presence of polycrystalline silicate grains in the ISM, which register
as crystalline structures using X-ray techniques (sensitive to short
range lattice order) and as amorphous in infrared observations (probing
long range lattice disorder). Further studies will be needed to 
reconcile these findings.

Because crystallization of silicates is inhibited by high energy
barriers, the presence of crystalline silicate features in a
mid-infrared spectrum 
is indicative of processing of grains in the circumstellar and/or 
interstellar medium. While in Galactic environments this points to
origins in both pre-main-sequence and post-main-sequence stars, in
active galaxies radiation from the accretion disk around supermassive 
black holes may be another source of energetic processing \citep{spoon06},
while shocks may play a role in interacting galaxies.

The complete absence of absorption features of crystalline silicates
in mid-infrared spectra of the Galactic ISM \citep{kemper04} implies a rapid
transformation of newly formed crystalline silicates into amorphous 
silicates in the interstellar medium \citep[$\leq$10$^8$ yr;][]{kemper04}. 
This amorphization may be achieved through sputtering of the dust
in strong shock waves resulting from supernova explosions 
\citep{jones94,jones96}, but would, however, likely affect amorphous
and crystalline silicates equally and thus leave their proportions
unchanged.
Cosmic rays, produced in the same supernova events, are 
thought to be more effective in reducing the crystalline silicate 
fraction, and may act on timescales of $\sim$10$^7$ years \citep{bringa07}.

For their sample of twelve ULIRGs \cite{spoon06} concluded that 
amorphization due to cosmic rays may lag in vigorous starburst 
environments, leaving enough freshly forged crystalline silicates 
in the ISM to be detectable in their galaxy spectra.
Our present far larger study puts the need for these special conditions
in doubt.

Our paper is organized as follows. 
Section 2 provides a brief description of how we processed spectra 
selected from the CASSIS repository for ingestion in the IDEOS database. 
Section 3 describes in detail our method and assumption for fitting 
the IDEOS SEDs to obtain observables. 
Section 4 details how we deployed the spectral decomposition tools
PAHFIT and QUESTFIT to obtain alternate sets of observables for our sample.
Sections 5 and 6 describe how we compute rest frame continuum fluxes
and synthetic photometry, respectively.
Section 7 explains our method for deriving silicate strengths.
Section 8 discusses what can be learned from mid-infrared diagnostic plots
containing thousands of IDEOS sources.
Section 9 analyses the crystalline silicate features that we detected.
Section 10 presents the IDEOS web portal.
Section 11 contains the discussion and our conclusions.

\section{The IDEOS spectra}\label{sec:ideosspectra}

%%%%%%%%%%%%%%%%%%%%%%%%%%%%%%%%%%%%%%%%%%%%%%%%%%%%%%%%%%%%%%%%%%
\begin{table*}
\centering
\caption{IDEOS sample statistics} \label{tab:ideosstats}
\begin{tabular}{lr}
\hline
\hline
CASSIS spectra in extragalactic or calibration programs (ECPs) & 5015 \\
CASSIS spectra in ECPs: no detection & 1263 \\
CASSIS spectra in ECPs: detection in single nod & 50 \\
CASSIS spectra in ECPs: Local Group galaxy or non-nuclear pointing & 200 \\
IDEOS galaxy spectra & 3558 \\
IDEOS galaxy spectra: appended with segments from other observations & 80 \\
IDEOS galaxies & 3335 \\
IDEOS galaxies with multiple observations of 1--4 segment(s) & 110 \\
IDEOS galaxies with averaged spectra from multiple observations & 25 \\
IDEOS BLAZARs with low and high-state spectra: 3C\,279 and 3C\,454.3 & 2\\
IDEOS galaxies at redshifts $\leq$0.1 & 1463\\
IDEOS galaxies at redshifts 0.1--0.2 & 456\\
IDEOS galaxies at redshifts 0.2--0.4 & 334\\
IDEOS galaxies at redshifts 0.4--0.6 & 166\\
IDEOS galaxies at redshifts 0.6--0.8 & 167\\
IDEOS galaxies at redshifts 0.8--1.0 & 197\\
IDEOS galaxies at redshifts 1.0--2.0 & 350\\
IDEOS galaxies at redshifts $\geq$2 & 204\\
IDEOS galaxies with a silicate strength (S$_{\rm sil}$) measurement & 2846\\
IDEOS galaxies with S$_{\rm sil}$$\geq$-1 & 2527\\
IDEOS galaxies with S$_{\rm sil}$$<$-1 & 319\\
\hline
\end{tabular}
\end{table*}
%%%%%%%%%%%%%%%%%%%%%%%%%%%%%%%%%%%%%%%%%%%%%%%%%%%%%%%%%%%%%%%%%%

The 3558 spectra that are part of IDEOS have been selected from a parent 
sample of 5015 spectra from CASSIS that met our initial selection
requirement of belonging to an extragalactic Spitzer-IRS observing,
science-demonstration, or calibration program. Of the 5015 spectra, 
1263 spectra did not result in a source detection, and another 50/5015 
were detected in only one nod\footnote{In standard 'staring mode' 
the observation is repeated after a small nod of the spacecraft to 
move the source from a position at 1/3 to 2/3 of the way along the 
length of the slit.} spectrum or only in a bonus\footnote{The bonus
order is really a repeat of a portion of the first order spectrum 
during an observation of the second order spectrum.} order. Of 
the remaining spectra, some 200 were discarded for mispointing, 
for a non-nuclear pointing (e.g. an observation of a supernova), 
for bad background-subtraction (in a crowded field), or for requiring
large and uncertain aperture corrections due to membership of the 
Local Group of galaxies. As is true for all CASSIS spectra, the IDEOS 
sample consists purely of spectra obtained in staring mode 
(as opposed to mapping mode).

In order to be able to measure observables from the IDEOS spectra, 
each extracted spectrum is first identified in the
NASA Extragalactic Database (NED), is assigned a redshift, and
has its spectral segments (orders) scaled and stitched. In
addition, for 80 incomplete spectra we were able to find 
and append additional
spectral segments from other observation of the same 
target. A detailed description of these steps is provided in Paper {\sc i}.

We spent great care to define meaningful fitting uncertainties for 
the IDEOS observables. It is important to realize, though,
that these uncertainties do not include uncertainties associated with
the flux calibration, slit losses, or stitching of the spectral segments. 
Uncertainties in the latter are compounded from segment LL1 to LL2 to
SL1 to SL2, where we assume the scaling factor of the LL1 segment to be
unity (see Paper {\sc i}). The segment-to-segment scaling 
uncertainties are generally small ($<$5\%) but can be 10\%--20\% 
for the LL2 to SL1 scaling for low S/N spectra of semi-extended 
sources, as it is hard to determine whether the flux in the
small order overlap is affected by excessive noise and spurious 
features or not. The IDEOS portal (see Sect.\,\ref{sec:portal}) 
provides information about unusual stitching uncertainties as 
part of the stitching metadata for each spectrum.

Among the 3558 IDEOS spectra we found 110 targets that had multiple
observations (380) of some or all spectral segments (SL2/SL1/LL2/LL1).
For 25 of these targets it was beneficial to the S/N of the spectrum 
to average these observations. Among them are 
Arp\,220, IRAS\,08572+3915NW, 3C\,273, IRAS\,07598+6508, and Mrk\,231.
For the remaining targets we identified the spectrum with the best S/N
and use that spectrum in scatter plots
as well as in statistical computations in this paper.
The total number of unique IDEOS targets is 3335.

Note that the IDEOS portal (Sect.\,\ref{sec:portal}) provides 
observables for all 3558 spectra, not just for the 3335 best versions.

\section{Spectral fitting} \label{sec:sedfitting}

We have used the Levenberg-Marquardt least-squares fitting software
package MPFIT \citep{markwardt09} and empirical spectral templates
to measure fluxes of emission
features, optical depths of absorption features, and flux densities 
of the continuum in six rest frame ranges of the mid-infrared spectrum. 
The ranges are modelled independently from one another, without 
any prior assumptions on the underlying continuum other than it being
a polynomial of a pre-defined order. The latter choice was made to
ensure the best possible fit to the individual features. We will 
refer to this approach as 'chunk fitting', and the code that
implements it as 'CHUNKFIT'.

The features fitted in the six ranges comprise various PAH emission bands,
fine-structure lines, pure rotational molecular hydrogen lines, 
crystalline silicate features, and absorption bands of water ice 
and aliphatic hydrocarbons. Not fitted are the 7.7 and 8.6\,$\mu$m 
PAH features, as the proximity of the 9.8\,$\mu$m silicate band
makes it hard to determine a local continuum. We also do not fit the
PAH plateau emission in the 5--18\,$\mu$m range. This is a consequence
of our choice to define the local continuum in this range using the 
method of \cite{hony01} and \cite{vermeij02}, which lumps in the PAH 
plateau emission with the continuum emission. As an alternative, in 
Sect.\,\ref{sec:seddecomp} we will use the method of 
\cite{smith07}, based on the PAH model of \cite{boulanger98}, which 
does separate out the PAH plateau emission from the continuum 
emission. A comparison of both approaches is offered by \cite{galliano08b}.

We further refrain from using silicate opacity profiles to model the
local continua in the CHUNKFIT ranges affected by the amorphous 
silicate features centered at 9.8\,$\mu$m and 18\,$\mu$m, as
these profiles may not be applicable to all galaxy-integrated spectra
and could result in bad fits to the local continua and thus
to the weak emission and absorption features we are trying to fit. 
We instead model the local continuum using polynomials, irrespective 
of whether the continuum is shaped by a silicate feature, or not.
In Sect.\,\ref{sec:silstrength} we describe the method we
use to infer the strength of the 9.8\,$\mu$m amorphous silicate feature.

In CHUNKFIT emission lines are modeled with gaussian profiles. An 
adequate choice given the gaussian shape of the IRS instrumental profile. 
Table\,\ref{tab:resolvingpower} lists the spectral resolving powers
from which our line widths derive. 

%%%%%%%%%%%%%%%%%%%%%%%%%%%%%%%%%%%%%%%%%%%%%%%%%%%%%%%%%%%%%%%%%%
\begin{table}
\centering
\caption{Spitzer-IRS characteristics} \label{tab:resolvingpower}
\begin{tabular}{lcrr}
\hline
\hline
Module & Wavelength Range & Resolving power & Slit width \\
       & [$\mu$m]         &                 & [$''$]     \\
\hline
SL2 & 5.13 -- 7.60   & 85--125 & 3.6 \\ % conforms to JVC
SL1 & 7.46 -- 14.29  & 61--120 & 3.7 \\ % conforms to Handbook
LL2 & 13.90 -- 21.27 & 82--125 & 10.5\\ % conforms to JVC
LL1 & 19.91 -- 39.90 & 58--112 & 10.7\\ % conforms to Handbook
\hline
\end{tabular}
\end{table}
%%%%%%%%%%%%%%%%%%%%%%%%%%%%%%%%%%%%%%%%%%%%%%%%%%%%%%%%%%%%%%%%%%

Fitting features in spectra with a range of signal-to-noise (S/N) has
proven challenging. In high S/N spectra, weak features
can only be fitted well if the order of the polynomial continuum 
is high enough to provide an accurate fit to the local continuum. 
In contrast, in spectra of low S/N the same high degree 
of freedom for the continuum spectral shape will result 
in wavy continua as MPFIT will attempt to fit some of the 
noise spurious features. This challenge is especially 
large in spectral ranges
that include the 9.7 and 18.5\,$\mu$m silicate features, where 
the continuum spectral structure can show strong curvature. We 
therefore devised S/N criteria for when to limit the polynomial 
order to mitigate waviness. These criteria are different for 
different spectral ranges and depend on S/N as well as on the 
strength of specific spectral features in these ranges.

To minimize the number of false detections of emission lines, 
it would seem prudent to fix the line center to the rest 
wavelength and to fix the width of the line profile to the 
instrumental resolution at the observed wavelength. In practice, 
however, MPFIT fits a higher fraction of the lines present
when the line center is allowed to shift by $\pm$600 km/s, and 
the line width is allowed to vary by $\pm$10\%.
Note that at the low spectral resolving power of the IRS
low-resolution modules line asymmetries and line broadening
associated with ionized gas outflows in AGN \citep{dasyra08} 
and ULIRGs \citep{spoon09a} cannot be resolved. It therefore is 
safe to assume that all emission lines have gaussian line profiles.
For PAH features we use a different approach to limit false
detections. Since a strong detection of an intrinsically weak PAH 
band likely indicates that a noise feature was fitted, we constrain
the allowed ratio of a faint to a strong PAH feature to an empirically 
determined range measured from a sample of high S/N PAH spectra.

%%%%%%%%%%%%%%%%%%%%%%%%%%%%%%%%%%%%%%%%%%%%%%%%%%%%%%%%%%%%%%%%%%
\begin{table}
\centering
\caption{Parameters in CHUNKFIT ranges} \label{tab:degreesoffreedom}
\begin{tabular}{lcc}
\hline
\hline
Range & Fit parameters & Degrees of freedom \\
\hline
5.39--7.25\,$\mu$m & $\leq$24 & 20--61 \\
8.7--10.3\,$\mu$m  & $\leq$19 & 13--47 \\
10.0--12.58\,$\mu$m& $\leq$21 & 14--48 \\
9.8--13.5\,$\mu$m  & $\leq$31 & 24--66 \\
13.0--15.4\,$\mu$m & $\leq$14 & 11--34 \\
14.4--16.2\,$\mu$m & $\leq$8  & 6--25 \\
14.5--21.0\,$\mu$m & $\leq$28 & 33--72 \\
19.0--36.5\,$\mu$m & $\leq$30 & 47--113 \\
\hline
\end{tabular}
\end{table}
%%%%%%%%%%%%%%%%%%%%%%%%%%%%%%%%%%%%%%%%%%%%%%%%%%%%%%%%%%%%%%%%%%

MPFIT does not compute upper limits for spectral features it deems 
unnecessary to use in a fit, nor does it check whether features 
that it does fit constitute true detections. Upon completion of a
CHUNKFIT model we therefore impose upper limits on a spectral 
feature if the height of that feature is less than three times 
the RMS noise divided by the square root of the 
number of spectral data points within the full width of the feature 
at 70\% of the peak flux ({\em FW70}). The latter factor
lowers the threshold for PAH features to be deemed detected, 
as the FW70 of a PAH feature is higher than for an emission line.
A complication in this effort arises from the fact that for most CASSIS
spectra the formal RMS uncertainty of the spectra is overestimated 
by up to a factor 3 when compared to the dispersion of flux values 
between nearby pixels in continuum regions of the spectrum. We
therefore base the criterion for when and how to impose upper limits 
on local rather than on formal noise measurements.

In the subsections below we describe the CHUNKFIT models for six 
rest frame ranges. The maximum number of free parameters and
the degrees of freedom in the models are listed in Table\,\ref{tab:degreesoffreedom}.
None of the features fitted are extinction 
corrected in the process.

\subsection{Features in the 5.39--7.25\,$\mu$m range} \label{sec:partial57}

For the 5.39--7.25\,$\mu$m spectral range our CHUNKFIT model comprises 
the following spectral components:
\begin{itemize}
\item A 6\,$\mu$m water ice absorption feature, derived
from the spectrum of NGC\,4418 \cite[see Appendix\,\ref{sec:appendix-a} for details;][]{spoon01}.
\item A 6.85\,$\mu$m aliphatic C-H deformation mode absorption feature, derived
from the spectrum of NGC\,4418 \cite[see Appendix\,\ref{sec:appendix-a} for details;][]{spoon01}
 \item The absorption-corrected 5.39--7.25\,$\mu$m continuum 
C$_{\rm cor}$($\lambda$),
represented by a third order polynomial function, and related to the local
absorbed continuum C$_{\rm abs}$($\lambda$) by 
C$_{\rm abs}$($\lambda$)=C$_{\rm cor}$($\lambda$) e$^{-\tau_{\rm ice}(\lambda)-\tau_{\rm 6.85}(\lambda)}$.
where $\tau_{\rm ice}$ is the optical depth in the 6.0\,$\mu$m water
ice feature and $\tau_{\rm 6.85}$ is the optical depth in the
aliphatic C-H deformation mode absorption feature.
\item PAH emission bands at 5.68, 6.04, and 6.22\,$\mu$m (PAH57, PAH60
and PAH62, hereafter). While the former bands are best represented by  
Gaussian profiles, the asymmetric profile of the 6.2\,$\mu$m feature 
\cite[e.g.][]{hony00} is best fitted by the asymmetric shape of a 
Pearson type-IV distribution (see Appendix\,\ref{sec:appendix-b}).
\item Emission lines of H$_{2}$ 0--0 S(7) at 5.51\,$\mu$m;
of H$_{2}$ 0--0 S(5) at 6.91\,$\mu$m; and of [Ar {\sc ii}] at 6.99\,$\mu$m,
all modeled by Gaussian profiles.
\end{itemize}

\begin{figure}[t!]
\includegraphics[scale=0.565]{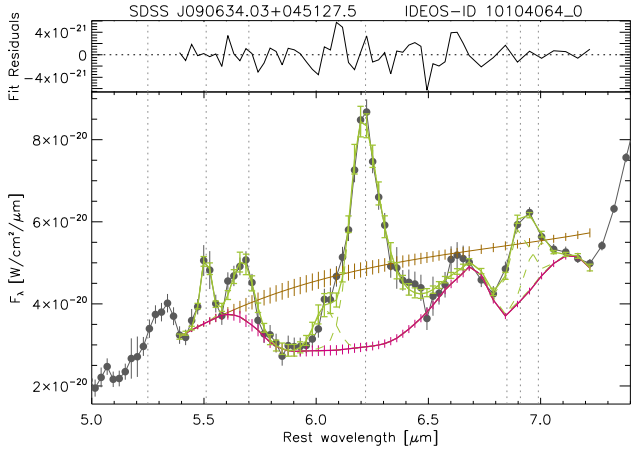}
\caption{Example of a fit to the 5.39--7.25\,$\mu$m spectrum of an
IDEOS source. The observed spectrum is shown as black connected 
dots, the fit to the data in green. Individual emission features 
are shown as green dashed curves on top of the local
absorbed continuum (pink). The absorption-corrected continuum inferred
from the fit is shown in brown. Vertical dotted lines denote commonly
detected features: 5.27$\mu$m PAH, 
5.51$\mu$m H$_2$ S(7), 5.70$\mu$m PAH, 6.22$\mu$m PAH, 
6.85 \& 7.25$\mu$m CH$_3$/CH$_2$ deformation modes, 
6.91$\mu$m H$_2$ S(5), 6.99$\mu$m [Ar {\sc ii}].
\label{fig:pah62_mpfit}}
\end{figure}

In our model, only the continuum component is subject to attenuation 
by water ice (6\,$\mu$m) and/or aliphatic hydrocarbons (6.85\,$\mu$m). 
This assumption is in line with spectral decomposition models 
for buried galactic nuclei \cite[e.g.][]{veilleux09} in which the 
emission lines and PAH features are assumed to have a circum-nuclear 
origin, and are hence unaffected by water ice and aliphatic 
hydrocarbon absorption occuring within the nucleus.
As a consequence, the equivalent width (EQW) of the PAH62 
feature is computed using the absorption-corrected rather than 
the absorbed 6.22\,$\mu$m continuum.
An example fit for a strongly ice and hydrocarbon absorbed
spectrum is shown in Fig.\,\ref{fig:pah62_mpfit}. Note 
that CHUNKFIT's measured optical depths will be lower limits to the
true optical depths in sources with strong circumnuclear PAH
emission. The emission of overlapping wings of the PAH features
will fill in the absorption features in these sources.

%%%%%%%%%%%%%%%%%%%%%%%%%%%%%%%%%%%%%%%%%%%%%%%%%%%%%%%%%%%%%%%%%%
%%%%%%%%%%%%%%%%%%%%%%%%%%%%%%%%%%%%%%%%%%%%%%%%%%%%%%%%%%%%%%%%%%

The number of free parameters compared to the number of spectral data
points in the 5.39--7.25\,$\mu$m range is large. To avoid reducing the
degrees of freedom of our model to below zero, the following lines 
are not included. [Mg {\sc vii}] at 5.50\,$\mu$m 
and [Mg {\sc v}] at 5.61\,$\mu$m. These magnesium lines are not commonly
detected as it takes 109\,eV and 186\,eV, respectively, to create Mg$^{4+}$ 
and Mg$^{6+}$. This means that these lines only arise in AGN and will
only be discernable in high S/N spectra. On top of that [Mg {\sc vii}] is 
severely blended with H$_2$ 0--0 S(7), which is much more commonly
detected in our sample. Adding [Mg {\sc v}] at 5.61\,$\mu$m
to our model would overcrowd the 5.4--6.0\,$\mu$m fit range,
effectively creating a three line blend. 
We further do not include the 6.11\,$\mu$m H$_2$ 0--0 S(6) 
line, which is generally faint.

To avoid false detections of PAH57 and PAH60 in low S/N spectra, 
we restrict the strength of the PAH57 and PAH60 bands to an 
empirically\footnote{Based on PAH57, PAH60 and PAH62 observations in a
sample of sources with PAH62 detections $\geq$25$\sigma$.} 
derived limit of 15\% of the much stronger PAH62 band.
To ensure realistic continuum fits also in low S/N spectra, we reduce 
the order of the polynomial continuum function from three to one in 
spectra that exhibit a S/N$<$6 at 6.6\,$\mu$m in the continuum.

A (moderately) deep depression between the PAH57 and PAH62 
features is a tell-tale signature of 6\,$\mu$m water ice absorption
\citep[e.g. Fig.\,\ref{fig:pah62_mpfit};][]{spoon01}.
Careful inspection of individual fits to sources with weak to moderate 
depressions between the PAH57 and PAH62 features revealed that 
CHUNKFIT in most cases refrained from fitting a water ice
feature, and instead prefers an unphysical concave continuum. 
In these cases (2/3 of the sources that we have identified ice in
by visual inspection) we intervene by not allowing CHUNKFIT to drop 
the water ice feature from consideration. In 10\% of these cases 
this is not enough, requiring another incentive to fit water ice: 
lowering the polynomial order of the ice-corrected continuum 
C$_{\rm cor}$ from three to one. Note that for sources for which
fitting a water ice feature is a subjective choice, as it is for
weak absorptions, this is reflected in the uncertainty in the 
fitted absorbed continuum C$_{\rm abs}$, ice-corrected continuum 
C$_{\rm cor}$, PAH62 flux, and PAH62 equivalent width, which are 
all elevated. In the IDEOS portal (see Sect.\,\ref{sec:portal})
we show examples of the effect of these interventions on the 
measured PAH62 quantities.

Finally, a point of caution. Measurements for the partially blended 
H$_2$ 0--0 S(5) and [Ar {\sc ii}] lines in sources with a strong
aliphatic hydrocarbon absorption band rely on the accuracy of the
adopted absorption profile of the aliphatic hydrocarbon absorption
band (see Appendix \ref{sec:appendix-a}). This is only partially 
reflected in the uncertainty of the fitted line fluxes.

\subsection{Features in the 8.7--10.3\,$\mu$m range} \label{sec:partial8710}

\begin{figure}[t]
\includegraphics[scale=0.565]{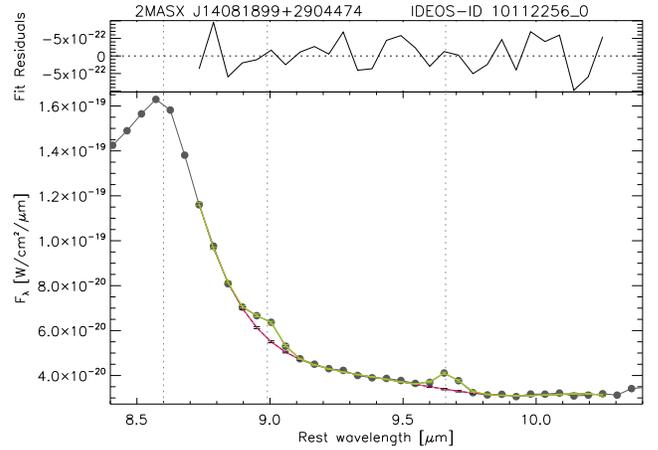}
\caption{Example of a fit to the 8.7--10.3\,$\mu$m spectrum of an
IDEOS source. The observed spectrum is shown in black, the fit 
to the data in green, and the fitted local continuum in pink. 
Vertical dotted lines denote commonly detected features:
8.6$\mu$m PAH, 8.99$\mu$m [Ar {\sc iii}], 9.66$\mu$m H$_2$ S(3).
\label{fig:ar3_h2s3_mpfit}}
\end{figure}

There are two emission lines in the 8.7--10.3\,$\mu$m range which 
cannot be observed in the IRS high-resolution modules in low redshift
sources, [Ar {\sc iii}] at 8.99\,$\mu$m and H$_2$ 0--0
S(3) at 9.66\,$\mu$m. For these we created a separate CHUNKFIT model, 
which comprises the following spectral components:

\begin{itemize}
\item The 8.7--10.3\,$\mu$m continuum, represented by a fifth order
polynomial function. 

\item Emission lines of [Ar {\sc iii}] at 8.99\,$\mu$m and H$_2$ 0--0
S(3) at 9.66\,$\mu$m, represented by Gaussian profiles.
\end{itemize}
An example of a model fit for the 8.7--10.3\,$\mu$m range 
is shown in Fig.\,\ref{fig:ar3_h2s3_mpfit}.

To ensure realistic continuum fits also in low S/N spectra, we lower
the order of the polynomial continuum function from five to two in 
spectra that exhibit a continuum S/N$<$7 in the 9\,$\mu$m range.

%%%%%%%%%%%%%%%%%%%%%%%%%%%%%%%%%%%%%%%%%%%%%%%%%%%%%%%%%%%%%%%%%%
%%%%%%%%%%%%%%%%%%%%%%%%%%%%%%%%%%%%%%%%%%%%%%%%%%%%%%%%%%%%%%%%%%

\subsection{Features in the 9.8--13.5\,$\mu$m range} \label{sec:partial1013}

For fitting the 9.8--13.5\,$\mu$m range we use a two-stage approach. 
First we model the features in the 10.0--12.58\,$\mu$m spectral range 
and then repeat the fitting for an expanded range of
9.8--13.5\,$\mu$m, using results from the 10.0--12.58\,$\mu$m model 
as constraints.

Our CHUNKFIT model for the 10.0--12.58\,$\mu$m range comprises the 
following spectral components:
\begin{itemize}
\item The 10.0--12.58\,$\mu$m continuum, represented by a third order
polynomial function. This continuum may exhibit strong curvature in
spectra that are strongly affected by silicate emission or absorption.
\item PAH emission bands at 10.64, 11.04, 11.25 and 12.00\,$\mu$m
(PAH107, PAH111, PAH112 and PAH120 hereafter). Like the PAH62 feature,
the PAH112 feature is represented by an asymmetric Pearson type-IV 
distribution profile (see Appendix \ref{sec:appendix-b}, to account for 
the asymmetric nature of the feature \cite[e.g.][]{hony00}. The 
other three bands are fitted by Gaussian profiles. 
\item Emission lines of [S {\sc iv}] at 10.51\,$\mu$m and H$_2$ 0--0
  S(2) at 12.28\,$\mu$m, represented by Gaussian profiles.
\end{itemize}

\begin{figure}[t]
\includegraphics[scale=0.565]{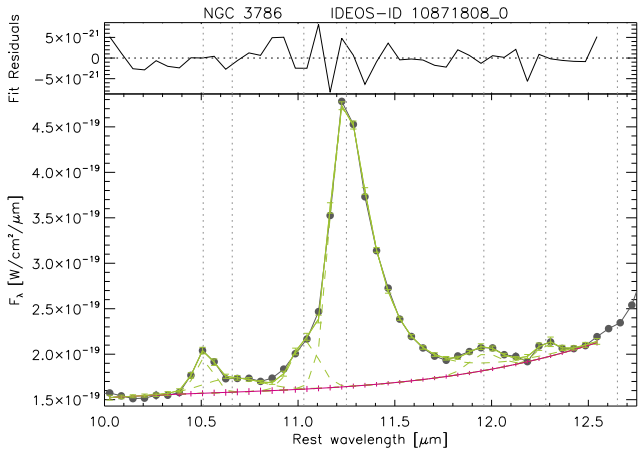}
\caption{Example of a 'stage-1' fit to the 10.0--12.58\,$\mu$m spectrum of an
IDEOS source. The observed spectrum is shown in black, and the fit 
to the data in green. Individual emission features are shown as 
green dashed curves on top of the local continuum (pink). 
Vertical dotted lines denote commonly detected features:
10.51$\mu$m [S {\sc iv}], 10.64, 11.04, 11.29, 11.98$\mu$m PAH, 
and 12.28$\mu$m H$_2$ S(2).
\label{fig:pah11_mpfit}}
\end{figure}

To ensure realistic continuum fits also in low S/N spectra, we lower
the order of the polynomial continuum function from three to one in 
spectra that exhibit a continuum S/N$<$7 in the 10.65--11.95\,$\mu$m 
range. If, however, the spectrum is PAH-dominated (i.e. the
equivalent width of the PAH112 feature is more than 0.6\,$\mu$m), 
the continuum S/N has to drop below 3 before a first order polynomial 
continuum is invoked, because PAH-dominated spectra have intrinsically
weaker continua in the 10.65--11.95\,$\mu$m range than
continuum-dominated spectra. An example model fit is shown in 
Fig.\,\ref{fig:pah11_mpfit}.

%%%%%%%%%%%%%%%%%%%%%%%%%%%%%%%%%%%%%%%%%%%%%%%%%%%%%%%%%%%%%%%%%%
%%%%%%%%%%%%%%%%%%%%%%%%%%%%%%%%%%%%%%%%%%%%%%%%%%%%%%%%%%%%%%%%%%

In the second stage we expand the wavelength coverage of our
CHUNKFIT model to
9.8--13.5\,$\mu$m to include the PAH feature at 12.65\,$\mu$m 
(PAH127) and the [Ne {\sc ii}] line at 12.81\,$\mu$m in our model.
The [Ne {\sc ii}] line is represented by a Gaussian profile and 
the asymmetric PAH127 feature by a Pearson IV profile 
(Appendix \ref{sec:appendix-b}).

Extension of the wavelength coverage requires the continuum 
to be fitted with a higher order polynomial function than for the 
10.0--12.58\,$\mu$m range, requiring a fourth to sixth order 
polynomial function, depending on the depth of the silicate 
feature. Only at low S/N we restrict the polynomial order to 
second order to avoid spectral artefacts and noise features 
to be fitted as if they were true continuum.

\begin{figure}[t]
\includegraphics[scale=0.565]{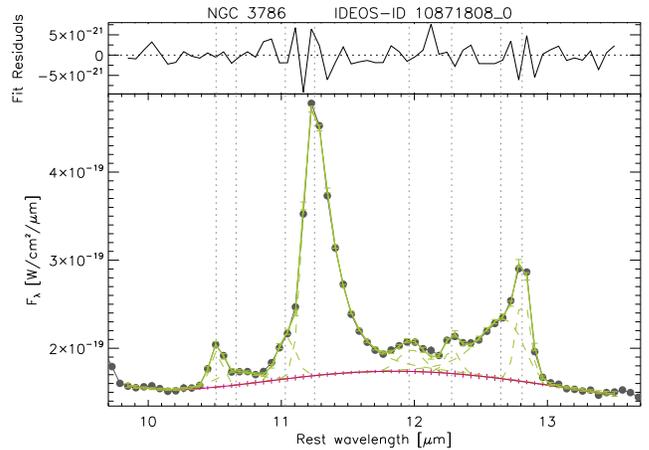}
\caption{Example of a 'stage-2' fit to the 9.8--13.5\,$\mu$m spectrum of an
IDEOS source. The observed spectrum is shown in black, and the fit 
to the data in green. Individual emission features are shown as 
green dashed curves on top of the local continuum (pink). 
Vertical dotted lines denote commonly detected features:
10.51$\mu$m [S {\sc iv}], 10.64, 11.04, 11.29, 11.98, 12.63$\mu$m PAH, 
12.28$\mu$m H$_2$ S(2), and 12.81$\mu$m [Ne {\sc ii}]. 
\label{fig:pah11_pah127_mpfit}}
\end{figure}

To ensure consistency between the two fitting stages,
features that were deemed non-detected in the first stage are
excluded in the second stage. An example model fit is 
shown in Fig.\,\ref{fig:pah11_pah127_mpfit}.

%%%%%%%%%%%%%%%%%%%%%%%%%%%%%%%%%%%%%%%%%%%%%%%%%%%%%%%%%%%%%%%%%%
%%%%%%%%%%%%%%%%%%%%%%%%%%%%%%%%%%%%%%%%%%%%%%%%%%%%%%%%%%%%%%%%%%

As can be clearly seen when comparing Figs.\,\ref{fig:pah11_mpfit}
and \ref{fig:pah11_pah127_mpfit}, the stage-2 continuum more closely
ressembles the spline continuum as intended by \cite{hony01} 
and \cite{vermeij02} than the stage-1 continuum does. Therefore,
barring a lack in spectral coverage beyond 12.58\,$\mu$m, the 
IDEOS observables for the features in this wavelength range are 
obtained from the stage-2 CHUNKFIT model. The only exception is 
the H$_2$ 0--0 S(2) line at 12.28\,$\mu$m, which sometimes is 
fitted more accurately in the 10.0--12.58\,$\mu$m (first stage) 
model, where the blue wing of the PAH127 feature is treated as 
continuum.

To avoid mistaking noise features and artefacts for faint PAH 
emission bands, we bootstrap the strength of the PAH107, PAH111, 
and PAH120 bands in our models to the much stronger PAH112 band. 
The maximum allowed relative strengths are listed in column three 
of Table\,\ref{tab:pahconstraints}. Only for the PAH127 
band, which can be stronger than the PAH112 band \citep{hony01}, 
the ratio exceeds unity. For the same reason we also restrict 
the widths of the PAH bands. The allowed FWHM ranges are shown 
in column four of Table\,\ref{tab:pahconstraints}.

Finally, our 9.8--13.5 and 10.0--12.58\,$\mu$m models do not include 
the hydrogen Hu-$\alpha$ recombination line at 12.37\,$\mu$m as it is 
too weak to be detected at the resolving power of IRS SL and LL.

%%%%%%%%%%%%%%%%%%%%%%%%%%%%%%%%%%%%%%%%%%%%%%%%%%%%%%%%%%%%%%%%%%
\begin{table}
%\centering
\caption{Constraints on 9.8--13.5\,$\mu$m PAH features} \label{tab:pahconstraints}
\begin{tabular}{lrrr}
\hline
\hline
Feature & $\lambda_0$ & F(PAH)/F(PAH112) & FWHM range \\
        &  ($\mu$m)   &                  &   ($\mu$m) \\
\hline
PAH107 & 10.64 & $\leq$0.10 & 0.141 -- 0.352 \\
PAH111 & 11.04 & $\leq$0.13 & 0.118 -- 0.165 \\
PAH112 & 11.29 & 1.0        & 0.241 -- 0.287 \\
PAH120 & 11.98 & $\leq$0.10 & 0.212 \\
PAH127 & 12.63 & $\leq$1.3  & 0.379 -- 0.412 \\
\hline
\end{tabular}
\tablecomments{Constraints were determined in a sample of sources
in which PAH112 was detected at $\geq$25$\sigma$.
Column 3 shows the maximum allowed flux in our model 
of a PAH feature relative to the flux of the PAH112 feature. 
Column 4 shows minimum and maximum FWHM allowed in our model.}
\end{table}
%%%%%%%%%%%%%%%%%%%%%%%%%%%%%%%%%%%%%%%%%%%%%%%%%%%%%%%%%%%%%%%%%%

\subsection{Features in the 13--15.4\,$\mu$m range} \label{sec:partial1315}

Longward of 13\,$\mu$m the most obvious features of interest are the 
[Ne {\sc v}] line at 14.32\,$\mu$m and the [Ne {\sc iii}] line at 
15.56\,$\mu$m. At the spectral resolution of the SL and LL IRS
modules, however, the [Ne {\sc v}] line is strongly blended with 
the [Cl {\sc ii}] line at 14.37\,$\mu$m (0.047\,$\mu$m separation
peak-to-peak at a spectral resolution of 
$\Delta\lambda$=0.06--0.12\,$\mu$m) 
and with the PAH feature at 14.22\,$\mu$m (PAH142) 
\citep{bernard-salas09,perez-beaupuits11}. 
It is therefore only possible to reliably measure the [Ne {\sc v}] line
from our spectra if the PAH142 feature and the [Cl {\sc ii}] 
line are not present.
Given the onset of the 18\,$\mu$m silicate feature around 15\,$\mu$m,
and the higher polynomial order required to fit the curvature of a
silicate-absorbed continuum beyond 15\,$\mu$m, we hoped to avoid 
extending the fitting range beyond 15\,$\mu$m. The number of free 
parameters in our model does, however, force us to extend of model 
to 15.4\,$\mu$m, just shy of the 15.56\,$\mu$m [Ne {\sc iii}] line.

%%%%%%%%%%%%%%%%%%%%%%%%%%%%%%%%%%%%%%%%%%%%%%%%%%%%%%%%%%%%%%%%%%
\begin{figure}[t]
\includegraphics[scale=0.565]{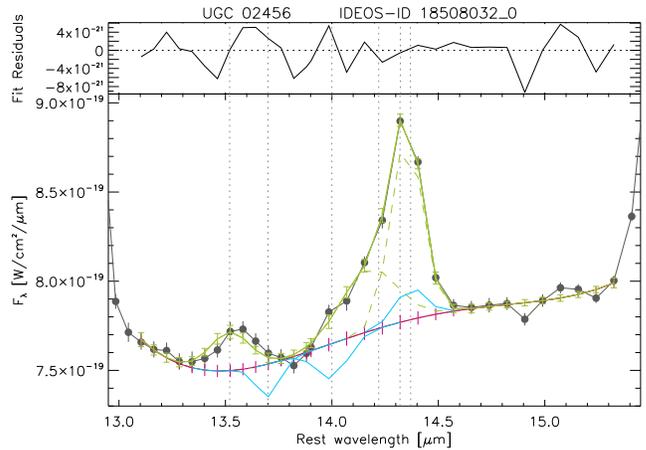}
\caption{Example of a fit to the 13.05--15.40\,$\mu$m spectrum of an
IDEOS source. The observed spectrum is shown in black, and the fit 
to the data in green. Individual emission features are shown as 
green dashed curves on top of the local continuum (pink). Upper 
limits for undetected model features are overplotted in cyan. 
Vertical lines denote commonly detected features:
13.52 \& 14.22$\mu$m PAH, 13.77$\mu$m C$_2$H$_2$, 14.02$\mu$m HCN, 
14.32$\mu$m [Ne {\sc v}], 14.37$\mu$m [Cl {\sc ii}]. 
\label{fig:ne5cl2_mpfit}}
\end{figure}
%%%%%%%%%%%%%%%%%%%%%%%%%%%%%%%%%%%%%%%%%%%%%%%%%%%%%%%%%%%%%%%%%%

Our CHUNKFIT model for this range comprises the following spectral 
components:
\begin{itemize}
\item The 13--15.4\,$\mu$m continuum, represented by a fourth order
polynomial function. This continuum traces the onset of the 18\,$\mu$m 
amorphous silicate feature, which may be in emission or absorption.
In spectra with continuum S/N$<$13 at 14\,$\mu$m we reduce the order 
of the polynomial continuum to second order.
\item Emission lines of [Ne {\sc V}] at 14.32\,$\mu$m, and 
[Cl {\sc ii}] at 14.37\,$\mu$m, represented by Gaussian profiles. 
\item PAH emission bands at 13.52, and 14.22\,$\mu$m, represented 
by Gaussian profiles.
\item Absorption profiles of C$_2$H$_2$ at 13.7\,$\mu$m and HCN gas 
at 14.0\,$\mu$m \citep{lahuis07}, represented by Gaussian profiles. 
We do not attribute a physical meaning to this choice of profile, nor 
to the measured optical depth. See \cite{lahuis07} for a proper 
model fit.
\end{itemize}

High-resolution Spitzer-IRS studies of starburst galaxies have shown
the [Cl {\sc ii}] line to be faint, and therefore hardly detectable 
in SL and LL spectra \citep{bernard-salas09}.
Therefore, if our initial model fit indicates both [Cl {\sc ii}] and 
[Ne {\sc v}] to be detected, we omit [Cl {\sc ii}] from the fitting 
parameters if [Ne {\sc v}] represents at least 30\% of the flux in 
the line blend, and then repeat the fit.
Likewise, in case PAH142 is blended\footnote{For a select 
few sources the 14.32\,$\mu$m [Ne {\sc v}] line can have a blue 
outflow wing extending to velocities of -3000 km/s or more 
\cite[e.g. IRAS\,13451+1232]{spoon09a,spoon09b}.
This would cause the [Ne {\sc v}] line profile to mimic the shape 
of a PAH14 band, with possible misidentification as a result. The 
number of sources with strong outflows in [Ne {\sc v}] is really 
small. All have been identified and specially processed.}
with [Ne {\sc v}], we replace the detection of [Ne {\sc v}] with an 
upper limit in case the flux ratio 
of [Ne {\sc v}]/PAH142 is below a certain value. For spectra with 
continuum S/N$<$50 we choose 0.83, and for spectra with continuum
S/N$>$50 it is 0.6. 
The above steps ensure that detections of 14.32\,$\mu$m [Ne {\sc v}] in 
IDEOS are true detections and not the result of misassignment of flux
in a line blend. Of course, any redshift error on the order
of 0.05$\mu$m ($\sim$1000\,km/s) will result in all line
flux to be assigned to just one of the lines in the blend.

\subsection{Features in the 14.5--21\,$\mu$m range} \label{sec:partial1421}

The 14.4--21\,$\mu$m spectral range contains three diagnostically
important emission lines, [Ne {\sc iii}] at 15.56\,$\mu$m, H$_2$ 0--0
S(1) at 17.03\,$\mu$m, and [S {\sc iii}] at 18.71\,$\mu$m. Both the 
[Ne {\sc iii}] and the [S {\sc iii}] line are well-separated from other
spectral features, unlike the H$_2$ 0--0 S(1) line, which, in star
forming galaxies, sits atop a conglomerate of PAH features, the
so-called 17\,$\mu$m PAH complex \citep{smith07}. The 14.4--21\,$\mu$m 
spectral range is also home to the O--Si--O bending vibration 
mode of amorphous silicates, which spans the entire range, 
and to the 16.1 and 18.6 \,$\mu$m absorption features due to
crystalline silicates \citep{spoon06}.

Our CHUNKFIT model for this range comprises the following spectral components:
\begin{itemize}
\item The 14.5--21\,$\mu$m continuum, represented by a fourth order
polynomial function. This continuum traces the smooth shape of the 
18\,$\mu$m amorphous silicate feature, which may be either in emission 
or absorption.
\item Emission lines of [Ne {\sc iii}] at 15.56\,$\mu$m, H$_2$
0--0  S(1) at 17.03\,$\mu$m, and [S {\sc iii}] at 18.71\,$\mu$m,
represented by a Gaussian profile. 
\item PAH emission bands at 15.9, 16.45, and 17.4\,$\mu$m, and a PAH
plateau centered at 17.0\,$\mu$m, all represented by Gaussian
profiles.
\item Absorption profiles of two crystalline silicate bands, 
at 16.1 and 18.6\,$\mu$m, fitted together
(see Appendix\,\ref{sec:appendix-c} and Fig.\,\ref{fig:crystsil}). 
\end{itemize}
The 16.1 and 18.6\,$\mu$m crystalline bands are only fitted in spectra 
that exhibit significant silicate absorption, S$_{\rm sil}<-0.5$. This 
avoids unphysical solutions in spectra dominated by PAH emission bands.
The bands are not fitted independent from each other as to limit the
number of free parameters, to avoid unphysical fits to the amorphous
silicate profile, and in recognition that their strengths appear 
correlated in deeply obscured sources \citep{spoon06}.

\begin{figure}[t!]
\begin{tabular}{c}
\includegraphics[scale=0.565]{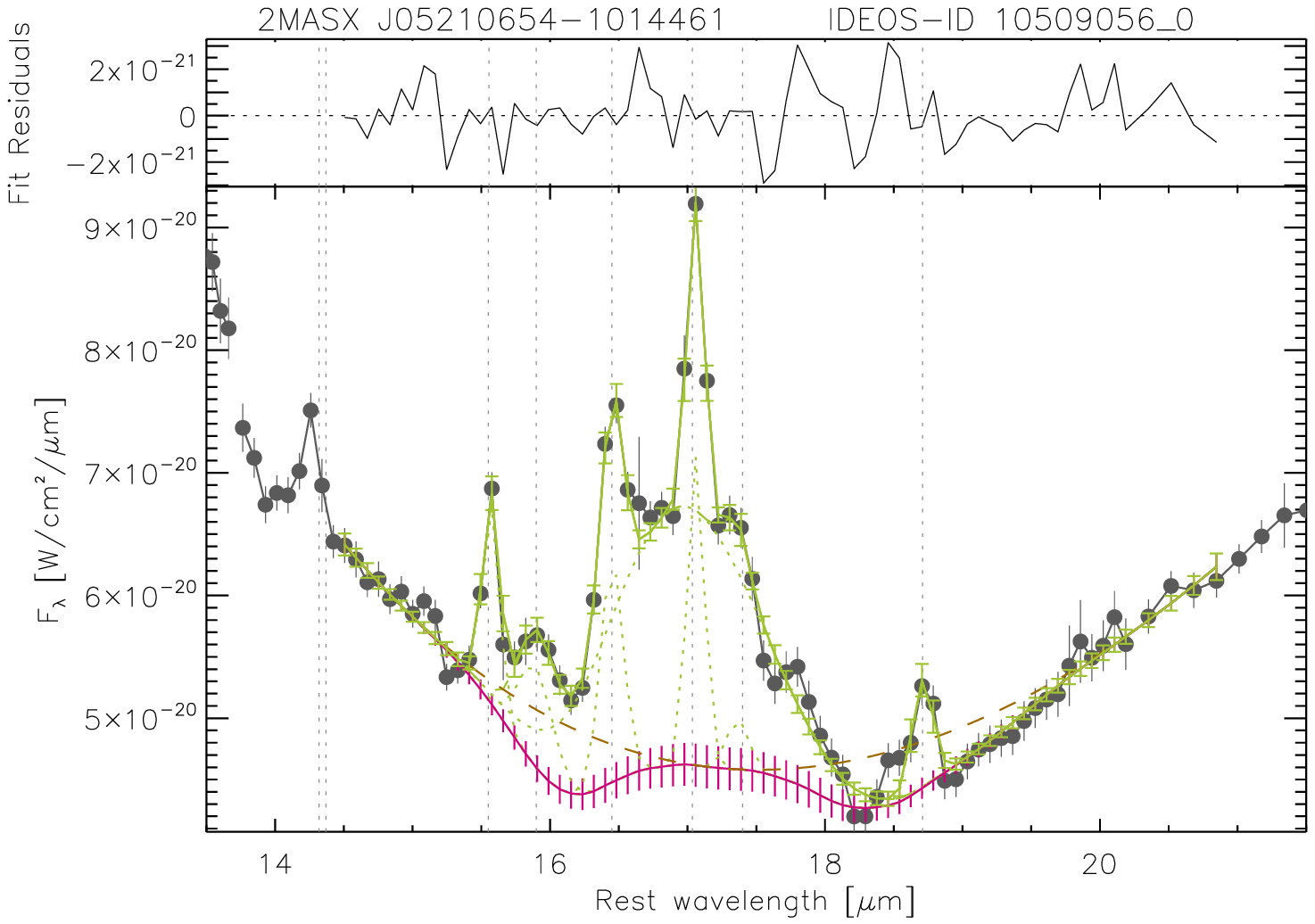}\\
\includegraphics[scale=0.565]{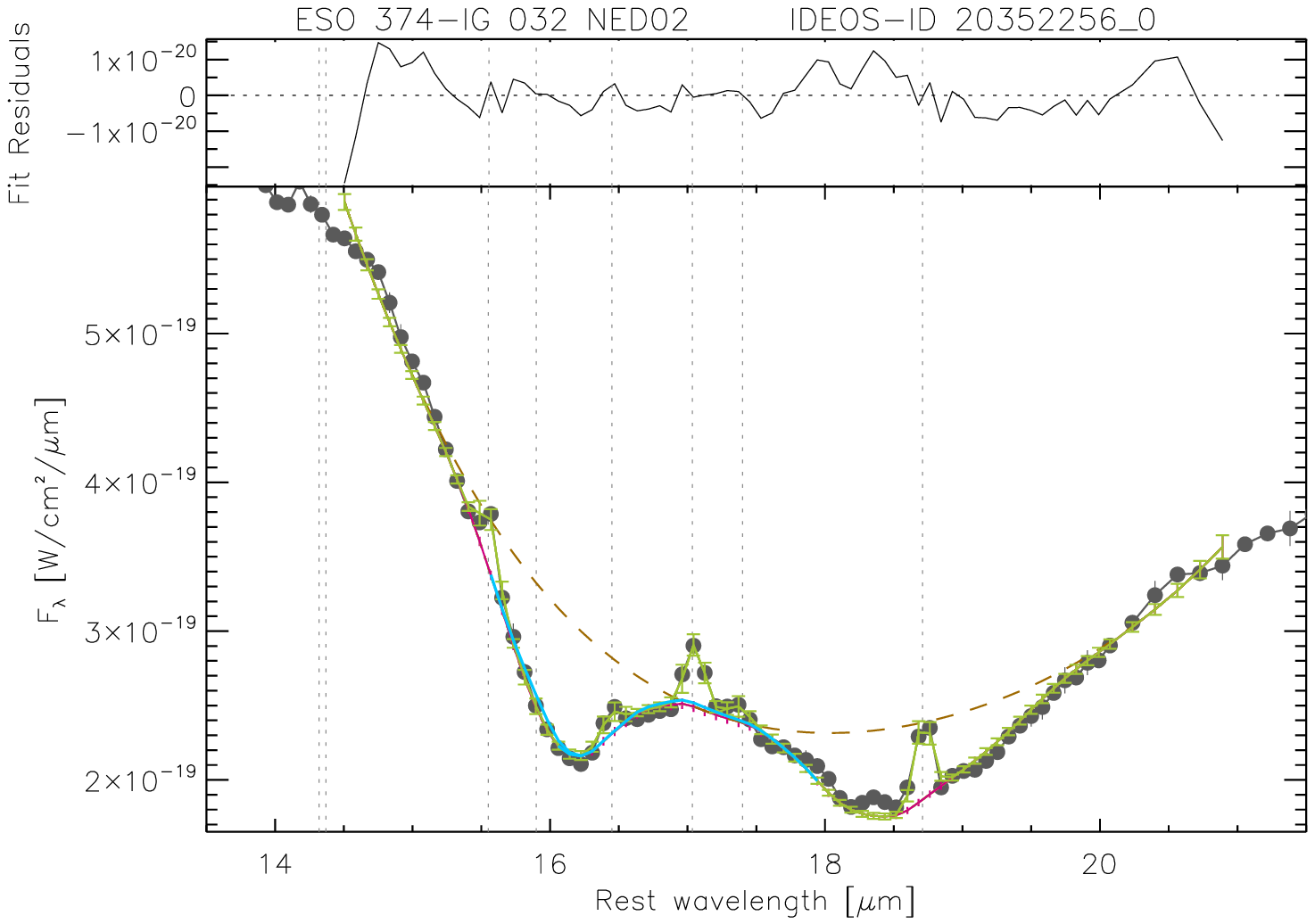}
\end{tabular}
\caption{Fits to the 14.4--21\,$\mu$m spectra of two IDEOS sources. 
In each fit the observed spectrum is shown in black and 
the fit to the data in green. The local continuum is plotted
in pink, and the continuum corrected for crystalline silicate 
absorption in brown.
Upper limits for undetected model features are overplotted in cyan. 
Vertical lines denote commonly detected features:
14.32$\mu$m [Ne {\sc v}], 14.37$\mu$m [Cl {\sc ii}], 15.56$\mu$m
[Ne {\sc iii}], 15.9$\mu$m \& 16.45$\mu$m PAH, 
17.03$\mu$m H$_2$ S(1), 17.40$\mu$m PAH, 18.71$\mu$m [S {\sc iii}].
\label{fig:ne3h2s1s3_mpfit}}
\end{figure}

Fits to the spectra of two dusty star forming nuclei are shown in 
Fig.\,\ref{fig:ne3h2s1s3_mpfit}. Both plots highlight the remarkable
good fit that the crystalline silicate template offers to almost all
spectra of deeply obscured galactic nuclei.

As can be seen in the upper panel of Fig.\,\ref{fig:ne3h2s1s3_mpfit},
in spectra with a strong 17\,$\mu$m PAH complex the accuracy of the
line flux measured for the [Ne {\sc iii}] line depends to some 
extent on the adopted profile for the 15.9\,$\mu$m PAH band. 
Similarly, for spectra with crystalline silicate absorption the 
line flux of the [S {\sc iii}] line is critically dependent on a
good fit to the crystalline-absorbed continuum around the line. Note 
that our CHUNKFIT model does not include the line blend of 
[P {\sc iii}] at 17.885\,$\mu$m 
and [Fe {\sc ii}] at 17.936\,$\mu$m, because these lines are generally
weak \citep{bernard-salas09}, and because these added free parameters
would decrease the degrees of freedom of the model to almost zero,
similar to the model for the 5.39--7.25\,$\mu$m range 
(Sect.\,\ref{sec:partial57}).

In spectra with continuum S/N$<$7 at 17\,$\mu$m we reduce the order 
of the polynomial continuum to third order, while in spectra with deep
silicate absorption features we increase it by one order.
Spectra that do not extend to at least 20.4\,$\mu$m cannot be fitted
with the model described above. For these spectra we perform a fit to
just the 14.4--16.2\,$\mu$m range to obtain a measurement of only the 
[Ne {\sc iii}] line.

%%%%%%%%%%%%%%%%%%%%%%%%%%%%%%%%%%%%%%%%%%%%%%%%%%%%%%%%%%%%%%%%%%
%%%%%%%%%%%%%%%%%%%%%%%%%%%%%%%%%%%%%%%%%%%%%%%%%%%%%%%%%%%%%%%%%%

\subsection{Features in the 19--36.5\,$\mu$m range} \label{sec:partial1936}

The 19--36.5\,$\mu$m spectral range is home to six emission lines of
great diagnostic importance. Two of these are strongly blended in 
low-resolution spectra, [O {\sc iv}]
at 25.89\,$\mu$m and [Fe {\sc ii}] at 25.99\,$\mu$m, at a separation
of only $\sim$1100 km s$^{-1}$ peak to peak. The 19--36.5\,$\mu$m range is
devoid of strong emission and absorption bands from PAHs and amorphous
silicates, but does exhibit features of crystalline silicates, either
in emission or absorption, centered at 23.3, 27.5 and 33.2\,$\mu$m 
\citep[e.g.][]{spoon06}, characteristic of Forsterite, the
magnesium-rich end member of the Olivines.

\begin{figure}[t!]
\includegraphics[scale=0.565]{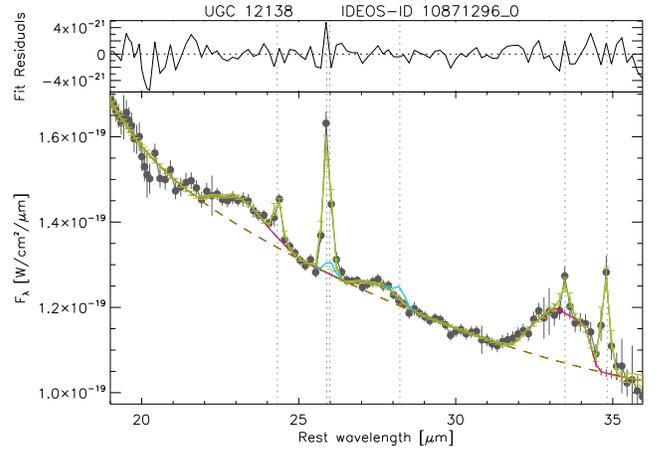}
\caption{Example of a fit to the 19--35.5\,$\mu$m spectrum of an
IDEOS source. The observed spectrum is shown in black, the fit 
to the data in green. The local continuum is shown in pink, and the 
continuum corrected for crystalline silicate emission in brown. 
Upper limits for undetected lines are shown in cyan.
Vertical lines denote commonly detected features:
24.32$\mu$m [Ne {\sc v}], 25.89$\mu$m [O {\sc iv}], 25.99$\mu$m
[Fe {\sc iii}], 28.22$\mu$m H$_2$ S(0), 33.48$\mu$m [S {\sc iii}],
and 34.82$\mu$m [Si {\sc ii}].
\label{fig:ll1range_mpfit}}
\end{figure}

Our CHUNKFIT model for this wavelength range contains the following 
spectral components:
\begin{itemize}
\item The 19--36.5\,$\mu$m continuum, represented by the following
polynomial continuum in ln($\lambda/\lambda_0$) space: 
F($\lambda$) = $g$\ exp[$a$\ (ln\ [$\lambda/\lambda_0$]) + $b$\ (ln\ [$\lambda/\lambda_0$])$^2$ + $c$\ ({\rm ln}\
[$\lambda/\lambda_0$])$^3$ + $d$\ (ln\ [$\lambda/\lambda_0$])$^4$],
where $\lambda_0$=25\,$\mu$m. 
If the spectrum does not extend to at least 29.3\,$\mu$m, d=0.
\item Emission lines of [Ne {\sc v}] at 24.32\,$\mu$m, [O {\sc iv}] 
at 25.89\,$\mu$m, [Fe {\sc ii}] at 25.99\,$\mu$m, H$_2$ 0--0
S(0) at 28.22\,$\mu$m, [S {\sc iii}] at 33.48\,$\mu$m, [Si {\sc ii}] 
at 34.82\,$\mu$m, and [Ne {\sc iii}] at 36.01\,$\mu$m, all 
represented by Gaussian profiles.
\item Emission/absorption profiles of Forsterite, which has features 
at 23.2, 27.6, and 33.2\,$\mu$m (see Appendix\,\ref{sec:appendix-c} and 
Fig.\,\ref{fig:crystsil}).
\end{itemize}
An example model fit is shown in Fig.\,\ref{fig:ll1range_mpfit} for a Seyfert
galaxy exhibiting high-ionization lines and crystalline silicates in emission.

Given the small velocity separation of the [O {\sc iv}] and [Fe {\sc ii}] 
lines (1150 km s$^{-1}$; spectral resolution 3,000--4,000 km s$^{-1}$), 
a small error in the wavelength calibration or in the redshift of 
the source may lead to an incorrect distribution of the line flux
between the two lines in the blend. Because of this, visual inspection 
of the model fit (Fig.\,\ref{fig:ll1range_mpfit}) is advised before 
using the computed [O {\sc iv}] and [Fe {\sc ii}] line fluxes.
Plots showing these model fits are available from the IDEOS portal
(Sect.\,\ref{sec:portal}).

%%%%%%%%%%%%%%%%%%%%%%%%%%%%%%%%%%%%%%%%%%%%%%%%%%%%%%%%%%%%%%%%%%
%%%%%%%%%%%%%%%%%%%%%%%%%%%%%%%%%%%%%%%%%%%%%%%%%%%%%%%%%%%%%%%%%%

\subsection{Upper and lower limits for equivalent widths of PAH features} \label{sec:lowerlimits}

For three of the PAH features fitted by our CHUNKFIT models, PAH62,
PAH11, and PAH127, we not only compute integrated fluxes but also 
equivalent widths. Below we describe how upper limits to the 
PAH flux and/or continuum flux affect the equivalent width that
we report.

In case a PAH feature is not (significantly) detected
(F$_{\rm PAH}$$<$3$\sigma$), also its equivalent width becomes 
an upper limit. 
If, instead, the underlying continuum, C, is not significantly 
detected (C$<$3$\sigma$), the equivalent width becomes a lower limit. 
If both the local continuum and the PAH flux are not significantly 
detected, the equivalent width will be reported as undefined.

\subsection{Compatibility of PAH flux measurements} \label{sec:pahcompatibility}

PAH flux measurements resulting from the CHUNKFIT models
presented in this section are compatible with PAH flux measurements 
from other methods in which the PAH feature flux is defined as 
the flux protruding above the local continuum. PAH profiles defined 
this way are approximately gaussian in shape at the resolving power 
of the IRS low-resolution modules.

In contrast, PAH flux and equivalent width measurements resulting 
from either the {\it Lorentzian} \citep{boulanger98} or the 
{\it Drude} \citep{smith07} PAH spectral decomposition method are not 
compatible with the aforementioned gaussian PAH profile assumption,
as Lorentzian and Drude shaped profiles  
display far broader wings than gaussian profiles do, resulting in a PAH
``continuum'' underneath the local/apparent continuum. PAH flux and equivalent
width measurements assuming Drude shape PAH profiles, resulting from
spectral decomposition, are presented in Sect.\,\ref{sec:seddecomp}.
A comparison of the diagnostic powers of both kinds of PAH profile
measurements is presented by \cite{galliano08b} and show the trends
to be the same.

\section{SED decomposition} \label{sec:seddecomp}

We have performed spectral decomposition on all IDEOS SEDs with 
sufficient coverage of the main PAH complex (6--12\,$\mu$m) 
to provide PAH flux and equivalent width measurements using the 
Drude profile PAH model. 
This provides access to a different set of PAH diagnostics in 
the literature than available for PAH measurements obtained 
using the gaussian PAH profile model (Sect.\,\ref{sec:sedfitting}).
For a comparison of both PAH models see \cite{galliano08b}.

With the goal of avoiding unphysical decomposition solutions
the implementation of the spectral decomposition differs depending
on the nature of the mid-IR SED. For spectra showing the signatures 
of a buried source, i.e. ice, aliphatic hydrocarbon, and silicate 
absorption features \cite[mid-IR classes 2A/B, 3A/B;][]{spoon07},
we used QUESTFIT \citep{veilleux09}, while for the remaining sources 
we used a modified version of PAHFIT \citep{smith07}. 
Both decomposition tools employ the same PAH profile shapes, thereby 
ensuring intercomparability of the PAH measurements between PAHFIT 
and QUESTFIT. Below we sketch the two decomposition methods in 
more detail.

\subsection{SED decomposition using PAHFIT} \label{sec:pahfit}

We have used a modified version of PAHFIT \citep{gallimore10} to
measure the fluxes and equivalent widths of the main PAH emission
bands in all but the most strongly obscured galaxies in the IDEOS
sample. 
PAHFIT \citep{smith07} uses MPFIT to find the best fitting combination 
of dust continua, stellar continuum, PAH emission bands, emission
lines, and dust extinction for a given Spitzer-IRS low-resolution 
spectrum. The 24 PAH features in PAHFIT are represented by Drude 
profiles, which, given their close spacing and broad wings, produce 
a PAH ``continuum'' underneath their peaks.

The modifications to PAHFIT made by \cite{gallimore10} consist of
a) excluding PAH and emission line features from extinction effects;
b) adding an optically thin warm dust emission component;
c) adding fine-structure emission lines of [Ne {\sc v}] and $[$Ne {\sc vi}$]$;
d) widening the choice of extinction laws to include the cold dust
model of \cite{ossenkopf92}, in addition to \cite{chiar06}
and an extinction law based on the silicate profile of \cite{kemper04}.

To fit the spectra of a sample as diverse as the IDEOS sample we 
made further modifications to the fitting code. We removed 
the faint H$_2$ 0--0 S(4) and S(6) lines from the model.
We raised the lowest 
allowed temperature for the optically thin warm dust emission 
component from 100 to 200\,K to prevent an 18\,$\mu$m silicate 
emission peak to be fitted by a cool dust black body component that
peaks around 18\,$\mu$m.
We also removed from each individual PAHFIT model any emission line 
that was not previously detected as part of the CHUNKFIT fitting 
effort (Sect.\,\ref{sec:sedfitting}). This reduces the number of 
free parameters at risk to be fitted to noise features or artefacts 
in low S/N spectra. Finally, we updated the spectral resolution
table to reflect our adopted values (Table\,\ref{tab:resolvingpower}).

Depending on the sign of the silicate strength of 
the spectrum (Sect.\,\ref{sec:silstrength}), PAHFIT
has either been run with the optically thin dust emission component
switched on, and the extinction on the star light and dust continua
switched off, or the opposite. This avoids run-away situations of
silicate emission and absorption components reaching unphysical levels 
to cancel each other out in otherwise featureless spectra.
Given the large differences in observed silicate emission profiles, 
for spectra with positive silicate strength we ran three PAHFIT 
models, each using a different choice of extinction law (see above)
to represent the silicate emission features. We then chose the best 
fitting model based on the reduced $\chi^2$ of the fit. 
For silicate absorption spectra, we only used two of our 
extinction laws: \cite{chiar06} and \cite{ossenkopf92}. 
Other than this, PAHFIT parameters were not tweaked to optimize 
fit results.

Given the large number of free parameters in any PAHFIT model, and the
absence of constraints on the ratios between the 24 PAH components, 
there is nothing preventing PAHFIT from finding a best fitting 
solution in an unphysical part of parameter space. While this will
not happen for PAH-dominated spectra, which PAHFIT was designed
to fit, there is no straight-forward automated iterative way 
to identify unphysical fit results and then improve upon them.
We therefore deem the PAH measurements from the CHUNKFIT fitting in 
Sect.\,\ref{sec:sedfitting} more robust, and provide
the PAHFIT results only as a service.

\subsection{SED decomposition using QUESTFIT} \label{sec:questfit}

Because of the limitations of PAHFIT stated above, IDEOS spectra 
dominated by strong absorption features of silicates,
ices and aliphatic hydrocarbons are not suitable candidates for
spectral decomposition by PAHFIT.  
QUESTFIT\footnote{The source code of QUESTFIT can be downloaded from 
GitHub: https://github.com/drupke/QUESTFIT} 
\citep{rupke21,veilleux09,schweitzer08}, on the other hand, is tailored to the
decomposition of deeply obscured sources, by including absorption 
by ices, aliphatic hydrocarbons, and crystalline silicates in 
the extinction model, and by limiting the number of 
free parameters governing the PAH contribution to just two: one 
per noise-free PAH emission spectrum included in QUESTFIT.
We have therefore used QUESTFIT to measure the
main PAH emission features of the almost 200 sources 
with mid-IR spectral classifications 2A/B, 3A/B 
\cite[][see also Sect.\,\ref{sec:spoondiagram}]{spoon07} that 
are classified as buried sources. 

Since the two noise-free PAH emission spectra used in QUESTFIT 
originate from PAHFIT \cite[their templates 3 and 4;][]{smith07}, 
the PAH fluxes can be decomposed into Drude profiles, fully 
comparable with those computed for non-buried sources by PAHFIT. 
Note that emission lines are not fitted by QUESTFIT and are therefore
masked out in the fitting process. QUESTFIT assumes all the extinction 
to occur in the nuclear continuum source, and the extinction in the 
circumnuclear star forming regions to be negligible \citep{veilleux09}.
This is reflected in the reported PAH fluxes and equivalent widths.

\subsection{Comparison of QUESTFIT and CHUNKFIT in the 
5.5--8\,$\mu$m range}

\begin{figure}[t]
\begin{center}
\includegraphics[scale=0.565]{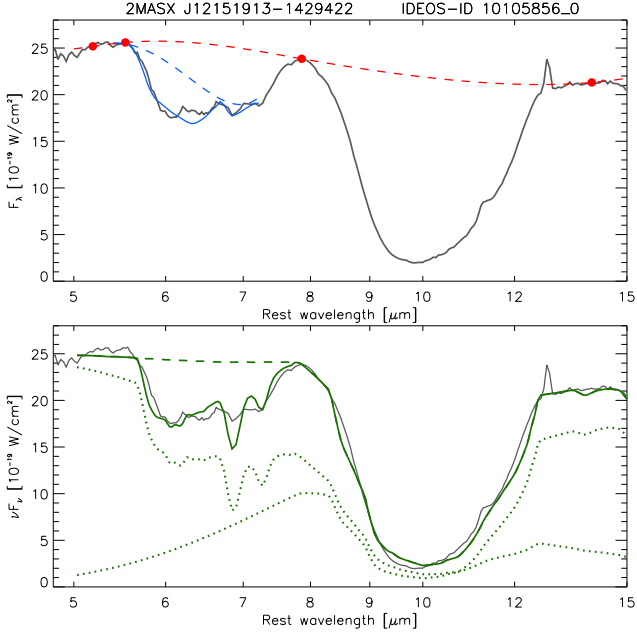}
\end{center}
\caption{Comparison of fits to the 5.5--8\,$\mu$m absorption
features in the ULIRG IRAS F12127--1412NE. The upper panel shows 
the spline-interpolated local continuum in red 
(see Sect.\,\ref{sec:underlyingcont}), and our CHUNKFIT model 
for the 5.37--7.25\,$\mu$m range (see Sect.\,\ref{sec:partial57}) 
in blue. The 5.7--7.25\,$\mu$m inferred continuum, corrected 
for ice and hydrocarbon absorption (see Sect.\,\ref{sec:partial57}), 
is shown as a blue dashed line. Clearly, the two dashed continua
are not in agreement. The lower
panel shows the QUESTFIT spectral decomposition of the continuum
(see Sect.\,\ref{sec:questfit}) in green. The best fit to the 
5.5--8\,$\mu$m absorption features requires one ice and hydrocarbon
absorbed component and one that lacks these features (dotted lines). 
The dashed green line is the implied absorption-corrected continuum, 
and is similar to the red dashed line in the upper panel.
\label{fig:icefits}}
\end{figure}

For some deeply obscured sources, like the ULIRG IRAS\,F12127--1412NE, 
we find substantial differences between the spectral decomposition results 
of QUESTFIT and the CHUNKFIT model for the 5.39--7.25\,$\mu$m 
range (Sect.\,\ref{sec:partial57}). As can be seen in the top panel 
of Fig.\,\ref{fig:icefits}, the continuum corrected for water ice
and amorphous hydrocarbon absorption (the blue dashed line) lies 
significantly below the spline interpolated continuum (the red
dashed line). The disparity appears to widen with increasing 
5.5-8\,$\mu$m wavelength. In contrast, the continuum components 
of the QUESTFIT spectral decomposition (bottom panel of 
Fig.\,\ref{fig:icefits}) do provide a good overall fit to the 
5.5--8\,$\mu$m range (except for the depth of the 6.85 and 
7.25\,$\mu$m features; see below).

The striking disparity cannot be explained by differences in 
the profiles of the 5.5--8\,$\mu$m absorption templates as these
are not that different from each other (compare the optical depth
profiles of NGC4418 and IRAS\,08572+3915 in Fig.\,\ref{fig:5-8um_bands}).
Instead, we attribute the superior fit by QUESTFIT to a substantial
contribution from a steeply rising continuum component devoid of
ice and hydrocarbon absorption, and less affected by silicate
absorption (bottom panel of Fig.\,\ref{fig:icefits}). Could this 
be the spectral signature of a secondary nucleus? Or the effect 
of a much thinner, less frosty cocoon across part of the central 
source -- a hole, if you will?

Inspection of the QUESTFIT result in the bottom panel of 
Fig.\,\ref{fig:icefits} further shows that the QUESTFIT model
could be improved by fitting the aliphatic hydrocarbon component
(the 6.85 and 7.25\,$\mu$m features) separate from the rest of 
the adopted absorption profile, as is done in the CHUNKFIT model.

Further examples of galaxies with similarly unusual 5.5--8\,$\mu$m 
optical depth profiles are IRAS\,00188--0856, F10398+3247, 
10485--1447, 13045+2353, F16156+0146NW, 17123--6245, 
20100--4156 and 23515--2917. All of these galaxies require 
secondary continuum components in their
QUESTFIT models that could be interpreted as signatures
of less than full coverage of the source by ice and 
hydrocarbon absorptions.

%%%%%%%%%%%%%%%%%%%%%%%%%%%%%%%%%%%%%%%%%%%%%%%%%%%%%%%%%%%%%%%%%%
%%%%%%%%%%%%%%%%%%%%%%%%%%%%%%%%%%%%%%%%%%%%%%%%%%%%%%%%%%%%%%%%%%

\section{Rest frame continuum flux densities} \label{sec:restcont}

Sampling of the rest frame continuum at various near and mid-infrared
wavelengths provides insight into the relative contributions of
stellar photospheric and hot and cool dust emission to the galaxy
spectrum.

We have used the IDEOS spectra to compute rest frame continuum flux
densities at seven feature-free wavelengths in the 3.7 to 30\,$\mu$m
range: at 3.7, 4.2, 5.5, 15.0, 24.0, and 30.0\,$\mu$m.
All but the 3.7 and 4.2\,$\mu$m rest frame fluxes are measured from 
the spectral fits for the wavelength ranges they are part of 
(Sect.\,\ref{sec:sedfitting}). For the remaining
two wavelengths the flux densities were computed in two steps. First, 
the average flux density in a narrow range of a few wavelength
elements around the central wavelength was measured. Then a 1$\sigma$ 
clipping was performed after which the first step was repeated. The 
average flux density and the uncertainty in the mean are the final
products.

\begin{figure*}[t]
\begin{center}
\includegraphics[scale=0.9]{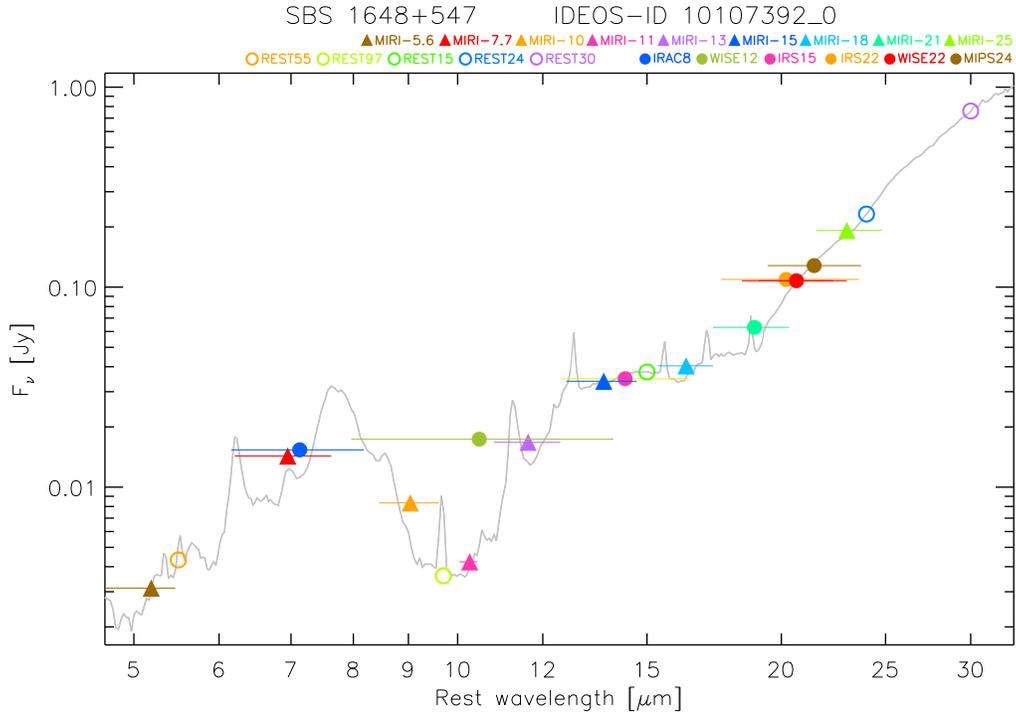}
\end{center}
\caption{Synthetic photometry and rest frame continuum flux densities
for galaxy SBS\,1648+547. The horizontal bars indicate the wavelength
ranges in which 80\% of the filter transmission occurs.
A legend is provided for the use of the 
symbols and color codes to distinguish the various photometric bands
and rest frame flux densities.
\label{fig:photometry}}
\end{figure*}

We include the rest frame continuum flux density measurements in 
Table\,\ref{tab:photometry}, along with two rest frame continuum 
flux density ratios: C(24\,$\mu$m)/C(5.5\,$\mu$m) and
C(30\,$\mu$m)/C(5.5\,$\mu$m). Fig.\,\ref{fig:photometry} shows
all rest frame continuum flux density measurements along with
synthetic photometry for galaxy SBS\,1648+547.

\section{Synthetic photometry} \label{sec:synphot}

The wide spectral range covered by the Spitzer-IRS low resolution
spectra allows us to compute synthetic photometry for our sources
in a large selection of IRAC, Spitzer, WISE and JWST-MIRI photometric bands.

We compute the synthetic photometric flux density as the photon-weighted
mean flux density over the bandpass of the filter, where the
normalization depends on the reference spectrum used for that filter.

For Spitzer-IRAC \citep{reach05} and Spitzer-IRS\footnote{See IRS Instrument Handbook
  version 5.0, Section 4.2.4} the reference spectrum is a power law 
F$_\nu(\nu)\sim\nu^{-1}$. The synthetic photometric flux density is 
thus defined as
\begin{equation}
\langle F_\nu(\lambda_{eff})  \rangle = \frac{\int F_\nu(\nu)
  (\nu_{eff}/\nu) R(\nu) d\nu}{\int
  (\nu_{eff}/\nu)^2 R(\nu) d\nu}
\end{equation}
where $\lambda_{eff}$ is the effective wavelength of the 
photometric band,  defined as:
\begin{equation}
\lambda_{eff}  = c \frac{\int \nu^{-2} R(\nu) d\nu}{\int \nu^{-1}
  R(\nu) d\nu} = \frac{\int R(\lambda) d\lambda}{\int \lambda^{-1}
  R(\lambda) d\lambda}
\end{equation}
and $R(\lambda)$=$R(\nu)$ is the filter transmission profile 
(in units of electrons per photon). For the IRAC-8, IRS-15 and 
IRS-22 photometric bands the effective wavelengths are 7.87, 
15.8 and 22.3\,$\mu$m, respectively.

For Spitzer-MIPS the reference spectrum is a T=10,000\,K black body 
\citep{engelbracht07}, and thus the synthetic photometric flux 
density is defined as
\begin{equation}
\langle F_\nu(\lambda_{\rm 0})  \rangle = \frac{\int F_\nu(\nu) (\nu_{\rm 0}/\nu) R(\nu) d\nu}{\int
  (\nu_{\rm 0}/\nu)  (\frac{B_\nu(\nu,T=10^4\,K)}{B_\nu(\nu_{\rm 0},T=10^4\,K)}) R(\nu) d\nu}
\end{equation}
where $\lambda_{\rm 0}$=c/$\nu_{\rm 0}$ is defined as:
\begin{equation}
\lambda_{\rm 0}  = \frac{\int c^2 \nu^{-3} R(\nu) d\nu}{\int c \nu^{-2}
  R(\nu) d\nu} = \frac{\int \lambda R(\lambda) d\lambda}{\int R(\lambda) d\lambda}
\end{equation}
resulting in $\lambda_{\rm 0}$=23.7\,$\mu$m for the MIPS-24 band.\\

For WISE \citep{wright10} the reference spectrum is a power law
F$_\nu(\nu)\sim\nu^{-2}$, and thus the synthetic photometric flux 
density is defined as
\begin{equation}
\langle F_\nu(\nu_{iso})  \rangle = \frac{\int F_\nu(\nu) (\nu_{iso}/\nu) R(\nu) d\nu}
{\int (\nu_{iso}/\nu)^3 R(\nu) d\nu}
\end{equation}
where $\lambda_{iso}$=c/$\nu_{iso}$ is the isophotal wavelength 
of the photometric band \citep{wright10}. For both band WISE-3
and WISE-4 discrepancies have been found between the pre-flight and 
on-sky performances, with red sources appearing too
faint in band WISE-3 and too bright in band WISE-4 \citep{wright10}. 
Following \cite{brown14}, we minimize the mismatch for WISE-4
by shifting the filter profile and isophotal wavelength upward by 3\%,
resulting in a revised isophotal wavelength of 22.8\,$\mu$m.
For band WISE-3 \cite{wright10} suggest that a 3--5\% downward shift 
of the filter profile might minimize the mismatch in that filter
band. Pending further investigation, we will adopt the pre-flight
filter profile and isophotal wavelength of 11.56\,$\mu$m for band
WISE-3.

%%%%%%%%%%%%%%%%%%%%%%%%%%%%%%%%%%%%%%%%%%%%%%%%%%%%%%%%%%%%%%%%%%
%%%%%%%%%%%%%%%%%%%%%%%%%%%%%%%%%%%%%%%%%%%%%%%%%%%%%%%%%%%%%%%%%%

For the MIRI Imager on the James Webb Space Telescope we adopt 
a T=5,000\,K black body as the reference spectrum as proposed by
\cite{glasse15}. Thus the synthetic photometric flux density 
is defined as
\begin{equation}
\langle F_\nu(\nu_{\rm 0})  \rangle = \frac{\int F_\nu(\nu) (\nu_{\rm 0}/\nu) R(\nu) d\nu}{\int
  (\nu_{\rm 0}/\nu)  (\frac{B_\nu(\nu,T=5000\,K)}{B_\nu(\nu_{\rm 0},T=5000\,K)}) R(\nu) d\nu}
\end{equation}
where $\lambda_{\rm 0}$=c/$\nu_{\rm 0}$ is defined as in Eq.\,5. 
The effective wavelengths for the MIRI filters as computed 
using Eq.\,5 are tabulated in Table\,\ref{tab:synphot}.

%%%%%%%%%%%%%%%%%%%%%%%%%%%%%%%%%%%%%%%%%%%%%%%%%%%%%%%%%%%%%%%%%%
\begin{table}{t}
\centering
\caption{Effective and pivot wavelengths for MIRI filters} \label{tab:synphot}
\begin{tabular}{lrr}
\hline
\hline
Band name & $\lambda_{eff}$ & $\lambda_{p}$\\
\hline
MIRI-770  &  7.67 &  7.64\\
%IRAC-8   &  7.87 &      \\
MIRI-1000 &  9.98 &  9.95\\
MIRI-1100 & 11.33 & 11.31\\
%WISE-12  & 11.56 &      \\
MIRI-1280 & 12.83 & 12.81\\
MIRI-1500 & 15.09 & 15.06\\
%IRS-15   & 15.8  &      \\
MIRI-1800 & 18.00 & 17.98\\
MIRI-2100 & 20.84 & 20.80\\
%IRS-22   & 22.3  &      \\
%WISE-22  & 22.8  &      \\
%MIPS-24  & 23.7  &      \\
MIRI-2550 & 25.40 & 25.36\\
\hline
\end{tabular}
\end{table}
%%%%%%%%%%%%%%%%%%%%%%%%%%%%%%%%%%%%%%%%%%%%%%%%%%%%%%%%%%%%%%%%%%

Unlike the flux calibration methods employed by infrared astronomers,
the HST method of flux calibration does not involve reference spectra,
color corrections and effective wavelengths. The synthetic photometric flux
density (Jy) is defined instead as a photon-weighted mean flux density
over the bandpass of the filter \citep{koornneef86,bohlin11}:

\begin{equation}
\langle F_\nu \rangle = \frac{\int F_\nu(\nu) \nu^{-1} R(\nu) d\nu}{\int \nu^{-1} R(\nu) d\nu}
\end{equation}

We have used this method to produce synthetic photometry for all nine
bands of the JWST MIRI imager \citep{glasse15}. The results are 
included in Table\,\ref{tab:photometry}.
In lieu of an effective wavelength, which depends on the reference
spectrum used, here we use the pivot wavelength $\lambda_p$ \citep{koornneef86} 
to associate the synthetic photometric flux to the filter. The pivot
wavelength is defined as 

 \begin{equation}
\lambda_p = \sqrt \frac{\int \lambda R(\lambda) d\lambda}{\int \lambda^{-1} R(\lambda) d\lambda}
\end{equation}

and the results for the various MIRI filters are tabulated in Table\,\ref{tab:synphot}.

\section{9.8\,$\mu$m silicate strength measurement} \label{sec:silstrength}

Following \cite{spoon07},
we define the strength of the 9.8\,$\mu$m silicate feature as:
\begin{equation}\label{eqn:silstrength}
S_{\rm sil} = \ln [ f_{\nu}(\lambda_{\rm peak}) / f_\nu^{\rm cont}(\lambda_{\rm peak})] 
\end{equation}
\noindent where f$_{\nu}$($\lambda_{\rm peak}$) is the flux density of the spectrum at 
the peak of the silicate feature, $\lambda_{\rm peak}$, and 
f$_\nu^{\rm cont}$($\lambda_{\rm peak}$) is 
the flux density of the underlying continuum at the same wavelength.

\subsection{The observed 9.1-12.65\,$\mu$m continuum}\label{sec:partial913}

\begin{figure}[t!]
\includegraphics[scale=0.565]{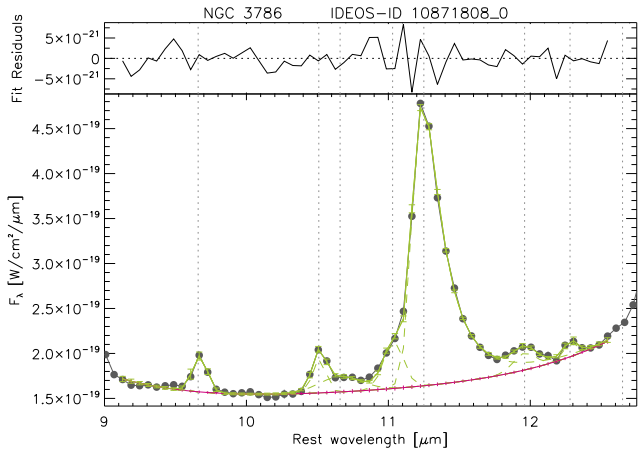}
\caption{Example of a fit to the 9.1--12.58\,$\mu$m spectrum of an
IDEOS source. The observed spectrum is shown in black, and the fit 
to the data in green. Individual emission features are shown as 
green dashed curves on top of the local continuum (pink). 
Vertical dotted lines denote commonly detected features:
9.66$\mu$m H$_2$ S(3), 10.51$\mu$m [S {\sc iv}], 
10.64, 11.04, 11.29, 11.98$\mu$m PAH, and 12.28$\mu$m H$_2$ S(2). 
\label{fig:sil_pah11_mpfit}}
\end{figure}

To determine the local continuum at $\lambda_{\rm peak}$, 
f$_{\nu}$($\lambda_{\rm peak}$), we use MPFIT to model the spectral 
structure in the 9.1--12.65\,$\mu$m range. For this we use the 
same approach as in Sect.\,\ref{sec:partial1013}, where we use
the results of the first stage (10.0--12.65\,$\mu$m) fit, to 
expand to a larger range that includes the H$_2$ 0--0 S(3) 
line at 9.66\,$\mu$m and the continuum down to 9.1\,$\mu$m.
The wider wavelength range requires the use of a fourth order
polynomial continuum, which is reduced to a third order 
polynomial continuum for spectra which are deemed noisy in 
the first stage fit. An example model fit is shown in 
Fig.\,\ref{fig:sil_pah11_mpfit}.

%%%%%%%%%%%%%%%%%%%%%%%%%%%%%%%%%%%%%%%%%%%%%%%%%%%%%%%%%%%%%%%%%%
%%%%%%%%%%%%%%%%%%%%%%%%%%%%%%%%%%%%%%%%%%%%%%%%%%%%%%%%%%%%%%%%%%

\subsection{The underlying continuum}\label{sec:underlyingcont}

The underlying continuum f$_\nu^{\rm cont}$ is an estimate of what the flux 
density of the spectrum would be in absence of silicate emission or
absorption. This is a somewhat subjective measurement that depends 
on our assumption on how the underlying continuum varies with
wavelength between "anchor" points sufficiently far from the silicate 
feature to be unaffected by it. The most common choices are either a 
spline or a power-law interpolation. 

\begin{figure}[t]
\includegraphics[scale=0.565]{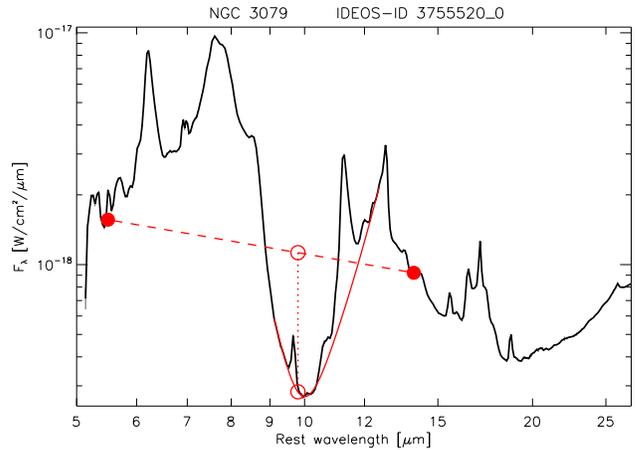}
\caption{Example of a power-law fit to the local continuum in the
5--14\,$\mu$m range to determine the underlying continuum
f$_{\nu}^{\rm cont}$ (dashed red line) in the 8--14\,$\mu$m range. 
Given that the 10\,$\mu$m silicate feature 
is in absorption, the silicate strength S$_{\rm sil}^{\rm powerlaw}$ 
is assessed at 9.8\,$\mu$m. 
The red continuous line is the local continuum,  
f$_{\nu}$, resulting from the MPFIT model for the 9.1--12.65\,$\mu$m
range described in Sect.\,\ref{sec:partial913}.
\label{fig:silfit_powerlaw}}
\end{figure}

\begin{figure}[t]
\includegraphics[scale=0.565]{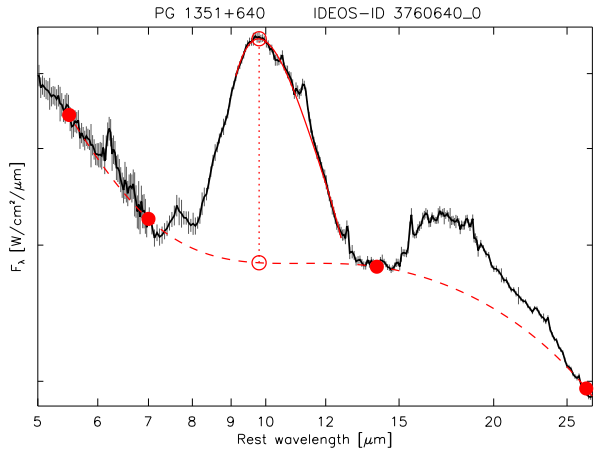}
\caption{Example of a spline fit to the local continuum in the
5--26\,$\mu$m range to determine the underlying continuum,
f$_{\nu}^{\rm cont}$, (dashed red line) valid for the 8--20\,$\mu$m 
range.
Given that the silicate features in this source are in emission, the
silicate strength S$_{\rm sil}^{\rm spline}$ would normally be assessed at
10.5\,$\mu$m. Here we assess it at 9.8\,$\mu$m, where visual
inspection shows the peak to reside.
The red continuous line is the local continuum,  
f$_{\nu}$, resulting from the MPFIT model for the 9.1--12.65\,$\mu$m
range described in Sect.\,\ref{sec:partial913}.
\label{fig:silfit_spline}}
\end{figure}

\begin{figure}[t]
\includegraphics[scale=0.565]{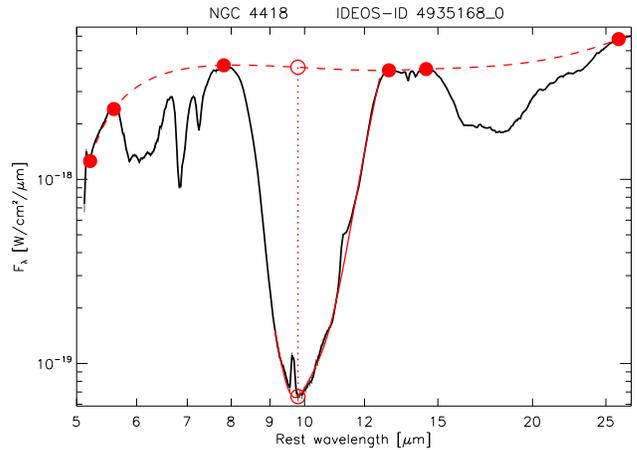}
\caption{Example of a spline fit to the local continuum of a deeply
  obscured source, NGC\,4418. Given the steep upturn at the shortest
  wavelengths and the presence of a deep ice and aliphatic hydrocarbon
  absorption complex in the 7\,$\mu$m range, an additional anchor
  point is needed at $<$5.5\,$\mu$m and the 7\,$\mu$m anchor point
  has to be shifted to 7.8\,$\mu$m. See Fig.\,\ref{fig:silfit_spline}
  for further details.
\label{fig:silfit_spline_n4418}}
\end{figure}

The diversity among the mid-infrared spectra in IDEOS implies that 
none of these interpolation methods provides optimal results in all
cases. We find that spline interpolation produces more realistic 
underlying continua in sources with little or no PAH emission, 
but power-law interpolation is more robust in sources with stronger 
PAH bands. We hence compute S$_{\rm sil}^{\rm spline}$ and
S$_{\rm sil}^{\rm powerlaw}$ for all sources and define the sample of sources
with strong PAH emission to be characterized by EQW(PAH11) $>$0.1\,$\mu$m and
S$_{\rm sil}^{\rm powerlaw}$$>$-2. The latter condition ensures that 
sources with weak
PAH11 emission and weak 11.25\,$\mu$m continuum emission, resulting from
strong silicate absorption, are not included with 
sources that have EQW(PAH11)$>$0.1\,$\mu$m due to strong 
PAH emission. For the power-law interpolation we need two anchors that 
are placed at 5.5 and 14.0\,$\mu$m \citep{spoon07}. See 
Fig.\,\ref{fig:silfit_powerlaw}. For the spline 
interpolation we take anchor points at several additional wavelengths:
7.0, 26.5, and 31\,$\mu$m. See Fig.\,\ref{fig:silfit_spline}.
For sources with insufficient spectral 
coverage at long wavelengths we replace the last two anchors by 25.0 
and 28.0\,$\mu$m, or 23.0 and 25.5\,$\mu$m. If the spectral coverage 
does not even reach 25.5\,$\mu$m, we replace the 14.0\,$\mu$m anchor 
by two at 13.5 and 15.0\,$\mu$m. For deeply obscured sources, like 
NGC\,4418 shown in Fig.\,\ref{fig:silfit_spline_n4418}, an additional
anchor around 5\,$\mu$m becomes necessary, while the anchor at
7.0\,$\mu$m is replaced by an anchor at 7.8\,$\mu$m.

%%%%%%%%%%%%%%%%%%%%%%%%%%%%%%%%%%%%%%%%%%%%%%%%%%%%%%%%%%%%%%%%%%
%%%%%%%%%%%%%%%%%%%%%%%%%%%%%%%%%%%%%%%%%%%%%%%%%%%%%%%%%%%%%%%%%%

%
Among the 2847 sources where the silicate feature is observed there are
616 sources for which the spectral coverage or 
SNR is insufficient\footnote{Our criterion is SNR$<$2 around the anchor point} 
to obtain either the 5.5 or 
the 14.0\,$\mu$m (13.5 and 15.0\,$\mu$m for splines) anchors from the
spectrum. We found that moving any of these anchors closer to the 
silicate feature can bias the results significantly, so we chose 
instead to estimate the flux density at 5.5 and 14.0\,$\mu$m by fitting 
the spectrum with a model that spans the 5--16\,$\mu$m range, and then 
measure the missing anchor fluxes on the model. 
For this purpose we use deblendIRS  \citep{hernan15}. 
The deblendIRS model uses only three physical components (stellar, 
ISM and AGN), each of them represented by an empirical template 
selected from a large library of observed spectra). The
best-fitting deblendIRS model reproduces the observed spectra with 
high accuracy (typical reduced $\chi^2\sim$1), and  spans the 5--16\,$\mu$m 
range irrespective of the wavelength coverage of the original
spectrum. To estimate the uncertainty in the extrapolation of the 
spectrum using the deblendIRS mode, we have compared in a random 
sample of 500 sources with sufficient spectral coverage and high S/N 
the actual 
fluxes at 5.5 and 15.0\,$\mu$m with those obtained from the deblendIRS 
model when the fitting range is reduced to 7.0--16.0\,$\mu$m and 
5.0--12.0\,$\mu$m, respectively. In both cases we obtain a 1-$\sigma$ 
dispersion of 25\%, with no significant bias.

The default anchor points for the spline or power law methods provide 
realistic underlying continua for most sources, but fail in the cases 
where the continuum has an unusual shape, like deeply obscured ULIRGs 
or quiescent galaxies, whose MIR spectra are dominated by the 
Rayleigh-Jeans tail of the 
stellar emission. For these sources we adjust manually the anchor 
wavelengths until a realistic continuum is obtained. 

%%%%%%%%%%%%%%%%%%%%%%%%%%%%%%%%%%%%%%%%%%%%%%%%%%%%%%%%%%%%%%%%%%
%%%%%%%%%%%%%%%%%%%%%%%%%%%%%%%%%%%%%%%%%%%%%%%%%%%%%%%%%%%%%%%%%%

\subsection{Shape of the 9.8\,$\mu$m silicate profile}

When the silicate feature appears in absorption it always peaks at 
$\sim$9.8\,$\mu$m. However, when in emission the peak is often broad and 
displaced to longer wavelengths. \citet{Hatziminaoglou15} reported 
that in a large sample of AGN, the peak of the silicate feature, when 
observed in emission, is at $\lambda$$>$10.2\,$\mu$m in 65\% of cases 
and at $\lambda$$>$10.6\,$\mu$m in 20\%. The exact peak wavelength of 
the broad silicate emission feature depends on the definition of the
underlying continuum, on the representation chosen 
($f_\nu$, $\nu f_\nu$, or $f_\lambda$), the temperature distribution
of the dust and composition of the emitting silicates, and, in 
low SNR spectra, on the random noise in the spectrum. 

It therefore matters what peak wavelength $\lambda_{\rm peak}$ is chosen.
To have a robust method that works also for low S/N spectra,
for silicate features found in absorption we will assume 
$\lambda_{\rm peak}$ to be at 9.8\,$\mu$m. When found in 
emission\footnote{45 out of 50 Monte Carlo simulations of the
silicate fits, as explained in Sect.\,\ref{sec:uncertainties}, 
have to produce a silicate emission feature.}
we will use 10.5\,$\mu$m, unless visual inspection shows the peak
to be at 9.8\,$\mu$m (e.g. PG1351+640 in Fig.\,\ref{fig:silfit_spline}). 
We interpolate the fitted 9--12.6\,$\mu$m continuum and the 
underlying continuum to $\lambda_{\rm peak}$ to obtain 
f$_{\nu}$($\lambda_{\rm peak}$) and f$_\nu^{\rm cont}$($\lambda_{\rm peak}$), 
and using Eq.\,\ref{eqn:silstrength}, S$_{\rm sil}$.

%%%%%%%%%%%%%%%%%%%%%%%%%%%%%%%%%%%%%%%%%%%%%%%%%%%%%%%%%%%%%%%%%%
%%%%%%%%%%%%%%%%%%%%%%%%%%%%%%%%%%%%%%%%%%%%%%%%%%%%%%%%%%%%%%%%%%

\subsection{Uncertainties}\label{sec:uncertainties}

We estimate the statistical uncertainty in S$_{\rm sil}$ using a Monte Carlo method. 
For every spectrum we obtain 50 copies, where gaussian noise has been
added to each pixel consistent with its flux uncertainty. We 
re-evaluate the underlying continuum and the fitted 
9--12.6\,$\mu$m continuum that best 
fits the silicate feature for each of these copies, and measure 
S$_{\rm sil}$ in all of them. We then calculate the uncertainty in 
S$_{\rm sil}$ as the standard deviation of the values obtained for 
the 50 copies. 

In sources with noisy spectra or very deep silicates, the flux of 
the fitted 9--12.6\,$\mu$m continuum at $\lambda_{\rm peak}$ is 
sometimes less than zero. For these sources we use the distribution 
of S$_{\rm sil}$ values in the 50 copies 
to give an upper limit for S$_{\rm sil}$ at the 95\% confidence level.

While the Monte Carlo method gives realistic statistical uncertainties 
for a given continuum interpolation method, it is important to keep 
in mind that the main uncertainty in the silicate strength may be
systematic in nature, associated with the choice of interpolation
method, or from the need to invoke deblendIRS.
Analysis of the silicate strength solutions from the spline and
powerlaw methods at the boundary of their validity ranges 
quantifies the systematic uncertainty as 0.2 in silicate strength.

\section{Diagnostic plots} \label{sec:diagnosticplots}

\subsection{Mid-Infrared spectral classification} \label{sec:spoondiagram}

Before the advent of Spitzer-IRS the study of spectral features in
galaxy spectra was limited to low-resolution spectra in the
5--11\,$\mu$m range (ISO-PHT-S) and 5--16\,$\mu$m range (ISO-CAM-CVF)
plus targeted high spectral resolution line observations between 
2 and 45\,$\mu$m (ISO-SWS).
Spitzer-IRS openend up the 5--37\,$\mu$m range for full range 
spectroscopy of thousands of galaxies in the Local Universe.
This enabled for the first time the use of the 9.8\,$\mu$m silicate 
feature as an obscuration diagnostic without the limitations 
imposed by the inability to properly define a local 5--14\,$\mu$m
continuum both in spectra obtained on the ground and from space.
The 'silicate strength', first defined in 2006, was subsequently 
used by \cite{spoon07} to classify galaxies based on their location
in the diagram that separates galaxies by the equivalent width
of the 6.2\,$\mu$m PAH feature (EQW62) and the silicate strength
(S$_{\rm sil}$).

\begin{figure*}[t]
\begin{center}
\includegraphics[scale=1]{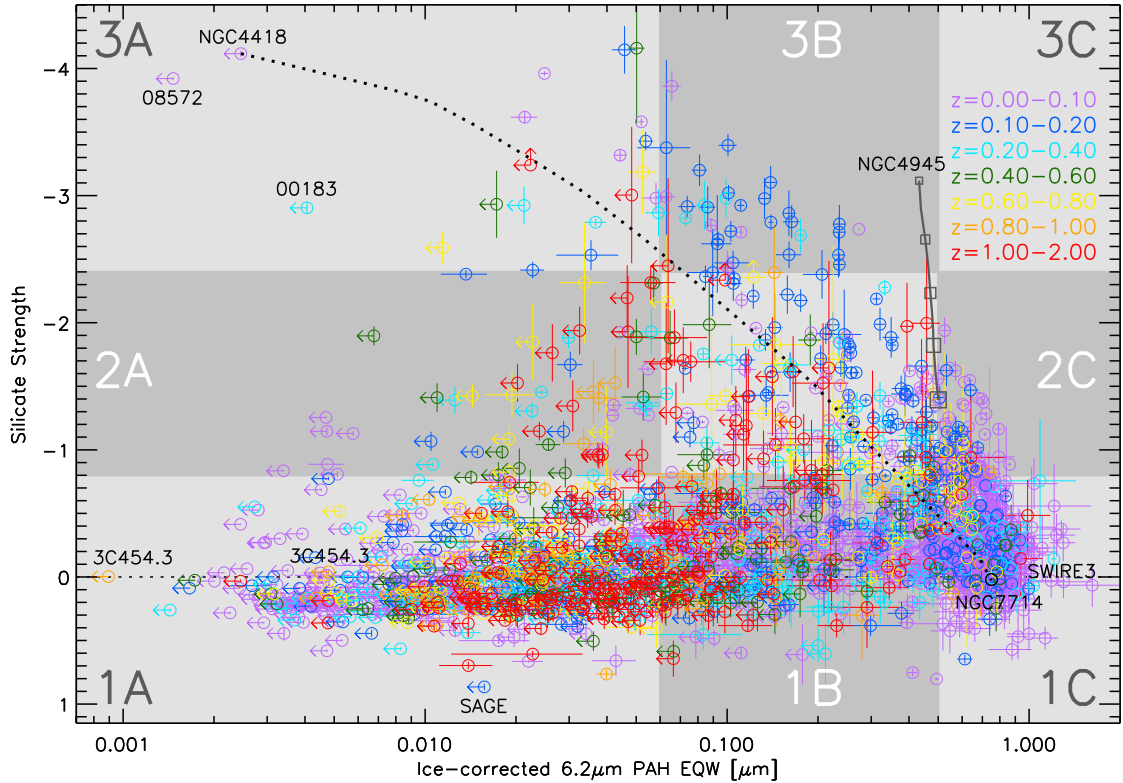}
\end{center}
\caption{Diagnostic diagram combining the ice-corrected equivalent
width of the 6.2\,$\mu$m PAH feature (Sect.\,\ref{sec:partial57})
with the silicate strength (Sect.\,\ref{sec:silstrength}). 
The 2524 galaxies are color-coded by redshift.
Also included are several Spitzer-IRS mapping spectra 
of NGC\,4945, centered on the deeply buried AGN and integrated  
over areas ranging from 5.5$"$$\times$5.5$"$ to 76$"$$\times$31.5$"$,
with the symbol size proportional to the area.
The diagram is subdivided in nine named quadrants, which form the basis
of our mid-infrared galaxy classification scheme. Galaxies in class
1A have a hot dust-dominated spectrum typical for AGN. Galaxies in
class 1C have PAH-dominated spectra typical for starburst galaxies.
Galaxies in class 3A have absorption-dominated spectra typical for
centrally heated dust shells or tori. Galaxies in between these
extremes show intermediate properties. The dotted line shows the
mixing line of galaxies with spectra intermediate between those 
of NGC\,4418 and NGC\,7714.
In the plot the label SAGE refers to SAGE1C\,J053634.78--722658.5, 
SWIRE to SWIRE3\,J105056.08+562823.0, 00183 to
IRAS\,F00183--7111, 08572 to IRAS\,08572+3915N. Galaxy 3C454.3 appears
twice, in its high and low states as discussed in the text.
\label{fig:forkdiagram}}
\end{figure*}

Fig.\,\ref{fig:forkdiagram} shows the distribution\footnote{1304 
galaxies at z=0--0.1; 386 at z=0.1--0.2; 243 at z=0.2--0.4;
142 at z=0.4--0.6; 125 at z=0.6--0.8; 81 at z=0.8--1.0; 243 at z=1.0--2.0}
of 2524 IDEOS
galaxies, color-coded by redshift bracket, over this diagram.
Galaxies appear to be confined within a wedge-shaped region 
demarkated by two prongs of a fork and three vertices:
\begin{itemize}
\item In the lower right we find galaxies dominated by exposed star
formation, as evidenced by strong PAH emission. The 9.8\,$\mu$m
silicate feature is weakly in emission or absorption.
\item In the lower left we find galaxies dominated by AGN-heated 
hot dust. These galaxies show very little or no sign of PAH 
emission (star formation) and show only weak\footnote{A weak
apparent silicate optical depth (silicate strength) does not 
rule out an appreciable silicate optical depth along the 
line of sight.} silicate emission or absorption.
\item In contrast, galaxies in the upper left are dominated by
deep absorption features of silicates. PAH emission features 
(the tell-tale signatures of exposed star formation) are generally 
faint or absent. Since the presence of
deep silicate absorption features requires a strong negative 
temperature gradient in the dust along our line of sight 
\cite[e.g.][]{sirocky08} and full and optically thick coverage 
of the power source, 
the power source in these galactic nuclei must be compact:
either an ultra-compact nuclear starburst or a supermassive
black hole, hidden in a dust cocoon or at the center of an 
edge-on torus.
\end{itemize}
The large differences in spectral appearance of galaxies
found in between these vertices makes a galaxy classification
scheme based on silicate strength and PAH62 equivalent width
useful.

Following \cite{spoon07}, we have classified\footnote{Compared 
to \cite{spoon07}, the class borders between classes A\&B and
B\&C have changed slightly as the result of the use of a 
Pearson IV profile to represent the PAH62 profile (see 
Appendix\,\ref{sec:appendix-b}} the IDEOS spectra into a grid
of 3-by-3 classes based on the measured ice-corrected PAH62 
equivalent width and the silicate strength. The classes
range from 1A to 3C and are overlaid in Fig.\,\ref{fig:forkdiagram}.
Even though this classification scheme is based on just two mid-infrared
observables, it better captures the essential differences 
among infrared galaxies than any of the other 2-dimensional 
mid-infrared diagnostic diagrams shown in
subsequent figures (Figs.\,\ref{fig:c30c55_silstrength}--\ref{fig:crystsil_silstrength}).

%%%%%%%%%%%%%%%%%%%%%%%%%%%%%%%%%%%%%%%%%%%%%%%%%%%%%%%%%%%%%%%%%%
%%%%%%%%%%%%%%%%%%%%%%%%%%%%%%%%%%%%%%%%%%%%%%%%%%%%%%%%%%%%%%%%%%

The large majority of the galaxies in this 'Fork Diagram' (84\%)
are scattered along the horizontal branch of the fork, which
connects classes 1A and 1C. Active galaxies with a Seyfert optical
classification are generally confined to classes 1A and 1B, and
the strongly AGN-dominated galaxies to class 1A. Starburst
galaxies are home to classes 1C, with the more dust enshrouded 
ones showing up in class 2C (e.g. M82). Normal star forming galaxies,
with spectra similar to the four noise-free template spectra
of \cite{smith07}, are found in class 1B (just over the border 
from class 1C) thanks to a stronger contribution of stellar 
photospheric emission to the 6.2\,$\mu$m continuum underneath
the 6.2\,$\mu$m PAH feature.
14\% of the galaxies in the Fork diagram are distributed over
classes 2B/2C/3A/3B, which, together with class 1C, form its 
diagonal prong, and which is demarcated by the dashed line in 
Fig.\,\ref{fig:forkdiagram} that represents the mixing line 
between the spectra of the buried nucleus of NGC\,4418 and the 
starburst galaxy NGC\,7714. 
Only 2\% of the galaxies in the Fork Diagram are found in 
class 2A, in between the two prongs of the fork. Compared
to the galaxies in class 3A, above them, their spectra show
shallower silicate absorption features. These shallower silicate 
features are most easily explained as resulting from dilution 
of the absorption spectrum by continuum emission,
either resulting from key hole openings\footnote{Only 5--10\% of 
the luminosity of a buried power source needs to be unveiled for
it to move all the way to the horizontal branch of the Fork 
Diagram \citep{marshall18}} in a dust cocoon
\citep{marshall18}, or from a glimpse into the central region
of a dust torus seen at an intermediate inclination 
\citep[A.\,Efstathiou priv. comm.]{rowan-robinson09}.

Classic galaxy evolution scenarios \citep{sanders96,hopkins06} 
predict merging galaxies to go through a phase of strong nuclear 
obscuration before a naked AGN emerges. In this scenario two
normal star forming galaxies would thus start their journey 
together in class 1C (or just across the border in class 1B), 
make their way up in the Fork Diagram,
before descending and ending up as a merged class 1A AGN. The 
large majority of IDEOS galaxies found in classes 2A-3B are
indeed caught as LIRGs and ULIRGs in interaction 
\citep[][and references there in]{armus20}.

With a sample of 2524 galaxies at hand it is interesting to point
out some interesting, in certain aspects extreme, sources 
in the Fork Diagram:
\begin{itemize}
\item The galaxy with the highest silicate strength (S$_{\rm sil}$=0.865
is SAGE1C\,J053634.78-722658.5. % 22435584_0
The galaxy was discovered in a survey of the Large Magellanic Cloud,
and has an infrared spectrum completely devoid of cold dust emission
associated with star formation \citep{hony11, vanloon15} and has hence
been referred to as a naked AGN.

\item The galaxy with the lowest EQW62 is the blazar 3C\,454.3. Spitzer-IRS
observed the galaxy 31 times between June 30, 2005 \cite[only weeks after the
historic outburst of early May 2005;][]{fuhrmann06}, and January 23,
2009, during which the upper limit for PAHEQW62 reached a lowest value
of 9$\times$10$^{-4}$\,$\mu$m on July 7, 2005. This event coincides with the 
highest measured 5.5\,$\mu$m continuum flux density of 516\,mJy, 
which is 30 times higher than its lowest measured value of 17\,mJy
on January 24, 2009. The slope of the powerlaw spectrum, as measured
from the 3.7 to 15\,$\mu$m continuum flux ratio ranged from 0.14 
to 0.25 in these 4.5 years.

\item The most deeply buried galaxies with the, by far, tightest upper 
limits for the presence of PAH62 emission are NGC4418 \citep{spoon01} 
and IRAS\,08572+3915NW \citep{spoon06} at S$_{\rm sil}$ -4.12, and
-3.92, and PAH62EQW$<$2.5$\times$10$^{-3}$\,$\mu$m and 
$<$1.5$\times$10$^{-3}$\,$\mu$m, respectively. Both galaxies are 
interacting, but neither is in the final stages of coalescence.
NGC\,4418 is connected to VV\,655 by a 50\,kpc gas bridge \citep{varenius17}, 
and IRAS\,08572+3915NW has a projected nuclear separation of 6\,kpc 
to IRAS\,08572+3915SE \citep{colina05}. Deep silicate features
can hence not be relied on as sign posts for the very final stages 
of a merger.

\item At a distance of 3.7\,Mpc, NGC\,4945 is the nearest galaxy
hosting a deeply buried AGN. Our line of sight
into the nucleus shows a wealth of ice absorption features 
\citep{spoon00,spoon03}, some of which (e.g. the 15\,$\mu$m 
CO$_2$ ice absorption feature \citep{perez-beaupuits11}) 
have not been detected in any other galaxy thus far. 
Within the IDEOS sample NGC\,4945 is unique for showing a 
strongly absorbed PAH emission spectrum rather than a 
strongly absorbed continuum spectrum at the position of
the AGN \citep{perez-beaupuits11}. This places the nucleus 
proper in an otherwise unpopulated section of the Fork Diagram.
Thanks to the availability of a Spitzer-IRS-SL map of the 
central 76$"$$\times$31.5$"$ (1.37kpc $\times$ 0.57kpc;
PI: H.W.W. Spoon), 
it is possible to quantify the effect of including a
larger and larger portion of the circumnuclear starburst 
ring and the galaxy disk in an extraction aperture. As shown in 
Fig.\,\ref{fig:forkdiagram}, the galaxy spectrum moves from
class 3B to 2B and 2C as the nuclear absorption spectrum gets 
more and more overwhelmed by a PAH emission spectrum associated 
with unobscured/exposed star formation from the galaxy disk.
The spectrum of NGC\,4945 as a whole would likely be 
classified as a class 1C starburst spectrum.
The example of NGC\,4945 illustrates to what extent the 
physical projected size associated with the Spitzer-IRS
SL slit differs between nearby and distant galaxies:
at z=0.05, the SL slit only includes the central 3\,kpc 
of a galaxy. 
Galaxies shown in purple in Fig.\,\ref{fig:forkdiagram} 
are thus repesented by their nuclear properties.

\item The highest redshift galaxy with a pure starburst 
spectrum in the fork diagram is SWIRE3\,J105056.08+562823.0 % 17419520_0
(z$_{\rm IRS}$=1.537) at S$_{\rm sil}$=-0.32 and
PAH62EQW=1.01\,$\mu$m. The source is labeled 'SWIRE3' 
in the diagram.

\end{itemize}

\subsection{Mid-Infrared continuum slope diagnostics} \label{sec:contslope_diagnostics}

%%%%%%%%%%%%%%%%%%%%%%%%%%%%%%%%%%%%%%%%%%%%%%%%%%%%%%%%%%%%%%%%%%
\begin{figure*}[t]
\begin{center}
\includegraphics[scale=0.8]{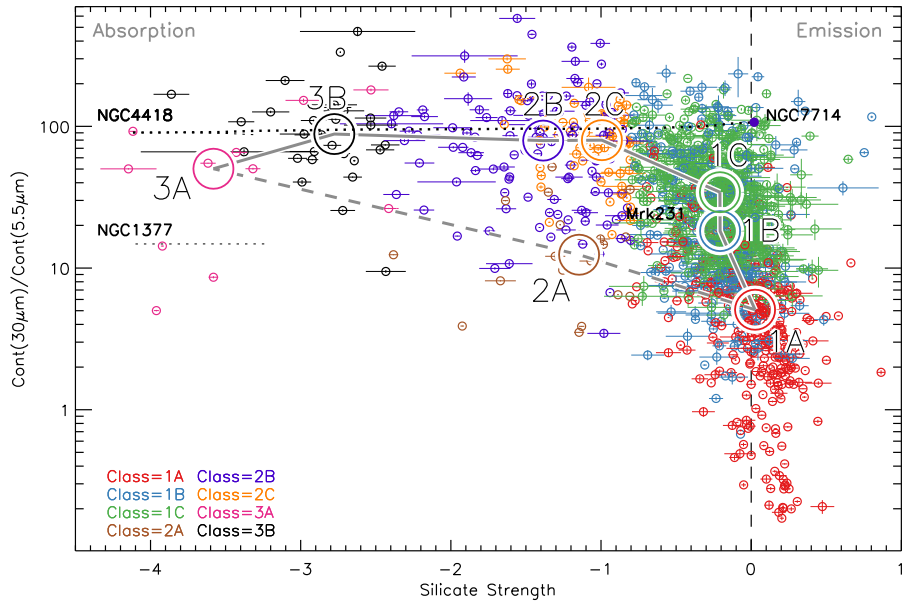}
\end{center}
\caption{Diagnostic diagram of the 30\,$\mu$m to 5.5\,$\mu$m
continuum ratio versus the 9.8\,$\mu$m silicate strength.
Large open circles denote the median locations of each of the
mid-IR classes (Sect.\,\ref{sec:spoondiagram}) in the diagram. 
The gray lines connecting classes 1C to 1A and 1C to 3A 
trace the two branches of the Fork 
Diagram (Fig.\,\ref{fig:forkdiagram}). 
The dashed gray line show the effect of a change in silicate
strength on sources that lack PAH contamination 
(Mid-IR classes 1A, 2A and 3A).
The dotted black line is the mixing line between the spectra
of NGC\,4418 and NGC\,7714.
\label{fig:c30c55_silstrength}}
\end{figure*}
%%%%%%%%%%%%%%%%%%%%%%%%%%%%%%%%%%%%%%%%%%%%%%%%%%%%%%%%%%%%%%%%%%

Our measurements of the rest frame continuum flux densities at 
various mid-infrared wavelengths allow us to use the ratio of
warm 30\,$\mu$m to hot 5.5\,$\mu$m rest frame continuum emission 
as an 
additional diagnostic. Exactly what the diagnostic power entails
depends on the adopted dust geometry around the heating sources:
\begin{itemize}
\item A galactic nucleus hidden within a dust shell will see a
decrease in the C(30\,$\mu$m)/C(5.5\,$\mu$m) ratio as the  
column density of the dust shell decreases.

\item A decrease in the covering factor of the obscuring shell 
around a galactic nucleus (the emergence of a ``keyhole'') will 
result in the decrease of the C(30\,$\mu$m)/C(5.5\,$\mu$m) ratio 
\citep{marshall18}.

\item A change in orientation of a non-clumpy torus from edge-on 
view to pole-on view will result in a decrease of the 
C(30\,$\mu$m)/C(5.5\,$\mu$m) ratio 
\citep[A.\,Efstathiou priv. comm.]{rowan-robinson09}.
\end{itemize}

Fig.\,\ref{fig:c30c55_silstrength} shows the distribution of the
IDEOS sources in a diagram of the C(30\,$\mu$m)/C(5.5\,$\mu$m) ratio
versus the silicate strength. Like in the Fork Diagram the majority
of the galaxies are found along two almost ortogonal branches. 
The vertical branch comprises sources that range from AGN to 
starburst galaxies (i.e. sources on the horizontal branch in
the Fork Diagram: classes 1A--1B--1C),
whereas sources on the horizontal branch are sources that range 
from enshrouded nuclei without significant circumnuclear star formation
to starburst galaxies with dusty nuclei (i.e. sources on the diagonal
branch in the Fork Diagram: classes 3A--3B--2B--2C--1C).

The dashed gray line shows the effect that a change in silicate
strength would have on the continuum slope for galaxies with similar 
low levels of spectral contamination by PAH emission 
(i.e. mid-IR classes 1A--2A--3A).
As can be seen in Fig.\,\ref{fig:c30c55_silstrength}, decreasing
the obscuration results in a decrease of the 
C(30\,$\mu$m)/C(5.5\,$\mu$m) ratio.

Taken at face value, it is remarkable that the median 
C(30\,$\mu$m)/C(5.5\,$\mu$m) ratio of the IDEOS sources
hovers around 50--100 over a large span in silicate strength
(-4 to -1) along the horizontal branch 
(classes 3A--3B--2B--2C--1C). Apparently, 
the increase in 5.5\,$\mu$m continuum emission
afforded by a lower obscuration level is compensated for
by an increase in warm 30\,$\mu$m continuum emission 
associated with an increased contribution of exposed
star formation along this branch.

%%%%%%%%%%%%%%%%%%%%%%%%%%%%%%%%%%%%%%%%%%%%%%%%%%%%%%%%%%%%%%%%%%
\begin{figure*}[t]
\begin{center}
\includegraphics[scale=0.8]{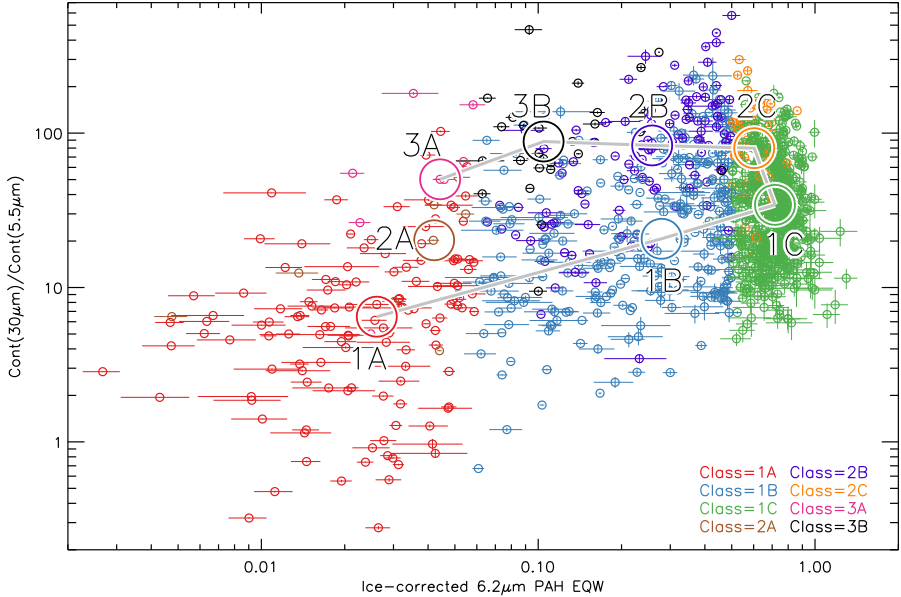}
\end{center}
\caption{The circles denote the median locations of each of the
mid-IR classes (Sect.\,\ref{sec:spoondiagram}) in the diagram. 
The gray connecting lines trace the two branches of the Fork 
Diagram (Fig.\,\ref{fig:forkdiagram}). 
The dashed gray line show the effect of a change in silicate
strength on sources that lack PAH contamination 
(Mid-IR classes 1A, 2A and 3A).
\label{fig:c30c55_pah62eqw}}
\end{figure*}
%%%%%%%%%%%%%%%%%%%%%%%%%%%%%%%%%%%%%%%%%%%%%%%%%%%%%%%%%%%%%%%%%%

Fig\,\ref{fig:c30c55_pah62eqw} is an adaptation of the Laurent Diagram
\citep{laurent00,peeters04}, using the wider wavelength coverage of
Spitzer-IRS. The Laurent Diagram was originally devised to quantify 
the contribution from AGN, PDR and HII regions to a galaxy spectrum
by delineating the PAH62 equivalent width and the 5--15\,$\mu$m
continuum slope. Buried nuclear activity was not considered 
back then, as there were very few galaxies for which the 
9.8\,$\mu$m silicate feature could be observed \citep{peeters04}. 

The diagram separates classic AGNs (mid-IR class 1A)
and starburst galaxies (mid-IR class 1C) into opposite corners. 
The diagonal thick gray line in the diagram connects these two 
extremes. The diagram is, however, less successful in separating 
out galaxies dominated to varying degrees by buried nuclear 
activity (mid-IR classes 3A-3B-2B-2C-1C). Both the Fork Diagram
(Fig\,\ref{fig:forkdiagram}) and the continuum slope versus
silicate strength diagram (Fig.\,\ref{fig:c30c55_silstrength})
do this far more effective.

\subsection{Mid-Infrared ionized gas excitation diagrams} \label{sec:midir_excitation_diagrams}

The mid-infrared fine-structure lines of ionized neon gas form an
excellent diagnostic for the excitation of the ionized gas.
The lines, [Ne {\sc ii}] at 12.81\,$\mu$m, [Ne {\sc iii}] at 
15.6\,$\mu$m, [Ne {\sc v}] at 14.32 \& 24.32\,$\mu$m, and
[Ne {\sc vi}] at 7.65\,$\mu$m, span a range of ionization potentials 
(21, 41, 97, and 127 eV), have critical densities $>$10$^{4.5}$ cm$^{-3}$,
do not suffer from strong differential extinction (due to
amorphous silicate resonances), and are insensitive to abundance 
uncertainties.

In our galaxy [Ne {\sc v}] and [Ne {\sc vi}] emission is detected only
from shocks associated with supernova remnants 
\cite[e.g. RCW\,103;][]{oliva99}. Their combined signal is not strong
enough to be detectable in nuclear or galaxy-integrated spectra like 
ours \citep{perez-beaupuits11}. The detection of [Ne {\sc v}] in an
IDEOS spectrum is hence a tell-tale sign for the presence of an AGN.
The inverse is not true, the absence of a [Ne {\sc v}] line detection
does not mean that an AGN is absent, as strong extinction in the 
line of sight to the narrow line region will decrease its equivalent
width.
The number of [Ne {\sc v}] detections (either at 14.32 or
24.32\,$\mu$m) in our sample is 390. Thus, at least 390/3335 galaxies 
in our sample host an AGN.

%%%%%%%%%%%%%%%%%%%%%%%%%%%%%%%%%%%%%%%%%%%%%%%%%%%%%%%%%%%%%%%%%%
\begin{figure*}[t]
\begin{center}
\begin{tabular}{c}
\includegraphics[scale=0.99]{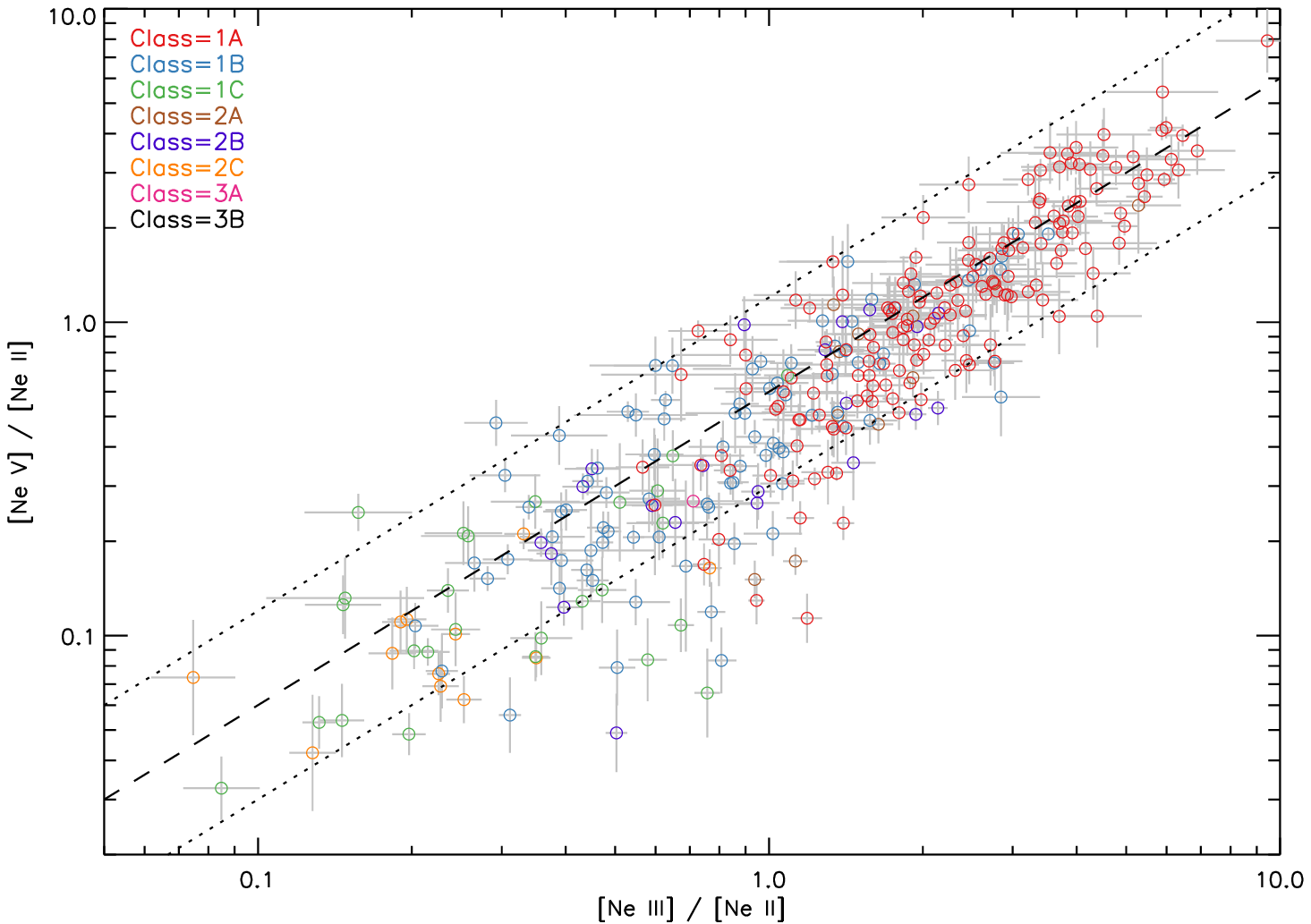} \\
\includegraphics[scale=0.99]{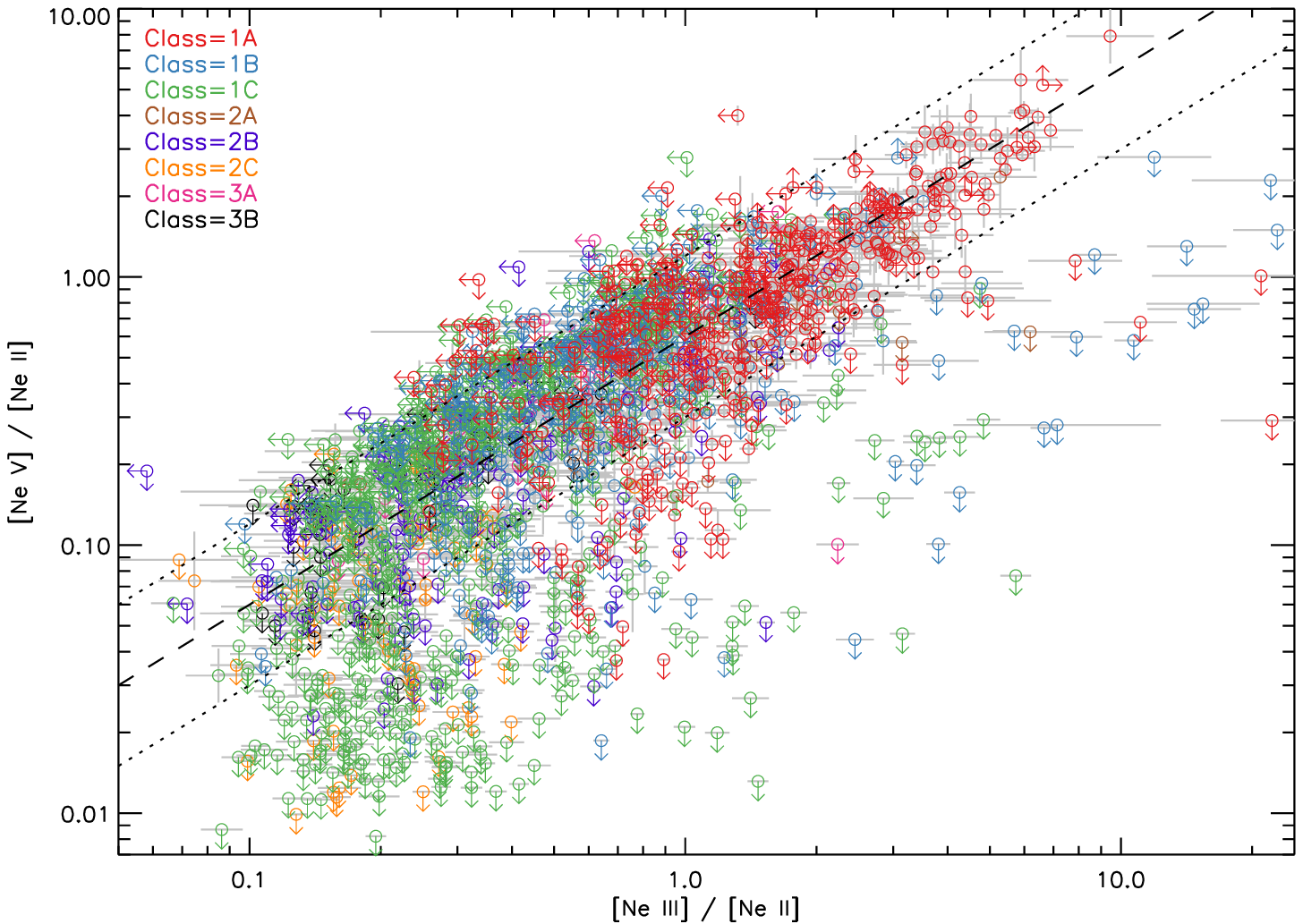}
\end{tabular}
\end{center}
\caption{Excitation diagram of ionized neon gas based on the
12.81\,$\mu$m [Ne {\sc ii}], 15.56\,$\mu$m [Ne {\sc iii}], and
14.32\,$\mu$m [Ne {\sc v}] lines.
In the upper panel we include just the 316 sources with detections for
all three lines, whereas in the bottom panel we also include the 1312 
sources for which one or both ratios are upper or lower limits. 
The dashed line indicates [Ne {\sc v}]/[Ne {\sc iii}]=0.6, and the
dotted lines a ratio of 0.3 and 1.2, respectively. The galaxies are 
color-coded by mid-infrared galaxy class as defined in 
Sect.\,\ref{sec:spoondiagram}. Drop-out sources found in the quadrant defined
by [Ne {\sc iii}]/[Ne {\sc ii}]$>$1 and [Ne {\sc v}]/[Ne {\sc iii}]$<$0.3
correspond to the low-metallicity sources found on the horizontal branch 
in Fig.\,\ref{fig:o4s3_ne3ne2}.
\label{fig:ne5ne2_ne3ne2}}
\end{figure*}
%%%%%%%%%%%%%%%%%%%%%%%%%%%%%%%%%%%%%%%%%%%%%%%%%%%%%%%%%%%%%%%%%%

%%%%%%%%%%%%%%%%%%%%%%%%%%%%%%%%%%%%%%%%%%%%%%%%%%%%%%%%%%%%%%%%%%
\begin{figure*}[t]
\begin{center}
\includegraphics[scale=0.99]{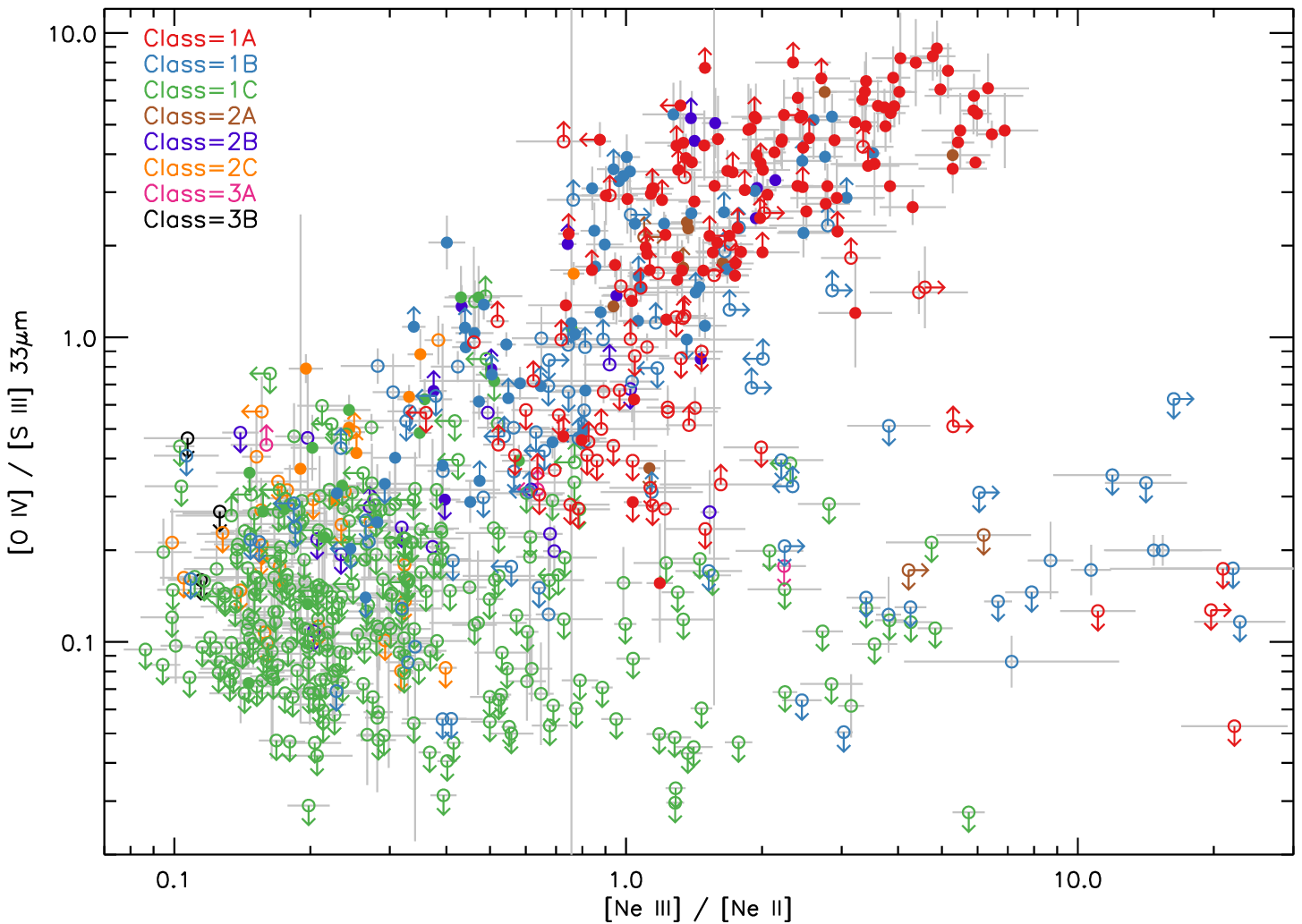}
\end{center}
\caption{Excitation diagram based on the 12.81\,$\mu$m [Ne {\sc ii}], 
15.56\,$\mu$m [Ne {\sc iii}], and
25.89\,$\mu$m [O {\sc iv}] and 33.48\,$\mu$m [S {\sc iii}] lines.
692 galaxies are plotted. Bonafide AGN, as confirmed by a detection 
of a [Ne {\sc v}] line, are shown as filled circles. 
The galaxies are color-coded by mid-infrared galaxy class as defined 
in Sect.\,\ref{sec:spoondiagram}.
\label{fig:o4s3_ne3ne2}}
\end{figure*}
%%%%%%%%%%%%%%%%%%%%%%%%%%%%%%%%%%%%%%%%%%%%%%%%%%%%%%%%%%%%%%%%%%

In the upper panel of Fig.\,\ref{fig:ne5ne2_ne3ne2} we show the
positions of 316 sources that have detections for all three neon 
lines, 12.81\,$\mu$m [Ne {\sc ii}], 15.56\,$\mu$m [Ne {\sc iii}], 
and 14.32\,$\mu$m [Ne {\sc v}]: all of them bonafide AGN. 
The source distribution is best described by a linear relation
characterized by [Ne {\sc v}]/[Ne {\sc iii}]=0.6, in good 
agreement with the results of \cite{gorjian07} for a sample
of Seyferts and 3C radio sources.
As the color-coding of the sources suggest, the highest
excitation AGNs have a mid-IR classification 1A (low PAH equivalent 
width and only weak silicate emission/absorption), whereas the 
lowest excitation AGNs are found among class 1C and 2C galaxies 
(dominated by a PAH emission spectrum). Intermediate
[Ne {\sc v}]/[Ne {\sc ii}] ratios are found among class 1B and 2B
sources.
The highest excitation source in our sample, as inferred from 
the [Ne {\sc v}]/[Ne {\sc ii}] ratio, is the nearby radio galaxy 
3C\,321. Its ([Ne {\sc iii}]/[Ne {\sc ii}] ratio is 9.4 and its
[Ne {\sc v}]/[Ne {\sc ii}] ratio is 7.9.

In the lower panel of Fig.\,\ref{fig:ne5ne2_ne3ne2} we also include
upper and lower limits for the line ratios, bringing the total
source count in the plot to 1628. Some of the upper limits for 
[Ne {\sc v}]/[Ne {\sc ii}] are clearly inconsistent
with membership of the diagonal band seen in the upper panel.
We identify these ``drop outs'' in the quadrant defined by 
[Ne {\sc iii}]/[Ne {\sc ii}]$>$1 and [Ne {\sc v}]/[Ne {\sc iii}]$<$0.3 
(the lower dotted line in Fig.\,\ref{fig:ne5ne2_ne3ne2})
with the low-metallicity galaxies found in Fig.\,8 of \cite{hao09}.
Among them are well-known sources like Haro\,11, NGC\,1140, Mrk1450, 
Mrk1499, and II\,Zw\,40.
Included among the drop outs are also Wolf-Rayet galaxies like
IRAS\,11485--2018.
Note that some of these galaxies have [Ne {\sc iii}]/[Ne {\sc ii}] 
ratios exceeding those for the most extreme AGNs by a factor 2 or more.
For example IRAS\,11485--2018: [Ne {\sc iii}]/[Ne {\sc ii}]=22$\pm$6.

Following \cite{hao09}, in Fig.\,\ref{fig:o4s3_ne3ne2} we plot the
[Ne {\sc iii}]/[Ne {\sc ii}] ratio versus the 
[O {\sc iv}]/[S {\sc iii}]\,33\,$\mu$m ratio. 
This separates galaxies with Blue Compact Dwarf
(BCD) properties along a horizontal ``drop out'' 
branch\footnote{The horizontal
branch is defined by [Ne {\sc iii}]/[Ne {\sc ii}]$>$1 and 
a ratio of [O {\sc iv}]/[S {\sc iii}]\,33\,$\mu$m to
[Ne {\sc iii}]/[Ne {\sc ii}] below 0.2} from galaxies 
which range from starburst to AGN dominated on
a diagonal branch. We identify the galaxies on the horizontal 
branch with the ``drop out'' sources in 
Fig.\,\ref{fig:ne5ne2_ne3ne2}.
Clearly, [O {\sc iv}]/[S {\sc iii}]\,33\,$\mu$m does a better
job at separating the low-metallicity drop-outs from other 
sources than [Ne {\sc v}]/[Ne {\sc ii}] does.
Note that, like in Fig.\,\ref{fig:ne5ne2_ne3ne2}, the 
[Ne {\sc iii}]/[Ne {\sc ii}] ratio reaches higher values among 
the galaxies on the horizontal branch than among galaxies on 
the diagonal branch.

%%%%%%%%%%%%%%%%%%%%%%%%%%%%%%%%%%%%%%%%%%%%%%%%%%%%%%%%%%%%%%%%%%
\begin{figure*}[t]
\begin{center}
\includegraphics[scale=0.99]{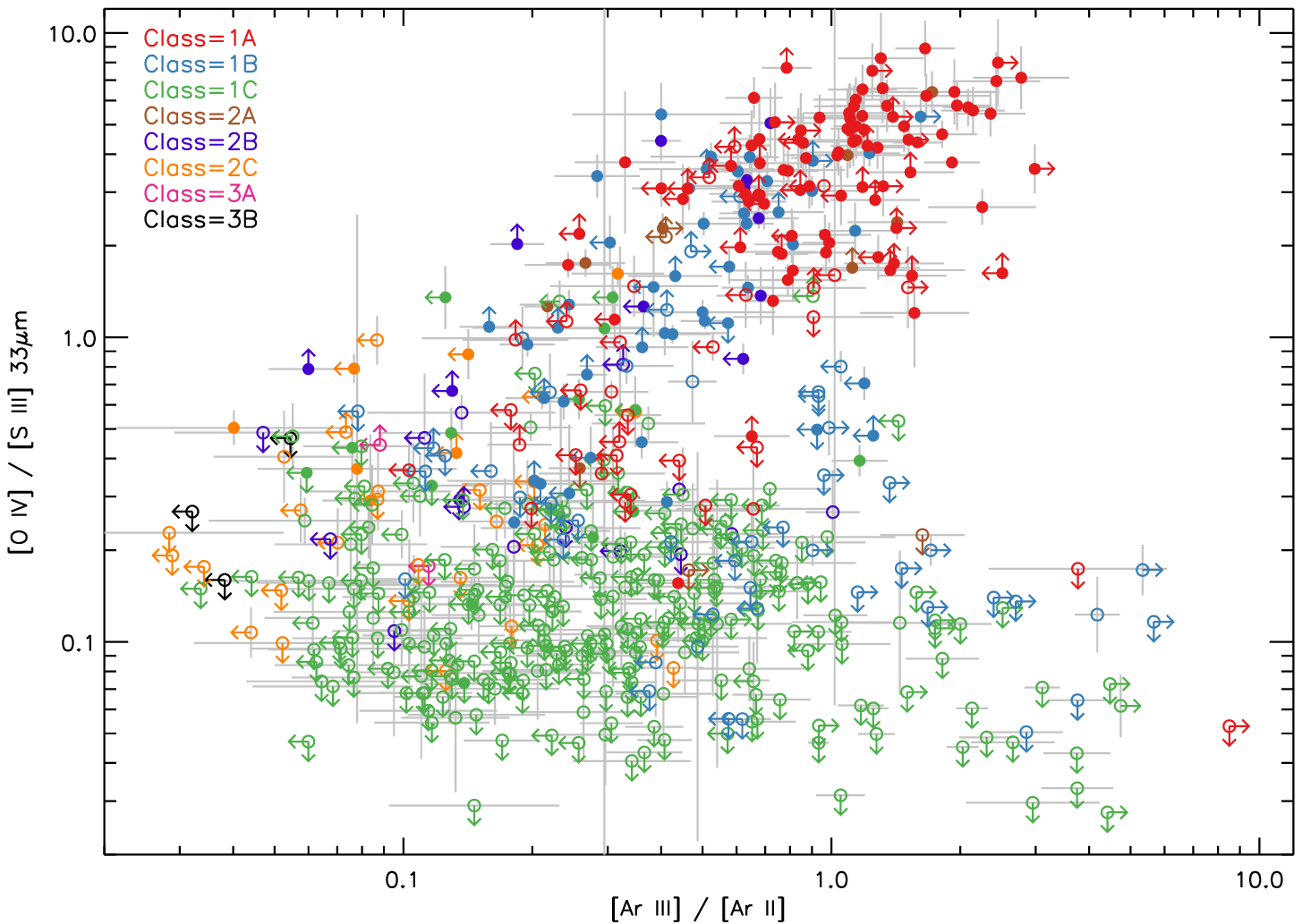}
\end{center}
\caption{Excitation diagram based on the 6.99\,$\mu$m [Ar {\sc ii}],
8.99\,$\mu$m [Ar {\sc iii}], 25.89\,$\mu$m [O {\sc iv}], and 
33.48\,$\mu$m [S {\sc iii}] lines.
565 galaxies with measurements for all four lines are plotted.
Bonafide AGN, as confirmed by a detection of a [Ne {\sc v}] line, 
are shown as filled circles. The galaxies are color-coded by
mid-infrared galaxy class as defined in Sect.\,\ref{sec:spoondiagram}.
\label{fig:o4s3_ar3ar2}}
\end{figure*}
%%%%%%%%%%%%%%%%%%%%%%%%%%%%%%%%%%%%%%%%%%%%%%%%%%%%%%%%%%%%%%%%%%

For galaxies at redshifts above 1.2 all three neon lines used in
the diagnostic diagram of Fig.\,\ref{fig:ne5ne2_ne3ne2} are 
redshifted out of the JWST-MIRI range. The only bright mid-infrared 
fine-structure lines left to probe the hardness of the radiation field
are the 6.99\,$\mu$m [Ar {\sc ii}], the 8.99\,$\mu$m [Ar {\sc iii}]
and the 7.65\,$\mu$m [Ne {\sc vi}]\footnote{We did not fit the
7.65\,$\mu$m [Ne {\sc vi}] line, as doing so would have required 
creating a CHUNKFIT model for a wavelength range in which the local
continuum is hard to define.} lines. These three lines can be 
detected with MIRI up to z=2.2. 

To assess whether the [Ar {\sc iii}]/[Ar {\sc ii}] ratio by itself suffices
as an AGN/starburst diagnostic, in Fig.\,\ref{fig:o4s3_ar3ar2},
we plot the [Ar {\sc iii}]/[Ar {\sc ii}] ratio versus the 
[O {\sc iv}]/[S {\sc iii}] ratio.
Like in Fig.\,\ref{fig:o4s3_ne3ne2}, galaxies are distributed along 
two prongs of a fork. The upper diagonal branch is populated with 
sources hosting an AGN, whereas the horizontal/downward tipping
branch we find star forming galaxies and low-metallicity galaxies.
This separation is clearer than in the previous two diagnostic
diagrams where most class 1C galaxies are found at the bottom
end of the diagonal branch along with the active galaxies higher up.
It is clear that, without measurement of the 7.65\,$\mu$m [Ne {\sc vi}] 
line (or coronal lines at shorter mid-infrared wavelengths), it is
impossible to determine from the [Ar {\sc iii}]/[Ar {\sc ii}] ratio alone
whether the galaxy is an AGN-starburst composite or purely star formation powered.
The argon line ratio by itself (just like the 
[Ne {\sc iii}]/[Ne {\sc ii}] ratio) is thus not a good AGN/starburst diagnostic.

\section{Crystalline silicates}\label{sec:crystsil}

\subsection{Crystalline silicate inventory}\label{sec:cryst_inventory}

%%%%%%%%%%%%%%%%%%%%%%%%%%%%%%%%%%%%%%%%%%%%%%%%%%%%%%%%%%%%%%%%%%
\begin{figure*}[t]
\includegraphics[scale=1]{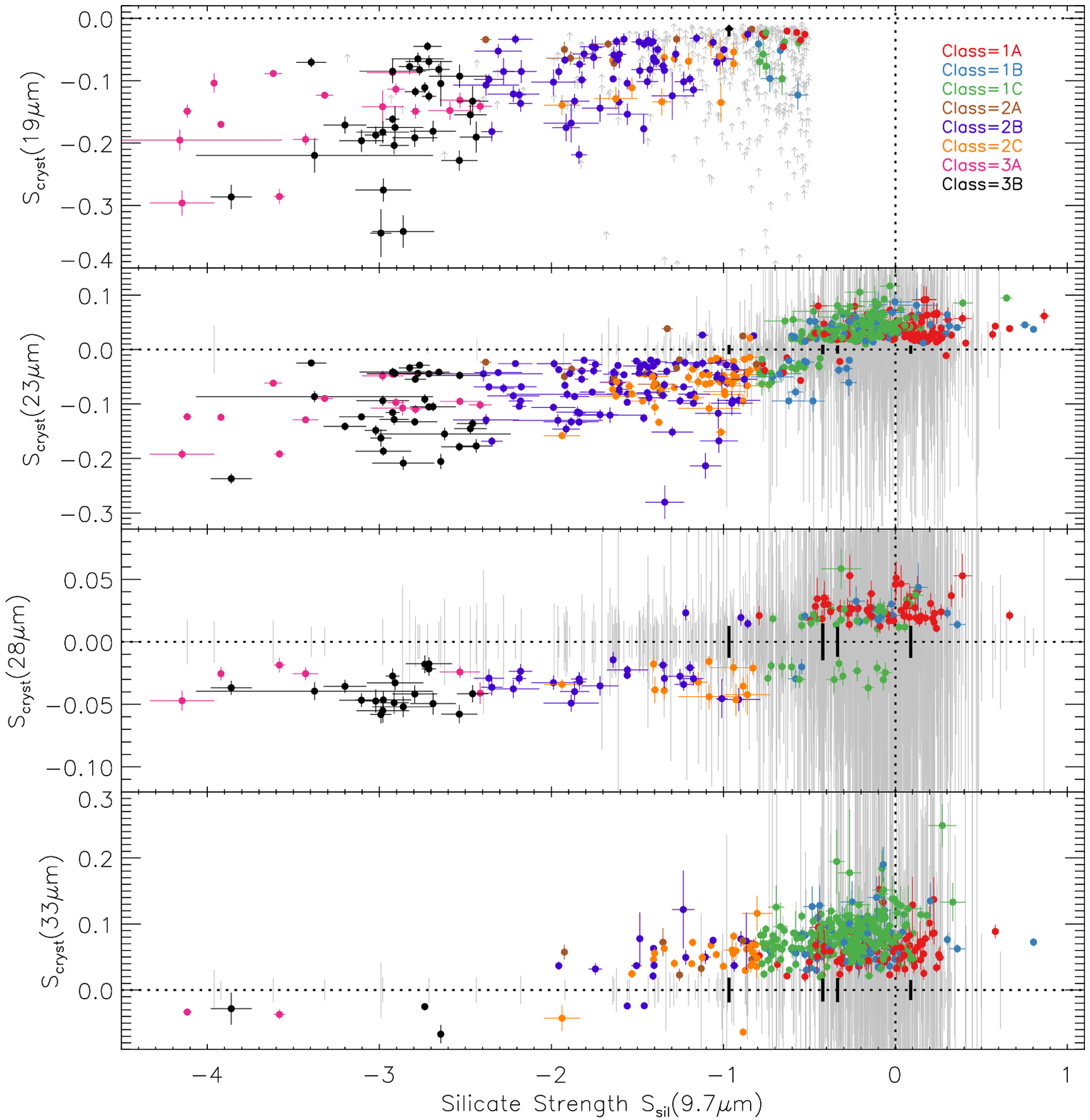}
\caption{Strength of four crystalline silicate bands as a function of 
the 9.7\,$\mu$m amorphous silicate strength. Detections of the
crystalline bands are shown as filled circles, non-detections in gray and black.
Since the 16 and 19\,$\mu$m bands are fitted as absorption features, and 
only in spectra with S$_{\rm sil}$$<$-0.5 (see Sect.\,\ref{sec:partial1421}),
a non-detection of the 19\,$\mu$m crystalline silicate band is
marked by a gray upward arrow.
The 23, 28 and 33\,$\mu$m crystalline silicate bands, on the other
hand, are allowed to be in emission or absorption 
(Sect.\,\ref{sec:partial1936}). 
Non-detections of these bands are shown as gray vertical 
bars stretching from -3$\sigma$ to +3$\sigma$.
The non-detections shown in black identify four sources 
(2MASS\,J19563578+1119050 at S$_{\rm sil}$=-0.97, ESO\,239-IG\,002 
at S$_{\rm sil}$=-0.42, II\,Zw40 at S$_{\rm sil}$=-0.34, and 
3C\,390.3 at S$_{\rm sil}$=0.09) that show
no trace of crystalline silicate features at any wavelength.
\label{fig:crystsil_silstrength}}
\end{figure*}
%%%%%%%%%%%%%%%%%%%%%%%%%%%%%%%%%%%%%%%%%%%%%%%%%%%%%%%%%%%%%%%%%%

As part of our SED fitting (Sect.\,\ref{sec:sedfitting}), we have 
detected {at a $>$95\% confidence level} emission and absorption 
features of crystalline silicates in 786/3335 IDEOS galaxies. 
These detections range from detections of a 
single band to detections of all\footnote{960/3335 galaxies have full 
coverage of this entire range. These galaxies necessarily reside at z$\leq$0.068.} 
five fitted bands in the 16--34\,$\mu$m range.
We find the detections not to be limited to a specific galaxy 
population. Crystalline silicates are, for instance, detected 
in low-metallicity galaxies (e.g. Haro\,11), but also in quasars
(e.g. 3C\,273), and in at 
least\footnote{Most spectra of early-type galaxies lack the 
crystalline-silicate-studded 23--34\,$\mu$m spectral range.}
one early-type galaxy, NGC\,1209.

We define the crystalline silicate strength in the same way as the
strength of the 9.8\,$\mu$m amorphous silicate feature (S$_{\rm sil}$;
Eq.\,\ref{eqn:silstrength}). A positive S$_{\rm cryst}$ indicates a
crystalline silicate feature seen in emission. 

Fig.\,\ref{fig:crystsil_silstrength} shows the strength of the 
crystalline silicate features as a function of S$_{\rm sil}$ for four
of the bands. The fifth band, the 16\,$\mu$m band, is not displayed,
as we impose a fixed ratio to the 19\,$\mu$m band (see Sect.\,\ref{sec:partial1421} 
and Appendix\,\ref{sec:appendix-c}). Shown in gray are the 
3\,$\sigma$ upper limits for non-detections of the crystalline
silicate features. Galaxies on the right (S$_{\rm sil}$$>$-0.8) are 
mostly AGNs (classes 1A and 1B in Fig.\ref{fig:forkdiagram})
and starburst galaxies (class 1C) and constitute 3/4 of the sources 
plotted in the figure. 
Towards the left the remaining sources are increasingly enshrouded.
The strongest feature in emission is the 33\,$\mu$m band, the
strongest features in absorption are the 16 \& 19\,$\mu$m bands.
Overall the detection rate of crystalline silicate bands is highest 
among the class 2A/B/C and 3A/B sources. For classes 1A/B/C only 17--26\% of 
galaxies have a detection of the 23\,$\mu$m feature, 4--14\% of 
the 28\,$\mu$m feature, and 33--56\% of the 33\,$\mu$m feature.
These percentages would be higher if the S/N of the spectra were higher.

All four panels in Fig.\,\ref{fig:crystsil_silstrength} show a clear trend 
of increasing crystalline silicate strength with increasing amorphous
silicate strength. This trend takes an almost linear form for the 19 and
23\,$\mu$m band, whereas for the other bands the slope reduces to
zero for the most deeply buried sources (S$_{\rm sil}<$-2).
Also notable is the shift in x-intercept of any fit to the observed
trend, shifting to more negative amorphous silicate strength as the
wavelength of the crystalline band plotted increases from 19 to 23, 28 
and 33\,$\mu$m.
The color coding by mid-IR class further reveals that 
class 2A sources (orange) display shallower 19 and 23\,$\mu$m 
crystalline bands than the class 2B and 2C sources in the same 
S$_{\rm sil}$ bracket (-2.4 to -0.8). We also see clear differences 
in the median S$_{\rm cryst}$(33\,$\mu$m) for class 1A and 1C 
sources: 0.057 and 0.080, respectively.

%%%%%%%%%%%%%%%%%%%%%%%%%%%%%%%%%%%%%%%%%%%%%%%%%%%%%%%%%%%%%%%%%%
\begin{figure*}[t]
\centering
\includegraphics[scale=0.8]{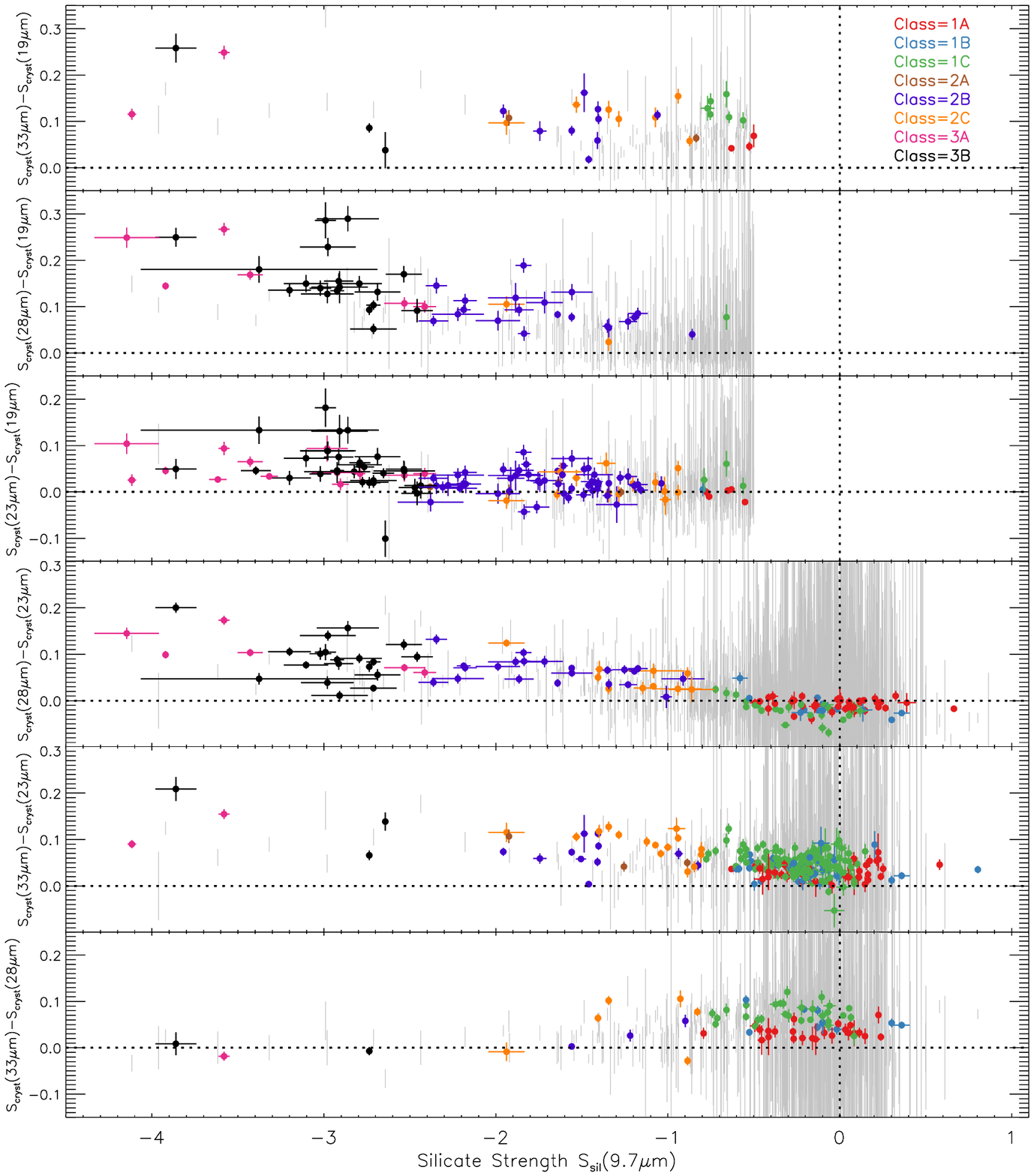}
\caption{Strength differences between the four crystalline silicate
bands as a function of the 9.7\,$\mu$m amorphous silicate strength.
Sources shown as filled circles have detections of both crystalline
bands. Sources shown as gray bars have limits on one or both
crystalline bands. Since we fit the 19\,$\mu$m crystalline band
only at S$_{\rm sil}<$-0.5, the strength differences in the upper
panels are only defined at S$_{\rm sil}<$-0.5.
\label{fig:crystsil_diff_silstrength}}
\end{figure*}
%%%%%%%%%%%%%%%%%%%%%%%%%%%%%%%%%%%%%%%%%%%%%%%%%%%%%%%%%%%%%%%%%%

Comparison of the strengths of the 19, 23, 28 and 33\,$\mu$m 
crystalline silicate bands on a source-to-source basis provides
another important insight: for every band pair the strength of 
the shorter wavelength feature is equal to or lower than that 
of the longer wavelength feature\footnote{S$_{\rm cryst}$(19\,$\mu$m)$\leq$S$_{\rm cryst}$(23\,$\mu$m)$\leq$S$_{\rm cryst}$(28\,$\mu$m)$\leq$S$_{\rm cryst}$(33\,$\mu$m)}. 
This difference is largest for the most deeply obscured sources 
(S$_{\rm sil}$$\sim$-4; class 3A) and gradually decreases as 
S$_{\rm sil}$ becomes less negative. Remarkably, this decrease 
appears to unfold in a 
similar way for galaxies on the mixing line towards classic AGNs 
(class\,3A$\rightarrow$2A$\rightarrow$1A) and towards classic 
starburst galaxies 
(class\,3A$\rightarrow$3B$\rightarrow$2B$\rightarrow$2C$\rightarrow$1C):
look for the few orange colored class 2A sources mixed in with the
turquoise and blue colored class 2B and 2C sources at intermediate
silicate strength (S$_{\rm sil}$=-2.4 to -0.8).
The differences in band strengths are smallest for the galaxies 
at the other extreme of the silicate strength scale: for galaxies 
on the classic AGN-starburst mixing line (1A--1B--1C) for which 
the 9.8\,$\mu$m silicate strength is close to zero and for 
which the 23, 28 and 33\,$\mu$m crystalline silicate bands are 
observed to be mostly in emission (Fig.\,\ref{fig:crystsil_silstrength}).

\subsection{Silicate features in spectra of centrally heated dust geometries}\label{sec:centrally_heated}

%%%%%%%%%%%%%%%%%%%%%%%%%%%%%%%%%%%%%%%%%%%%%%%%%%%%%%%%%%%%%%%%%%
\begin{figure}[t]
\includegraphics[scale=0.57]{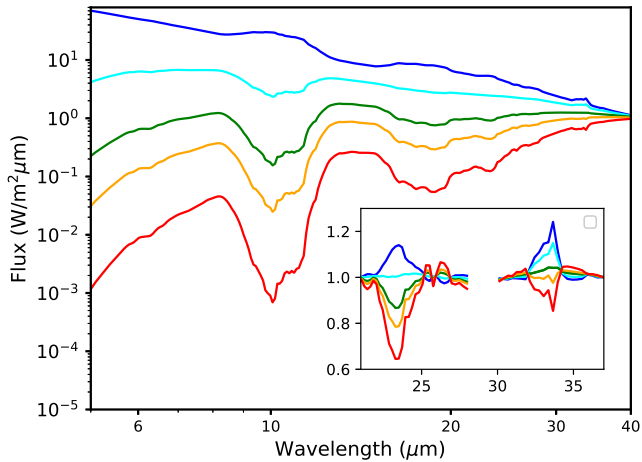}
\caption{Simulated mid-infrared spectra of five centrally heated dust
shells. The dust mass increases by a factor 100 from the top to the
bottom spectrum (blue to red). 
Clearly visible at various wavelengths are the radiative transfer 
effects caused by enhanced opacity associated with the presence of 
amorphous and crystalline silicates at those wavelengths. 
The strongest of these opacity enhancements, centered 
at 9.8 and 18\,$\mu$m, are associated with vibration modes of 
amorphous silicates. Narrower features, centered at 11, 16, 19, 
23, 28 and 33\,$\mu$m are vibration modes of crystalline silicates 
(forsterite).
At the lowest dust column (blue spectrum), all silicate features 
are in emission, as the optical depth does not surpass unity
at any wavelength. As the dust column increases, features switch 
from emission to absorption at wavelengths where the local optical 
depth exceeds unity. The highest optical depth is measured
in the red spectrum: $\tau_{9.8}$=85. 
The inset shows the continuum-normalized spectral profiles of the
23 and 33\,$\mu$m Forsterite features. The 23\,$\mu$m band switches 
from emission to absorption at a lower dust column density than 
required for the 33\,$\mu$m band to become optically thick. 
\label{fig:model_spectra}}
\end{figure}
%%%%%%%%%%%%%%%%%%%%%%%%%%%%%%%%%%%%%%%%%%%%%%%%%%%%%%%%%%%%%%%%%%

The above observations are consistent with an origin for the
amorphous and crystalline silicate features in a centrally
heated dust geometry, either spherical in nature or arranged
in a torus geometry. 

In Fig.\,\ref{fig:model_spectra} we show simulated 5--40\,$\mu$m 
spectra for five centrally heated dust shells with dust masses 
increasing by a factor 100 from the top to the bottom spectrum.
The model spectra have been computed using the 2D version of the
MCMAX code \citep{min09}, with the following parameters. For the 
central heating source we adopt a 2500\,K stellar SED.
The gas density falls off as r$^{-1}$, and is parametrized as in
\cite{meixner02}. The dust mixture is composed of 75\% MgFeSiO$_4$
amorphous dust (Jena database optical constants; \cite{dorschner95}), 
20\% am-Carbon (optical constants of \cite{zubko96}), and 
5\% forsterite Mg$_2$SiO$_4$ (optical constants of \cite{suto06}).
The optical constants are temperature dependent between 50 and
295\,K. We adopt a dust size of 0.1\,$\mu$m and a DHS grain shape
distribution \citep{min05} with f$_{\rm max}$=0.8.
As can be seen in Fig.\,\ref{fig:model_spectra}, any departure
from a flat or monotonously changing opacity curve as a function 
of wavelength will result in spectral structure to appear in 
the emerging spectrum of the centrally heated dust structure.
For the model spectra in Fig.\,\ref{fig:model_spectra}, the
inclusion of minerals in the dust composition has the effect
of increasing the dust opacity around the central wavelengths
of the vibration modes of the included amorphous and crystalline 
silicate species. Radiative transfer then dictates that, if the
dust structure is optically thin at all mid-infrared wavelengths, 
all mineral features will appear as emission features in the 
emerging spectrum (the dark blue spectrum in Fig.\,\ref{fig:model_spectra}).
Increasing the amount of dust around the heating source will
result in the wavelength range to which the 9.8\,$\mu$m mode
of amorphous silicate contributes to become optically thick
first, resulting in the feature to switch from an emission to
an absorption feature (the light blue spectrum in 
Fig.\,\ref{fig:model_spectra}) thanks to the negative temperature 
gradient in the centrally heated dust structure. Residing deep 
within the 9.8\,$\mu$m amorphous silicate opacity profile, 
also the 11\,$\mu$m forsterite band will be in the optically 
thick regime and, consequently, appear in absorption too. 
A further increase in the amount of dust around the heating 
source will bring the broad 18.5\,$\mu$m amorphous silicate 
band, along with the narrower 16 and 19\,$\mu$m crystalline
silicate bands, into absorption as the optical 
depth at these wavelengths surpasses unity
(the green and orange spectra in Fig.\,\ref{fig:model_spectra}).
Next are the 23 and 28\,$\mu$m crystalline silicate bands.
The last\footnote{Besides the silicate features also
the dust continuum becomes optically thick.} 
mid-infrared feature to go into absorption is the 
33\,$\mu$m crystalline silicate band. This is shown
by the red spectrum in Fig.\,\ref{fig:model_spectra}.

In the absence of PAH emission, the shortest wavelength
silicate bands will be the most sensitive probes of the 
obscuring dust column. In practice, however, only above 
20\,$\mu$m PAH emission associated with circumnuclear 
star formation will not fill in the silicate 
emission and absorption features. This thus leaves the 
23, 28 and 33\,$\mu$m forsterite bands as the only 
suitable probes of nuclear obscuration in galaxy-integrated
spectra.

Fig.\,\ref{fig:model_spectra} also clearly illustrates
another radiative transfer effect: the redistribution
of infrared radiation to longer wavelengths as the dust
column increases. From the top to the bottom model SED
the C(30\,$\mu$m)/C(5.5\,$\mu$m) continuum slope changes
by several orders of magnitude. Unfortunately, stellar
photospheric emission and PAH emission at 5.5\,$\mu$m 
may contaminate the accuracy of this diagnostic by 
boosting the emission at 5.5\,$\mu$m.

\subsection{Tracers of nuclear obscuration in dusty galactic nuclei}\label{sec:tracers_nucobs}

\begin{figure}[t!]
\centering
\begin{tabular}{c}
\includegraphics[scale=0.43]{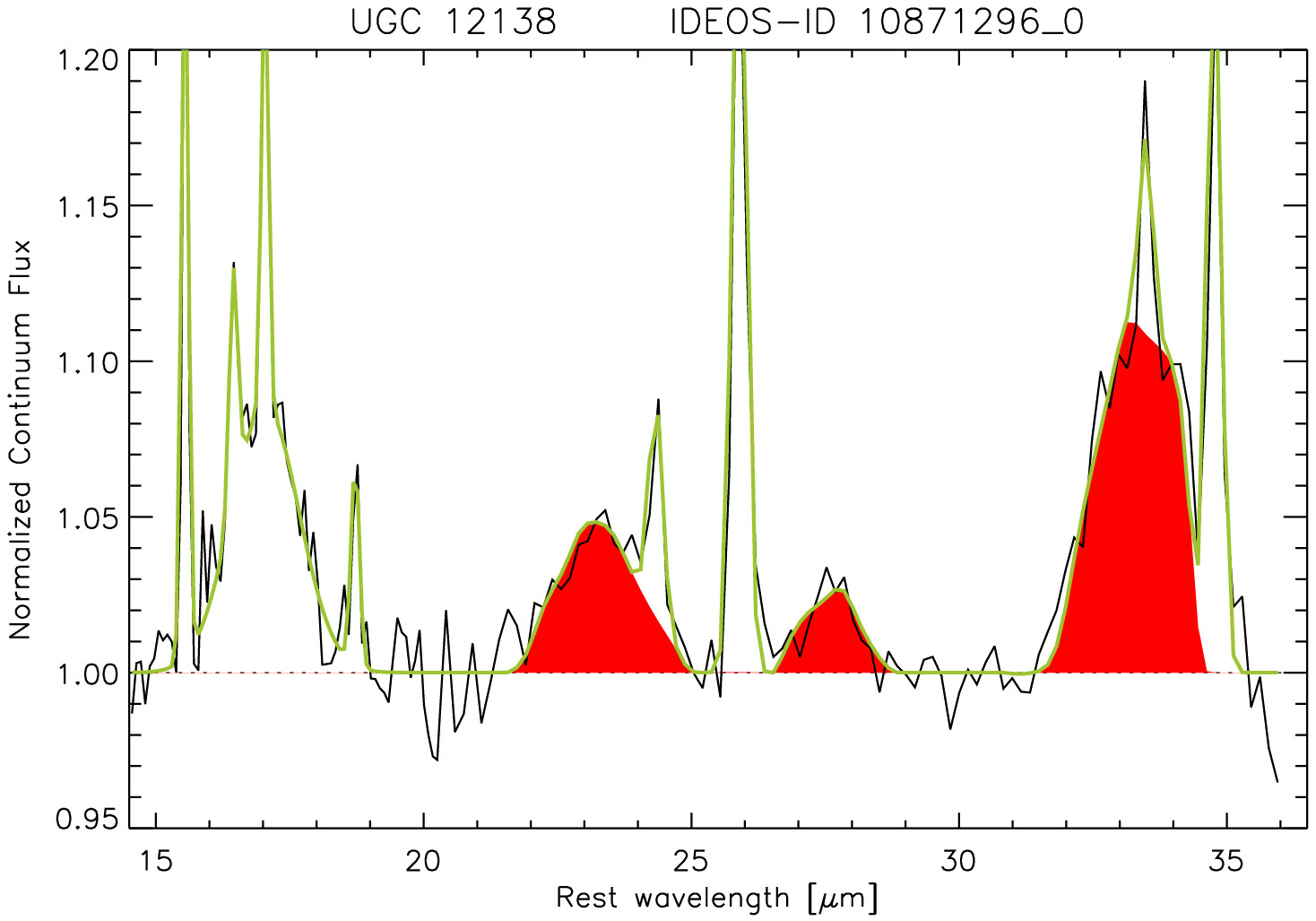}\\
\includegraphics[scale=0.43]{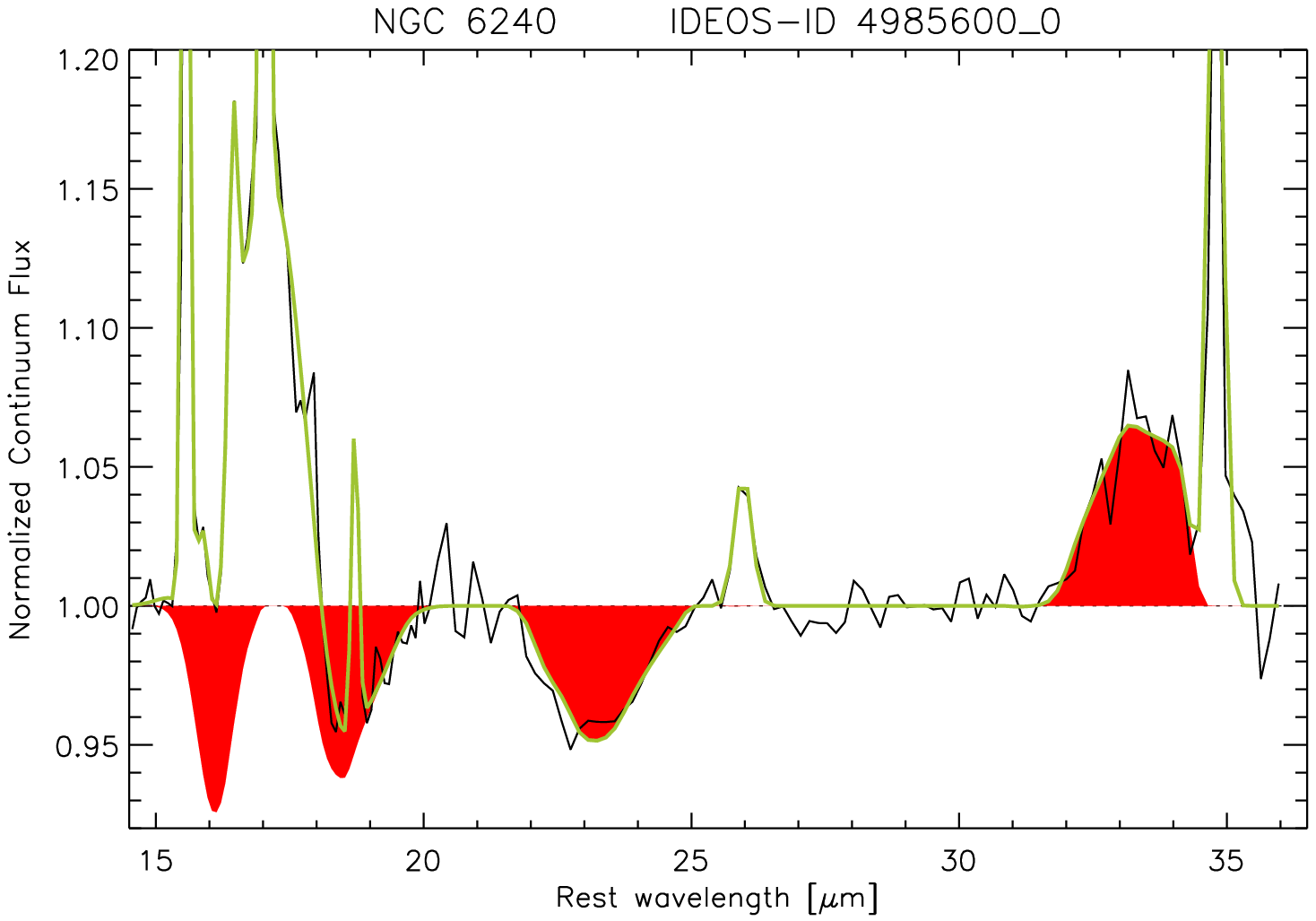}\\
\includegraphics[scale=0.43]{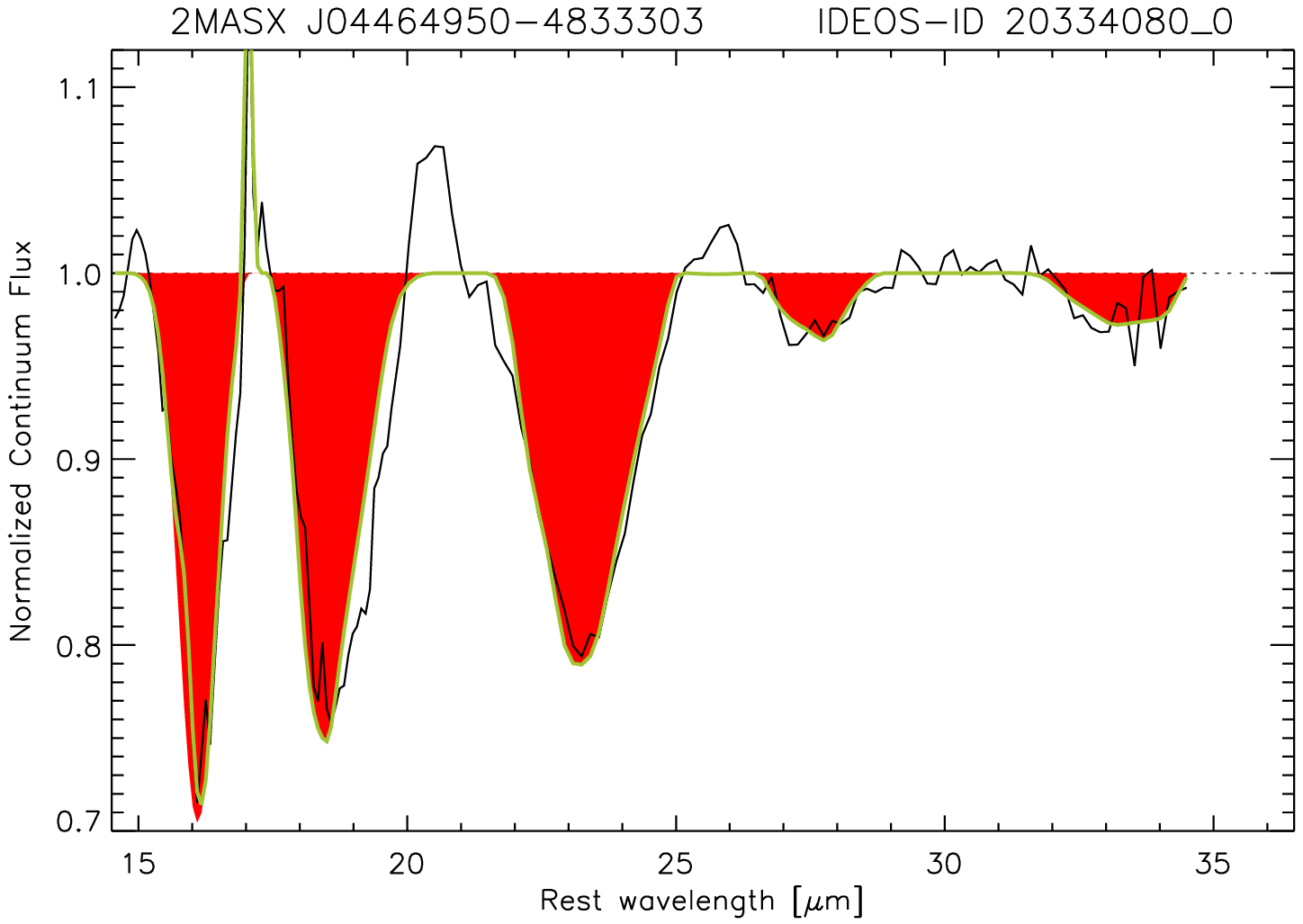}
\end{tabular}
\caption{16--33\,$\mu$m crystalline silicate spectra arranged in
order of increasing obscuration.
Each spectrum is normalized by dividing the observed spectrum by the
fitted continuum. In the top spectrum (S$_{\rm sil}=$-0.13) all fitted 
crystalline bands (red) are in emission. In the bottom spectrum 
(S$_{\rm sil}=$-3.86) all are in absorption. The spectrum in the 
middle panel (S$_{\rm sil}=$-1.41) shows the 33\,$\mu$m crystalline 
feature in emission and the 23\,$\mu$m crystalline
feature in absorption, while the feature in between, the 28\,$\mu$m 
crystalline band, is undetected. 
\label{fig:silicate_spectrum}}
\end{figure}

Assuming the nuclei of dusty IDEOS sources to ressemble centrally 
heated dust geometries, we can use the 23\,$\mu$m 
and 33\,$\mu$m forsterite bands to devise a crude measure of nuclear 
obscuration:
\begin{itemize}
\item 23\,$\mu$m and 33\,$\mu$m bands in emission: low obscuration. 
\item 23\,$\mu$m band in absorption and the 33\,$\mu$m band 
in emission: intermediate obscuration.
\item 23\,$\mu$m and 33\,$\mu$m bands in absorption: high obscuration. 
\end{itemize}
These three regimes are illustrated in Fig.\,\ref{fig:silicate_spectrum}
by the 15--35\,$\mu$m continuum-normalized spectra of three galaxies
with decreasing silicate strengths (-0.13, -1.41, and -3.86).
Note the absence of the 28\,$\mu$m feature in the middle
spectrum. In this source the dust continuum at 28\,$\mu$m 
is likely in between optically thin and thick, rendering 
the feature undetectable despite a non-zero column of 
forsterite. This same effect can be seen among the model spectra 
in Fig.\,\ref{fig:model_spectra}, where the 33\,$\mu$m forsterite 
feature is indiscernible in the orange spectrum, which represents 
the second highest dust column density in the plot.
Unfortunately, the proximity of the 33\,$\mu$m forsterite band
to the long wavelength spectral cut-off of IRS-LL1 limits the use 
of this diagnostic to sources at z$<$0.068. This thus
excludes\footnote{IRAS\,01003--2238 (z=0.118), IRAS\,F01166--0844SE
(z=0.118), IRAS\,03158+4227 (z=0.134), IRAS\,09039+0503 (z=0.125), 
IRAS\,09539+0857 (z=0.129), IRAS\,F21329--2346 (z=0.125), and
IRAS\,F16156+0146NW (z=0.132) all show what looks like a blue wing 
of a 33\,$\mu$m crystalline absorption feature right up to the 
end of the spectrum. If real, this, in combination 
with our detection of the 23\,$\mu$m crystalline absorption feature,
would indicate that these sources harbor a highly obscured nucleus.} 
most of the local ULIRG population.

%%%%%%%%%%%%%%%%%%%%%%%%%%%%%%%%%%%%%%%%%%%%%%%%%%%%%%%%%%%%%%%%%%
%%%%%%%%%%%%%%%%%%%%%%%%%%%%%%%%%%%%%%%%%%%%%%%%%%%%%%%%%%%%%%%%%%

In our sample of 232 sources with detections for both the 23 
and the 33\,$\mu$m crystalline band we find no sources for 
which the 23\,$\mu$m crystalline band is in emisson and the 
33\,$\mu$m feature is in absorption. Since this combination 
would be unphysical for a centrally heated dust structure, 
this means that, in principle, all these 232 sources could 
harbor a centrally heated dust structure.

Further insights can be gathered from overlaying our 23 and
33\,$\mu$m crystalline silicate diagnostic onto the Fork Diagram
(Sect.\,\ref{sec:spoondiagram}) and onto the diagram that delineates 
the slope of the 5.5--30\,$\mu$m continuum and the 9.8\,$\mu$m 
amorphous silicate strength (Sect.\,\ref{sec:contslope_diagnostics}). 
As discussed in Sect.\,\ref{sec:spoondiagram}\,\&\,\ref{sec:contslope_diagnostics},
these diagrams delineate three types of sources:
\begin{itemize}
\item Galaxies dominated by a centrally heated dust structure with 
low EQW(PAH62) and mostly cool mid-infrared colors (class 3A) 
\item Galaxies dominated by a classic AGN 
with similarly low EQW(PAH62) and warm mid-infrared colors (class 1A) 
\item Classic starburst galaxies with high EQW(PAH62) and
cool mid-infrared colors (class 1C).
\end{itemize}

\begin{figure*}[t]
\begin{tabular}{lr}
\includegraphics[scale=0.57]{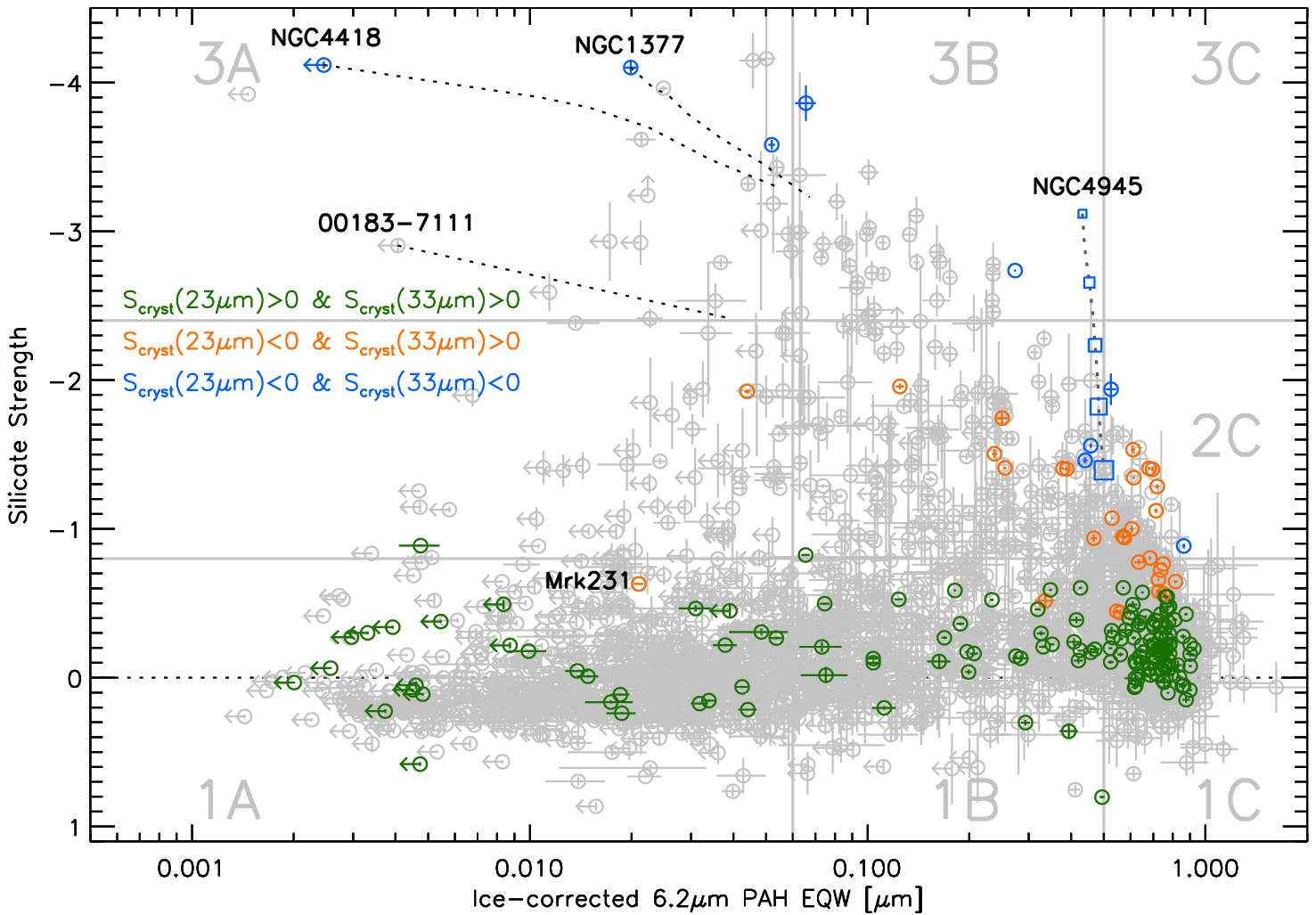} &
\includegraphics[scale=0.592]{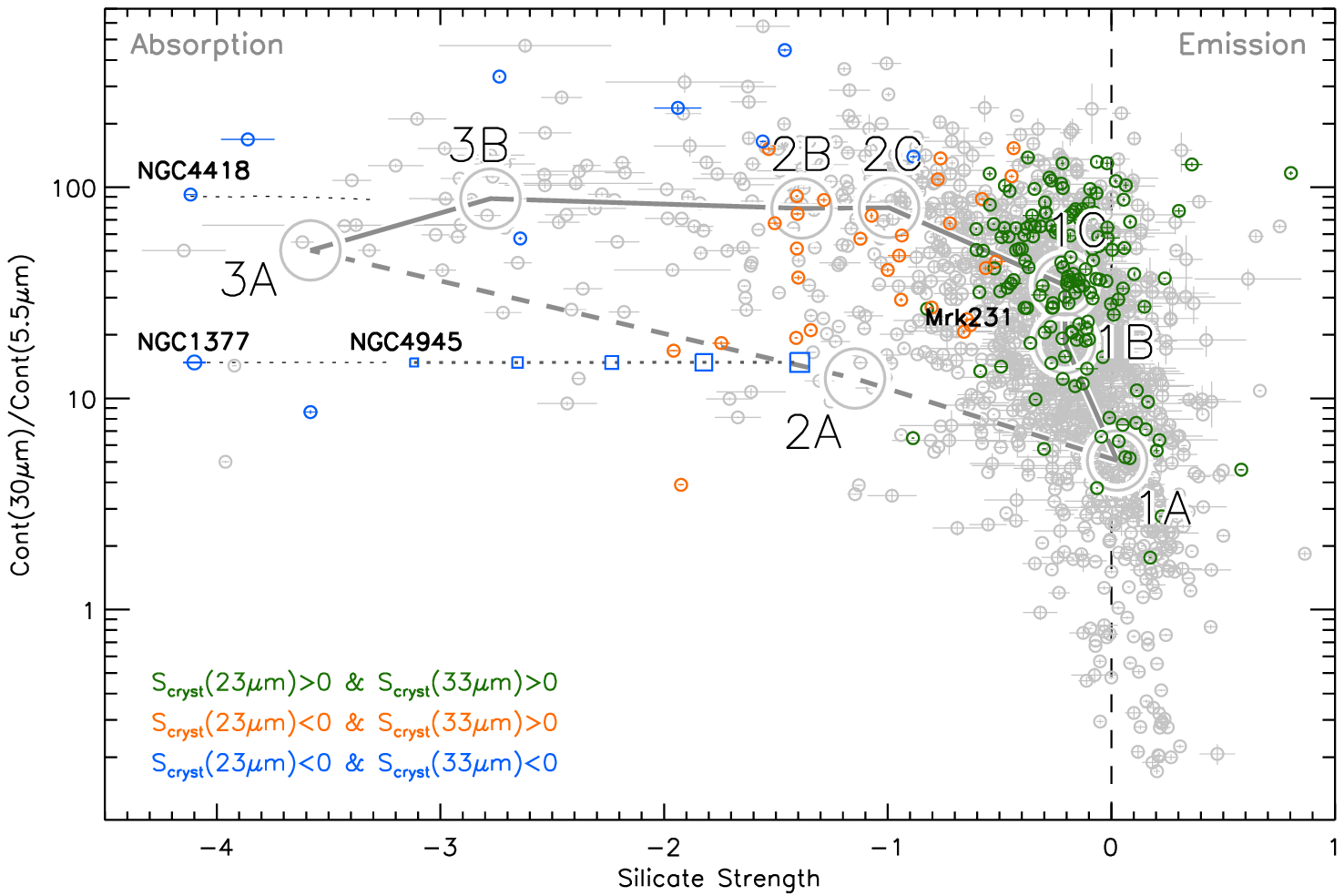}\\
\includegraphics[scale=0.57]{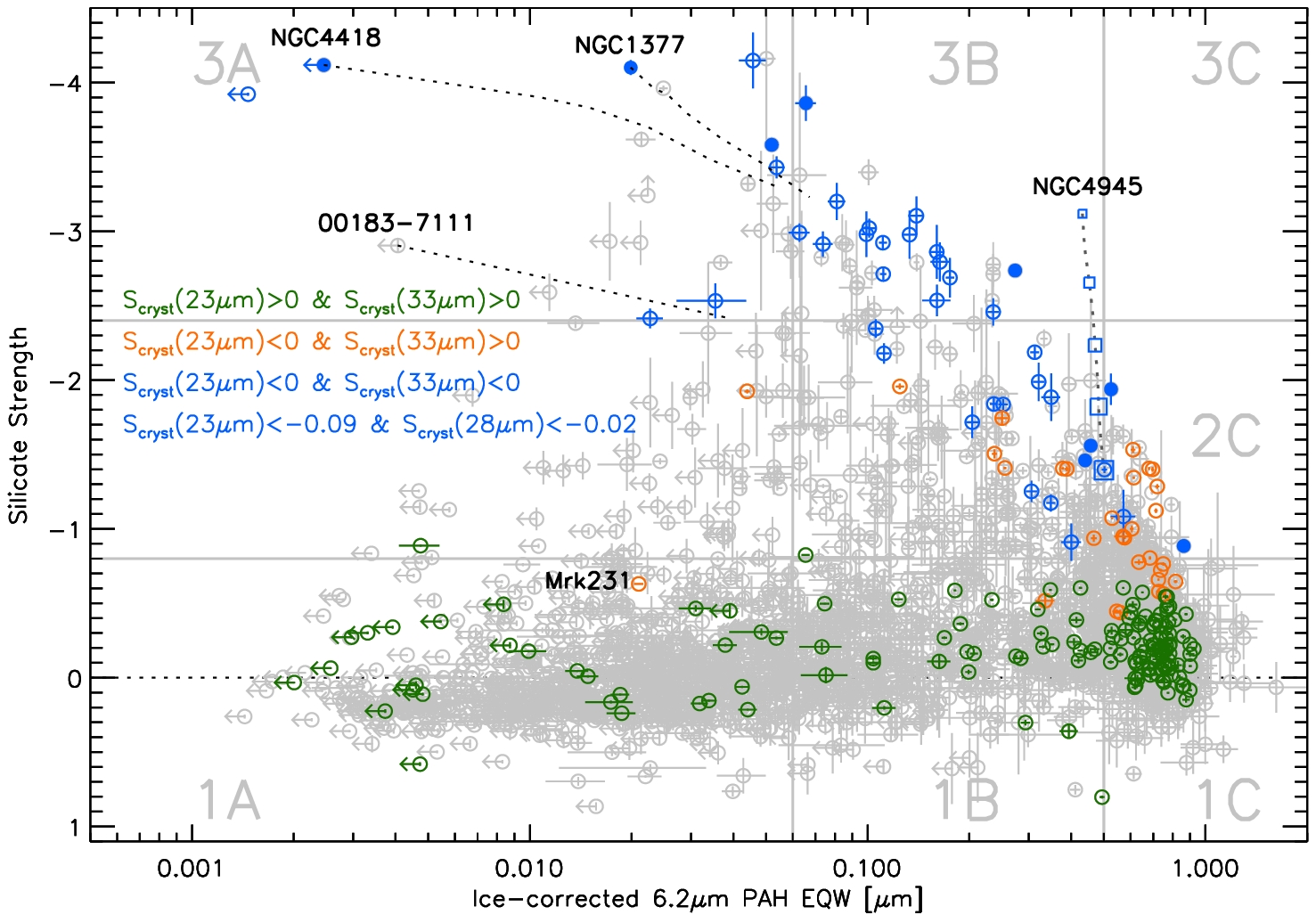} &
\includegraphics[scale=0.592]{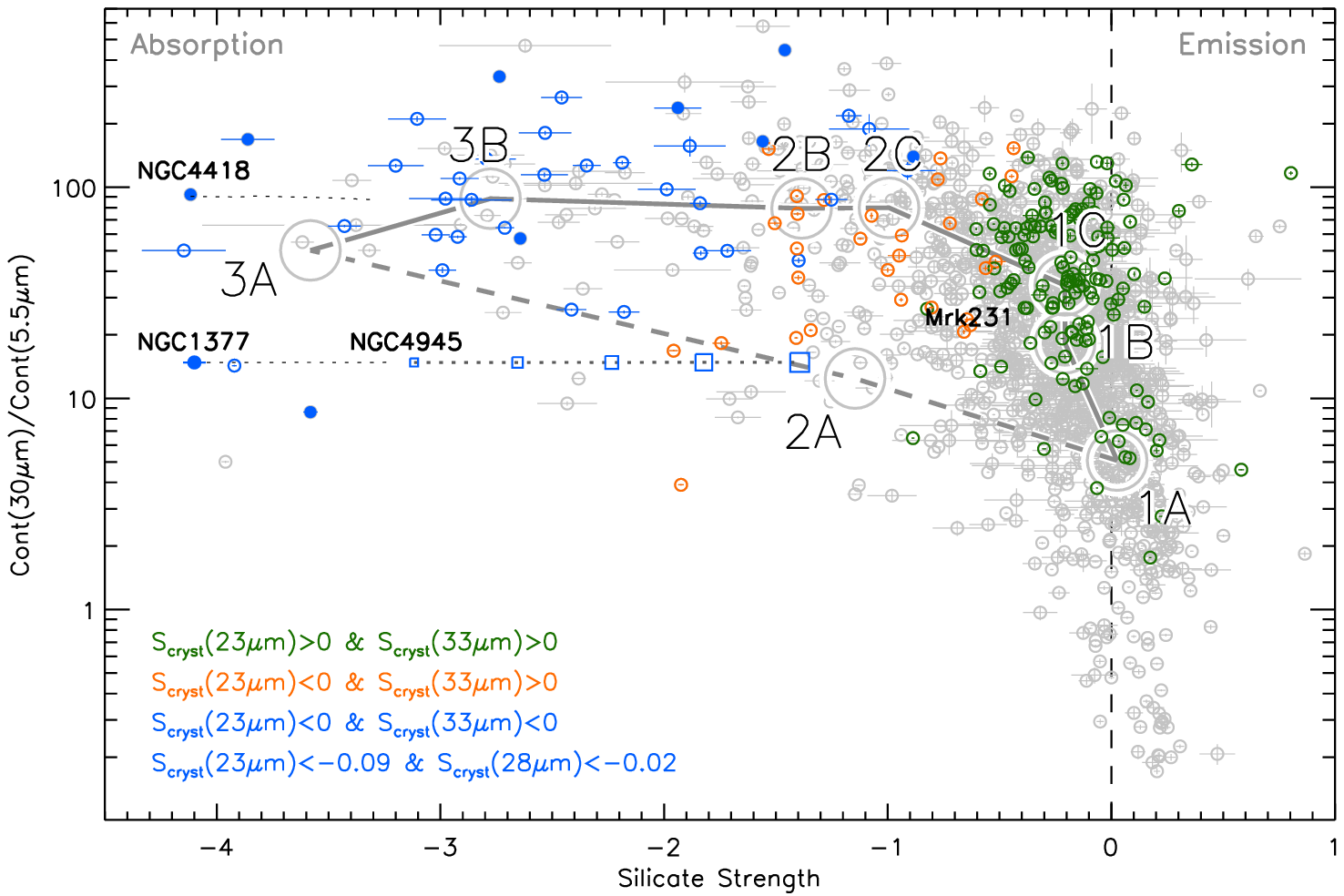}
\end{tabular}
\caption{Left: Diagnostic diagram combining the ice-corrected equivalent
width of the 6.2\,$\mu$m PAH feature (Sect.\,\ref{sec:partial57})
with the silicate strength. Sources plotted as gray circles are the
same sources as in Fig.\,\ref{fig:forkdiagram}.
Overplotted in color are sources with particular combinations of 
crystalline silicate strengths.
Green denotes S$_{\rm cryst}$(23\,$\mu$m)$>$0 \& S$_{\rm cryst}$(33\,$\mu$m)$>$0.
Orange denotes S$_{\rm cryst}$(23\,$\mu$m)$<$0 \& S$_{\rm cryst}$(33\,$\mu$m)$>$0.
Blue filled circles denotes S$_{\rm cryst}$(23\,$\mu$m)$<$0 \& 
S$_{\rm cryst}$(33\,$\mu$m)$<$0, whereas blue open circles denote 
S$_{\rm cryst}$(23\,$\mu$m)$<$-0.09 \& S$_{\rm cryst}$(28\,$\mu$m)$<$-0.02.
There are no occurences of 
S$_{\rm cryst}$(23\,$\mu$m)$>$0 \& S$_{\rm cryst}$(33\,$\mu$m)$<$0.
The dotted lines extending from NGC\,4418, NGC\,1377, and
IRAS\,F00183--7111 in the direction of class 1C show the 
effect of adding circumnuclear star formation (NGC\,4945 disk 
template) to the spectra of these sources. The lines end where
the circumnuclear star formation reaches 20\% of the total
5.5--13\,$\mu$m flux. 
The connected blue squares show the effect of larger and larger
spectral apertures centered on the nucleus of NGC\,4945.
Right: Diagnostic diagram of the rest frame 30\,$\mu$m to 5.5\,$\mu$m
continuum ratio versus the 9.8\,$\mu$m silicate strength.
Sources and lines shown in gray are the same as in 
Fig.\,\ref{fig:c30c55_silstrength}. Color-coding is the same 
as in the left panel.
\label{fig:forkdiagram_c30c55_silstrength_crystdiff}}
\end{figure*}

If we color-code galaxies with 23\,$\mu$m \& 33\,$\mu$m bands 
that are both in emission as green, galaxies with the 23\,$\mu$m 
band in absorption and the 33 \,$\mu$m band in emission as
orange, and galaxies with the 23\,$\mu$m \& 33\,$\mu$m bands 
both in absorption as blue, the upper panels of 
Fig.\,\ref{fig:forkdiagram_c30c55_silstrength_crystdiff} 
show the blue (high obscuration) and green (low obscuration) 
galaxies to be clearly separated, with the orange (intermediate
obscuration) sources found in between.

Since our sample of high-obscuration sources is quite small,
we can use the sample's S$_{\rm cryst}$(23$\mu$m) and S$_{\rm cryst}$(28$\mu$m)
characteristics to find additional sources beyond the
redshift cut-off for measuring the 33\,$\mu$m forsterite band
(z$>$0.068). Using the criterion S$_{\rm cryst}$(23$\mu$m)$<$-0.09 and 
S$_{\rm cryst}$(28$\mu$m)$<$-0.02 we find 32 additional 
high-obscuration galaxies up to z=0.257, which we plot along
with the other high-obscuration sources as blue symbols in 
the lower panels of 
Fig.\,\ref{fig:forkdiagram_c30c55_silstrength_crystdiff}.
This further populates the upper envelope of the diagonal branch 
in the Fork Diagram and the left half of the C(30\,$\mu$m)/C(5.5\,$\mu$m)
versus silicate strength diagram, while still showing a clear
separation between blue and green sources in both.

%%%%%%%%%%%%%%%%%%%%%%%%%%%%%%%%%%%%%%%%%%%%%%%%%%%%%%%%%%%%%%%%%%
%%%%%%%%%%%%%%%%%%%%%%%%%%%%%%%%%%%%%%%%%%%%%%%%%%%%%%%%%%%%%%%%%%

For sources along the mixing line 3A$\rightarrow$2A$\rightarrow$1A,
for which the obscuration decreases (silicate strength becomes 
less negative), this separation between high and low obscuration 
sources is more or less expected because emission components 
associated with star formation (e.g. PAH features) do not
complicate the interpretation.
For sources along the mixing line 
3A$\rightarrow$3B$\rightarrow$2B$\rightarrow$2C$\rightarrow$1C 
(which traces the diagonal branch in the Fork Diagram), however, 
the observed separation between blue and green sources is less 
straightforward to interpret due to the increasing importance 
of exposed (as opposed to buried) star formation towards class 1C. 

If circumnuclear star formation were negligible among the sources
with crystalline silicate diagnostics in 
Fig.\,\ref{fig:forkdiagram_c30c55_silstrength_crystdiff}, the
interpretation of the clear separation between blue and green 
sources and the presence of orange sources dotted in between 
would be straightforward: the nuclear power source 
becomes less and less buried along the mixing line from 
class 3A to class 1C.
In a more realistic scenario, however, we recognize that the
physical projected size covered by the Spitzer-IRS slit 
expands with redshift and will contain some or all of the
galaxy disk in the majority of our spectra. The 
23--33\,$\mu$m crystalline signatures of the nuclear SED
will thus be polluted with optically thin, warm dust continuum 
emission associated with circumnuclear star formation. 
\cite{diaz-santos10} 
have shown that these contributions can be significant.
For ULIRGs up to 30\% of the 5--15\,$\mu$m emission may 
originate from outside the nucleus \citep{marshall18}.

We have used the IRS mapping observations of the nearby
Seyfert-2 galaxy NGC\,4945 \citep[Spoon et al. in prep.;][]{perez-beaupuits11}
to create a 5--36\,$\mu$m template spectrum of cicumnuclear star formation. 
The spectrum comprises all spatial positions in a LL1-gridded map 
where S$_{\rm sil}$$>$-0.6, thus excluding the deeply buried nucleus. 
While the nuclear spectrum clearly shows the 23 and 33\,$\mu$m forsterite
bands in absorption, the circumnuclear spectrum is devoid
of crystalline silicate bands.
{\it If the absence of crystalline features in this galaxy disk 
spectrum is representative for galaxy disks in general\footnote{We
also inspected spectra taken of H{\sc ii} regions in the disk 
of M\,101 \citep{gordon08} and found no crystalline silicate features.
Even if we had, the contribution of H{\sc ii} regions to the
total disk spectrum would be small.} , 
circumnuclear star formation will not change the diagnostic
power of the 23 and 33\,$\mu$m forsterite bands}, as adding
continuum to the nuclear SED will only weaken the nuclear
crystalline features, but not flip them from emission to
absorption, or vice versa. The main result will be a
displacement of the source in the Fork Diagram towards 
class 1C as the silicate feature fills up and PAH emission 
features strengthen compared to the continuum.
This effect is illustrated by the dotted tracks originating
from the buried nuclei of NGC\,4418, NGC\,1377 \citep{roussel06}, 
and IRAS\,F00183--7111 \citep{spoon04} in all panels of
Fig.\,\ref{fig:forkdiagram_c30c55_silstrength_crystdiff},
and by the track for different sized apertures centered on
the buried AGN in NGC\,4945.

In this interpretation, the lone blue point among a sea of 
orange in the Fork Diagram, 
close to the border of quadrants 2C and 1C (at EQW(PAH62)=0.86\,$\mu$m
and S$_{\rm sil}$=-0.88), has to be a galaxy with a deeply
buried nucleus, which has been displaced in the direction
of class 1C by a strong contribution of circumnuclear
star formation. Indeed, archival HST imaging of IRAS\,17578--0400N
shows complex extended emission around the galaxy nucleus 
(Armus, priv. comm.). And HCN mm-observations by \cite{falstad21}
result in the nucleus to be classified as a 'CON', a compact 
obscured nucleus. 
Circumnuclear star formation is thus able to disguise a buried
nucleus at wavelengths below 20\,$\mu$m, but not above. The 
classification into blue, orange and green points thus seems to
be a better indication for the level of dust obscuration in
a nucleus than the position of a source in the Fork Diagram.
Only if aperture sizes were chosen to exclude the galaxy disks
the Fork Diagram diagnostic features and the crystalline 
diagnostics would likely be in agreement.

Since in the Fork Diagram there are no orange or green 
labeled sources identified below S$_{\rm sil}$=-2, it may be 
reasonable to regard all galaxies below this limit 
as hosting 
a deeply obscured nucleus. This would increase the total 
count of confirmed deeply obscured nuclei from 40 to 
89.

Also note the position of the intermediate dust enshrouded
nucleus of Mrk\,231 (color-coded orange) in the left panels of 
Fig.\,\ref{fig:forkdiagram_c30c55_silstrength_crystdiff},
close to the locus of the classical AGNs on the horizontal
branch. Unlike Mrk\,231, classical AGN do not show evidence for
centrally heated dust geometries as revealed by their assigned
green color-code. The simplest explanation for the proximity of
Mrk\,231 to the classical AGN is that the dust structure 
which covers the central power source has a 
small opening through which unattenuated continuum leaks 
out and fills up the 9.8\,$\mu$m silicate absorption feature. 
As modeled by \cite{marshall18}, such a ``keyhole'' can be fairly
small: less than 10\% of the surface area of the mid-infrared shell.
The additional continuum does not affect the crystalline silicate
diagnostics, as it cannot flip the 23\,$\mu$m forsterite feature 
from absorption to emission as required to turn the source from
orange to green in
Fig.\,\ref{fig:forkdiagram_c30c55_silstrength_crystdiff}. 
The nucleus of Mrk\,231 is thus more deeply 
obscured than the silicate strength indicates.
Interestingly, this conclusion is supported by our spectral 
decomposition using QUESTFIT (Sect.\,\ref{sec:questfit}), 
which requires one deeply obscured component and one unobscured 
component for the best fit. Further evidence 
in support of this conclusion will be presented below.

%%%%%%%%%%%%%%%%%%%%%%%%%%%%%%%%%%%%%%%%%%%%%%%%%%%%%%%%%%%%%%%%%%
\begin{figure}[t]
\begin{center}
\includegraphics[scale=0.57]{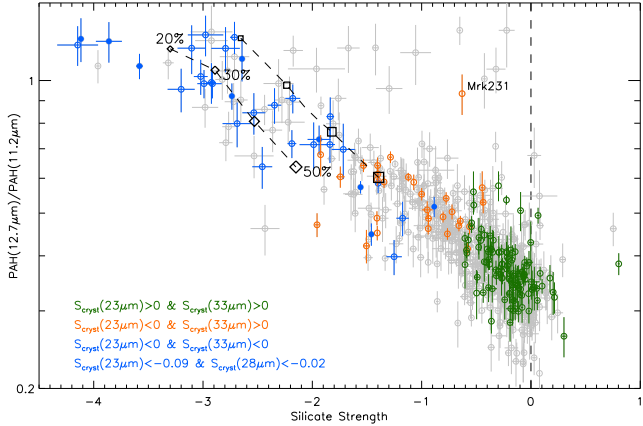}
\end{center}
\caption{Ratio of the 12.7\,$\mu$m to the 11.2\,$\mu$m PAH feature
as a function of silicate strength for sources with 8$\sigma$
detections of both features. Sources shown in blue, orange
or green have detections for both the 23 and 33\,$\mu$m crystalline 
silicate bands. Color-coding is the same as in 
Fig.\ref{fig:forkdiagram_c30c55_silstrength_crystdiff}. 
Overlaid are two tracks showing the effect of increased galaxy
disk contribution. The track with squares superimposed shows the effect
of increasing the aperture size centered on the nucleus of
NGC\,4945 from 13$"$$\times$13$"$ to 76$"$$\times$31.5$"$.
The other track (diamonds) shows the effect of adding disk emission 
(using the NGC\,4945 disk template) to the spectrum of NGC\,4418.
The percentages listed alongside the latter track indicate the 
circumnuclear contribution as part of the total 5.5--13\,$\mu$m 
emission for that track.
\label{fig:pahratio_silstrength}}
\end{figure}
%%%%%%%%%%%%%%%%%%%%%%%%%%%%%%%%%%%%%%%%%%%%%%%%%%%%%%%%%%%%%%%%%%

Another probe of the obscuration in a galactic nucleus is 
the ratio of the 12.7 and 11.2\,$\mu$m
PAH fluxes. This ratio is sensitive to the silicate column 
density through the effects of differential extinction in 
the steep red wing of the 9.8\,$\mu$m silicate feature. 
This aspect was explored by \cite{hernan20} using a 
subsample of IDEOS sources. In the absence of obscuration 
the PAH127/PAH112 ratio has a median value of 0.38 with 
a 5\% dispersion (related to PAH excitation conditions in 
individual galaxies). As shown in Fig.\,\ref{fig:pahratio_silstrength},
the PAH112 feature weakens relative to the PAH127 feature
as the silicate strength becomes more negative. The highest
observed PAH127/PAH112 ratio in our sample, and the highest
permitted ratio in CHUNKFIT (Sect.\,\ref{sec:partial1013}), 
occurs among the most deeply buried sources: 
PAH127/PAH112=1.3. While the large majority of galaxies
scatters around the diagonal line in the plot, there are 
some clear outliers: galaxies for which the silicate 
strength is less negative than based on just the PAH ratio.
For some of these galaxies, mostly classic AGN, the
high ratio reflects intrinsic conditions in the galaxy, 
while in other cases the PAH112 and PAH127 features were
observed in different spectral orders (e.g. SL1 and LL2),
which may result in spectral stitching-related fitting anomalies.
Colored data points in Fig.\,\ref{fig:pahratio_silstrength}
are sources for which our 23 \& 33\,$\mu$m and 23 \& 28\,$\mu$m 
crystalline silicate diagnostics are available.
Again we see a clear separation between the sources
color-coded in blue and green, with intermediate
obscuration nuclei (color-coded in orange) found in 
between. This distribution suggests that the silicate 
strength, the crystalline silicate diagnostics, and the 
PAH127/PAH112 ratio largely trace each other.

%%%%%%%%%%%%%%%%%%%%%%%%%%%%%%%%%%%%%%%%%%%%%%%%%%%%%%%%%%%%%%%%%%
\begin{figure*}[t]
\begin{tabular}{lr}
\includegraphics[scale=0.57]{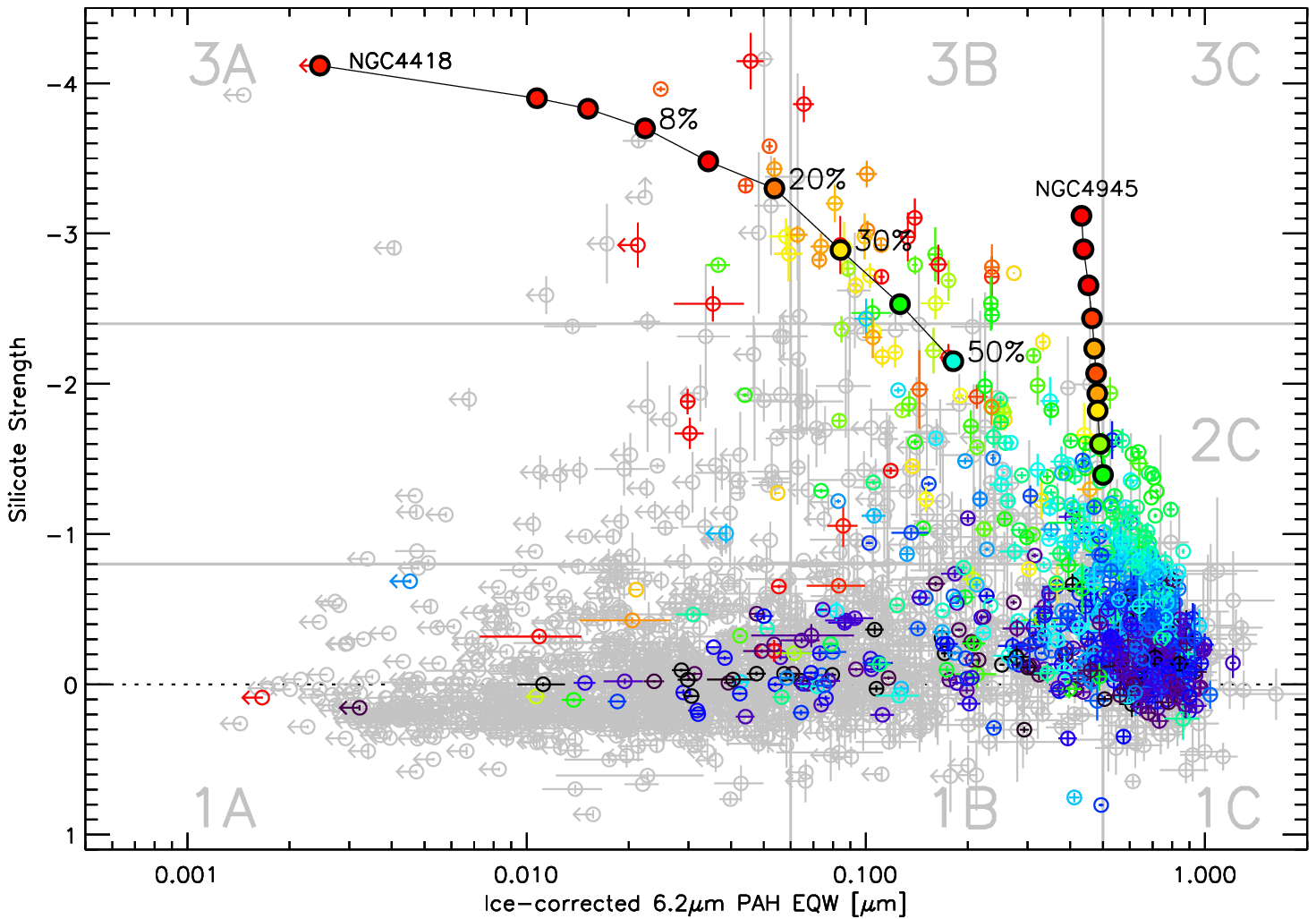} &
\includegraphics[scale=0.595]{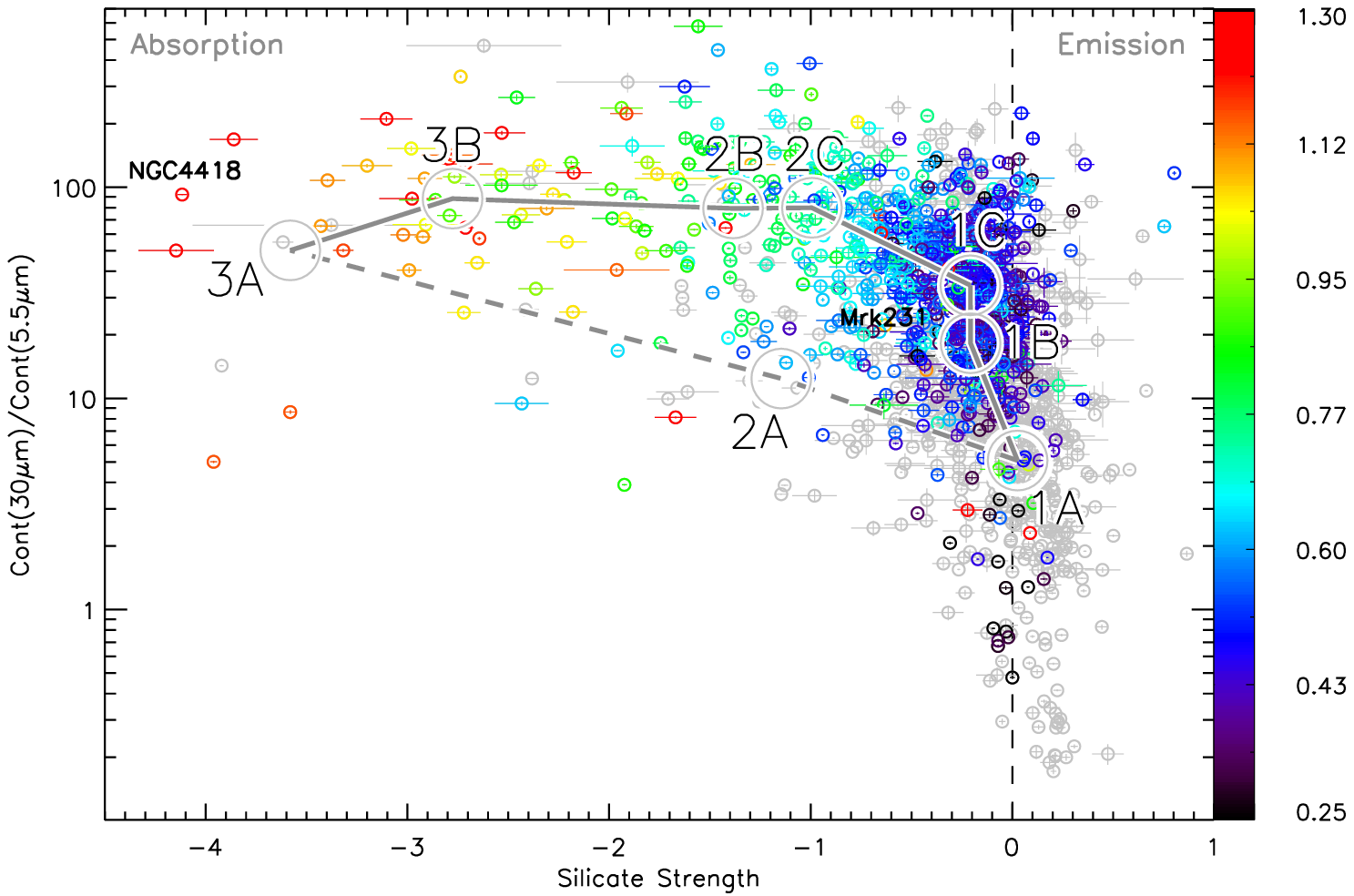}
\end{tabular}
\caption{Left: Diagnostic diagram combining the ice-corrected equivalent
width of the 6.2\,$\mu$m PAH feature (Sect.\,\ref{sec:partial57})
with the silicate strength. The sources plotted in gray are the
same sources as in Fig.\,\ref{fig:forkdiagram}.
Overplotted in rainbow colors (see legend in right panel) are 
sources with detections of both
the PAH11 and the PAH127 feature. Red denotes a PAH127/PAH11 ratio 
around 1.3, while black denotes a PAH127/PAH11 ratio below 0.25.
Overplotted are two tracks. The track for NGC\,4945 shows how
the PAH127/PAH11 ratio changes as the aperture centered on the
nucleus increases in size from 5.5$"$$\times$5.5$"$ to 
76$"$$\times$31.5$"$. The other track shows how
the PAH127/PAH11 ratio changes when more and more disk emission
(NGC4945 template) is added to the spectrum of NGC\,4418. The 
percentage listed indicates what fraction of the total 5.5--13\,$\mu$m
flux is comprised of disk emission.
Right: Diagnostic diagram of the rest frame 30\,$\mu$m to 5.5\,$\mu$m
continuum ratio versus the 9.8\,$\mu$m silicate strength.
Sources and lines shown in gray are the same as in 
Fig.\,\ref{fig:c30c55_silstrength}. The color-coding is the
same for both panels.
\label{fig:forkdiagram_c30c55_silstrength_pahratio}}
\end{figure*}
%%%%%%%%%%%%%%%%%%%%%%%%%%%%%%%%%%%%%%%%%%%%%%%%%%%%%%%%%%%%%%%%%%

Overlaid on the plot are two tracks. The first one, connecting 
the four dark squares, show where the spectra of different 
sized apertures (13$"$$\times$13$"$ to 76$"$$\times$31$"$) 
centered on the buried nucleus of NGC\,4945 fall on 
the plot.
The other track shows the effect of adding circumnuclear 
star formation (using the NGC\,4945 circumnuclear template)
to the spectrum of NGC\,4418. Both tracks clearly show that,
as more circumnuclear star formation is added to the spectrum
of a buried source, the source shift towards a lower PAH ratio
and a less negative silicate strength without leaving the
distribution. The PAH127/PAH112 ratio by itself is thus not a 
good tracer of nuclear obscuration unless the spectral slit is 
small enough to isolate the nucleus. 
The ability of sources to shift to the lower right along 
the diagonal may explain the presence of several blue sources 
among the orange ones. For instance IRAS\,17578--0400N at 
S$_{\rm sil}$=-0.88 and PAH127/PAH112=0.51, discussed before.

The figure also shows Mrk\,231 to be located far to the
right of the main correlation. For the observed PAH127/PAH112
ratio of 0.93 to fit in with the main trend, the silicate 
strength would have to be much more negative: -2 to -2.5 rather
than -0.63. As discussed above, unobscured mid-infrared emission 
protruding from a key hole opening in the obscuring shell 
may fill in the silicate feature and cause Mrk\,231 to 
shift to the right.

Fig.\,\ref{fig:forkdiagram_c30c55_silstrength_pahratio}
shows the Fork Diagram and the 30-to-5.5\,$\mu$m versus
silicate strength diagram color-coded by the value of
the PAH127/PAH112 ratio. Along the class 
3A$\rightarrow$3B$\rightarrow$2B$\rightarrow$2C$\rightarrow$1C
mixing line the PAH ratio shows a clear progression 
towards a lower ratio. When interpreted as an extinction 
probe, this suggests a decrease of nuclear obscuration
from class 3A to class 1C.
Overlaid on the Fork Diagram are two tracks showing the 
effect on the PAH ratio of adding circumnuclear star formation. 
One track for NGC\,4945, 
using aperture sizes increasing from 5.5$"$$\times$5.5$"$ 
to 76$"$$\times$31.5$"$ centered on the buried nucleus,
the other showing the effect of adding circumnuclear star 
formation to the deeply obscured nucleus of NGC\,4418.
As can be clearly seen, the color-coding of the tracks
blends in perfectly with the galaxy-integrated observations 
underneath. This confirms that the presence of circumnuclear 
star formation affects the PAH ratio in the same way as a 
decrease of nuclear obscuration does. Thus, unlike the 
crystalline silicate diagnostic, the PAH127/PAH112 ratio 
by itself cannot produce conclusive evidence for a decrease 
of nuclear obscuration from class 3A towards class 1C. It 
may at best be supportive.

\subsection{The crystallinity of the ISM in dusty galactic nuclei}\label{sec:crystallinity}

\begin{figure}[t]
\centering
\includegraphics[scale=0.57]{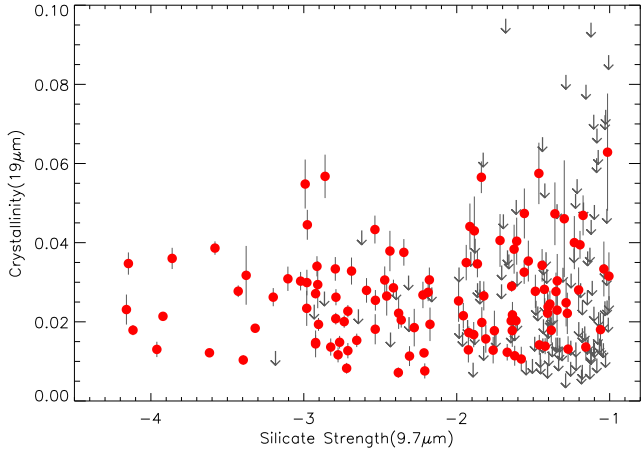}
\caption{Crystallinity of the interstellar medium in galaxies with
measurements of the 9.7\,$\mu$m amorphous silicate absorption
feature and the 18.5\,$\mu$m crystalline silicate absorption 
feature. Upper limits are shown as gray arrows.
\label{fig:crystallinity}}
\end{figure}

More than a decade ago, \cite{spoon06} used the strengths of the 16\,$\mu$m
crystalline silicate feature and 9.7\,$\mu$m amorphous silicate
feature to infer the crystallinity of the ISM in a sample of ULIRGs.
The method assumes that the feature arises from a column of cold dust
in front of a warm central source. Under these conditions the radiative 
transfer equation lacks an emission term.
Here we prefer to use the 19\,$\mu$m crystalline band instead, because
it is more frequently detected and modeled in Galactic sources (see 
Appendix\,\ref{sec:appendix-c}).
Following \cite{kemper04} we define the crystallinity of the ISM as 

\begin{equation}\label{eqn:crystallinity}
C_{\rm ISM} = \frac{N_{\rm cryst}}{N_{\rm cryst}+N_{\rm am}}
\end{equation}

where N$_{\rm cryst}$ and N$_{\rm am}$ are the column densities of 
crystalline and amorphous silicates, respectively. For sources 
with amorphous silicates firmly in absorption (S$_{\rm sil}<$-1), 
this becomes

\begin{equation}
C_{\rm ISM}(19\mu m) = \frac{S_{\rm cryst}(19\mu m)/\kappa_{\rm
    cryst}(19\mu m)}{S_{\rm cryst}(19\mu m)/\kappa_{\rm
    cryst}(19\mu m)+S_{\rm sil}/\kappa_{\rm am}(9.7\mu m)}
\end{equation}

where $\kappa_{\rm am}$(9.7$\mu$m) 
is the peak mass absorption coefficients of amorphous 
silicates at 9.7$\mu$m, 2.4$\times$10$^3$ cm$^2$ g$^{-1}$ \citep{dorschner95},
and $\kappa_{\rm cryst}$(19$\mu$m) is the
baseline-subtracted\footnote{To determine the peak mass absorption
coefficients of the 16--34\,$\mu$m forsterite features, we fitted a 
spline function to the continuum component underlying the features 
and subtracted it from the mass absorption spectrum.} peak mass
absorption coefficients of Forsterite at 19$\mu$m, 4.8$\times$10$^3$ cm$^2$ g$^{-1}$. 
The latter was computed with the grain model of \cite{min05}, using 
the optical constants of \cite{suto06} for Forsterite at T=200K, 
and adopting a distribution of hollow spherical grains with 
0.1\,$\mu$m radius and a maximum vacuum fraction of 80\%.
As can be seen in Fig.\,\ref{fig:crystallinity}, this 
results in 19$\mu$m-crystallinities between 0.7\% and 6\% 
for obscured nuclei (S$_{\rm sil}$$<$-1) with a detection 
of the 19$\mu$m crystalline silicate absorption 
feature (114/254 sources), and in upper limits 
as low as 0.3\% for sources with a non-detection of the feature 
(140/254 sources).
On average the 19$\mu$m crystallinity of these obscured nuclei 
(S$_{\rm sil}$$<$-1) is 2.7\%, with both the 
most uncertain crystallinities and the strictest upper limits 
occurring among the least obscured sources. 

The large spread in crystallinities seen in Fig.\,\ref{fig:crystallinity} 
may very well explain why among sources with deep silicate absorption
features (S$_{\rm sil}$$<$-2) and non-zero crystallinities we see so
many non-detections of 28 and 33$\mu$m crystalline bands in 
Fig.\,\ref{fig:crystsil_silstrength}.
Examples are 
IRAS\,02530+0211 (S$_{\rm sil}$=-3.6; C$_{\rm ISM}$(19$\mu$m)=1.2\%) % 6652160_0
and IRAS\,00188--0856 (S$_{\rm sil}$=-2.7; C$_{\rm ISM}$(19$\mu$m)=0.8\%), % 4962560_0
which have 3--4 times lower crystallinities than for instance 
IRAS\,04454--4838 (S$_{\rm sil}$=-3.9; C$_{\rm ISM}$(19$\mu$m)=3.6\%). % 20334080_0

Finally, are there any galaxies at S$_{\rm sil}$$<$-1 that lack detectable crystalline
silicates to strict limits? To investigate this question we have five
crystalline silicate bands at our disposal. We consider a band
undetected if S$_{\rm cryst}$(19$\mu$m)$>$-0.016, $|$S$_{\rm cryst}$(23$\mu$m)$|<$0.010, 
$|$S$_{\rm cryst}$(28$\mu$m)$|<$0.015, or $|$S$_{\rm cryst}$(33$\mu$m)$|<$0.02.
Among the 70 sources 
at S$_{\rm sil}$$<$-1 with coverage of all five cystalline silicate bands
we find none that meets the strict upper limits set for a
non-detection of all five bands.

\subsection{The detection rate of crystalline silicates in other galaxies}\label{sec:cryst_det_rate}

%%%%%%%%%%%%%%%%%%%%%%%%%%%%%%%%%%%%%%%%%%%%%%%%%%%%%%%%%%%%%%%%%%
%%%%%%%%%%%%%%%%%%%%%%%%%%%%%%%%%%%%%%%%%%%%%%%%%%%%%%%%%%%%%%%%%%

For the 2526 galaxies in our sampe that are not (deeply) enshrouded
(S$_{\rm sil}$$>$-1) our method is not suitable to infer a
crystallinity as the features it relies on are not optically thick.
We can determine, though, whether in a spectrum crystalline silicates 
are present or absent to meaningful limits by taking stock of the
measured crystalline silicate strengths in the 23, 28 and 33\,$\mu$m
bands\footnote{The 16 and 19\,$\mu$m bands are not included, since 
they are only fitted in spectra with S$_{\rm sil}$$<$-0.5.}.
While we won't be able to infer a column density without generating 
a full radiative transfer model for these galaxies, this exercise may 
provide insight into the prevalence of crystalline silicates and the 
conditions under which they can survive.

Apart from galaxies with detections of one, two, or all three bands,
there are only four galaxies among 874 galaxies at S$_{\rm sil}$$>$-1
with coverage of all three bands and the high S/N in their
22--34\,$\mu$m  spectra required to get below these strict limits
that do not show any evidence for the 
presence of these bands. These galaxies are  
2MASS\,J19563578+1119050 % IDEOSID=20313856_0
(S$_{\rm sil}$=-0.97), ESO\,239-IG002 % IDEOSID=20318208_0 
(S$_{\rm sil}$=-0.42), II\,Zw\,40 %9007616_0
(S$_{\rm sil}$=-0.34), and 3C\,390.3 % IDEOSID=4673024_0
(S$_{\rm sil}$=+0.09). All are marked by black error bars in 
Fig.\,\ref{fig:crystsil_silstrength}.
For three of these, their silicate strength in the -0.3 to -1.0 
range may imply that the 23, 28 and 33\,$\mu$m bands are simply
not fully sensitive to what would be a detectable column density
at a different silicate strength. This leaves 3C\,390.3 as the
only source for which the strict non-detection of crystalline 
silicate bands may be significant. That is, if synchrotron emission
at 33\,$\mu$m can be ruled out as a source of dilution of the 
33\,$\mu$m forsterite band of this broad-line radio galaxy.

%%%%%%%%%%%%%%%%%%%%%%%%%%%%%%%%%%%%%%%%%%%%%%%%%%%%%%%%%%%%%%%%%%
\begin{figure}[t]
\centering
\includegraphics[scale=0.57]{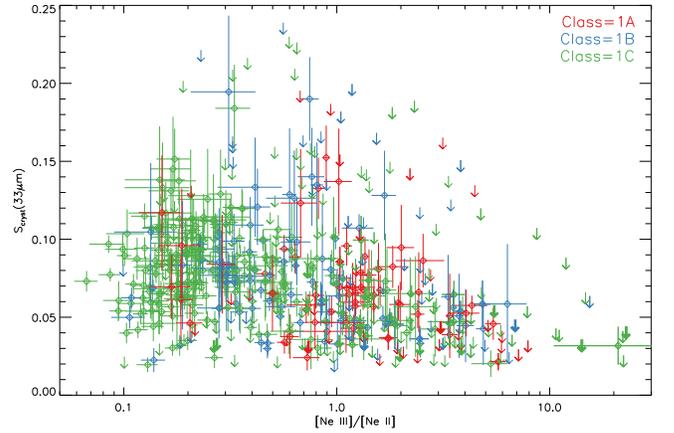}
\caption{Strength of the 33\,$\mu$m crystalline silicate emission band
as a function of the hardness of the radiation field in galaxies at
S$_{\rm sil}$$>$-0.8. Thick upper limit symbols denote the detection
of the 23\,$\mu$m forsterite band either in emission or absorption.
\label{fig:cryst33ne3ne2}}
\end{figure}
%%%%%%%%%%%%%%%%%%%%%%%%%%%%%%%%%%%%%%%%%%%%%%%%%%%%%%%%%%%%%%%%%%

Is there more that we can learn from the detectability of crystalline
silicate features in sources at S$_{\rm sil}$$>$-1 ?
All 2120 galaxies on the horizontal branch of the Fork Diagram 
(Fig.\,\ref{fig:forkdiagram}; classes 1A/B/C; S$_{\rm sil}$$>$-0.8) have
one thing in common: the 33\,$\mu$m crystalline silicate band is
expected to be seen in emission (see Fig.\,\ref{fig:crystsil_silstrength}).
Whether the band is discernable is not just a matter of column density or
temperature of the crystalline silicates, though, but also depends on the
strength of the local 33\,$\mu$m continuum and the strengths of the
33.48\,$\mu$m [S {\sc iii}] and 34.82\,$\mu$m [Si {\sc ii}] emission
lines. With these caveats in mind, we looked for trends among the 
834 class 1A/B/C sources at redshifts below z=0.068 for which the 33\,$\mu$m 
crystalline silicate band was in range.

Fig.\,\ref{fig:cryst33ne3ne2} shows the 33\,$\mu$m crystalline silicate
strength as a function of the hardness of the radiation field as
measured by the [Ne {\sc iii}]/[Ne {\sc ii}] ratio. As can be clearly
seen, the upper envelope of S$_{\rm cryst}$(33$\mu$m) decreases
towards a higher neon ratio with no distinction between mid-infrared
galaxy class (as shown by color coding).
Above [Ne {\sc iii}]/[Ne {\sc ii}]=6.5 there is only one more 
galaxy with a detection of the 33$\mu$m forsterite band: the
low-metallicity galaxy SHOC\,391 %18890752_0
at [Ne {\sc iii}]/[Ne {\sc ii}]=22.
This is, however, not the only galaxy above 
[Ne {\sc iii}]/[Ne {\sc ii}]=6.5 with a detection of forsterite.
As shown by thick upper limit symbols, there are two sources
with detections of the 23$\mu$m forsterite band, {\it in absorption}:
the low-metallicity galaxies SBS\,1533+574B % 8996352_0
and IRAS\,11485--2018. % 18891008_0
A similar trend can be seen by using [O {\sc iv}]/[Ne {\sc ii}] or 
[O {\sc iv}]/[S {\sc iii}]\,33\,$\mu$m instead,
with one important difference: since all low-metallicity galaxies have
upper limits for the [O {\sc iv}] line, as is true for all galaxies powered
by stellar sources of ionizing radiation (see e.g. Fig.\,\ref{fig:o4s3_ne3ne2}),
these galaxies appear along with other star forming galaxies
towards the extreme left in those diagrams.

Since a silicate strength is measured relative to the
local continuum, it is possible that the observed decrease in 
S$_{\rm cryst}$(33$\mu$m) with increasing [Ne {\sc iii}]/[Ne {\sc ii}]
can be fully attributed to an increase in the 33\,$\mu$m flux density.
The comparison of the panchromatic SED modeling of \cite{galliano08a}
to the mid-IR spectral decomposition of \cite{galliano08b}, 
for 30 nearby galaxies, indeed shows that an increase in 
[Ne {\sc iii}]/[Ne {\sc ii}] is accompanied by an increase 
in 33\,$\mu$m dust emission relative to the total IR emission.
Since it is beyond the scope of the current paper to include 
far-infrared data, we cannot determine whether the crystallinity
decreases as the hardness of the ionizing SED increases, or 
whether the 33\,$\mu$m crystalline silicate feature gets 
overwhelmed by the 33\,$\mu$m dust continuum.

Attempts to find correlations of S$_{\rm cryst}$(33$\mu$m) with the
30\,$\mu$m to 5.5\,$\mu$m continuum slope and with EQW(PAH62) 
have come up empty. In the absence of a mid-infrared diagnostic
that allows us to distinguish between quiescent and vigorous 
star formation among IDEOS galaxies we also cannot address the 
question whether the higher input of stardust in a galaxy would 
have a measurable effect on the global dust content.

For the IDEOS sample as a whole it thus appears that crystalline
silicates are a common component of the ISM of galaxies --- or, 
taking into account the non-detection of crystalline silicates in 
the spectra of the disks of NGC\,4945 and M\,101 
(Sect.\,\ref{sec:tracers_nucobs}), of galactic nuclei.

\section{IDEOS portal} \label{sec:portal}

The results of our spectral fitting and decomposition can be examined
in detail via the IDEOS web portal \url{ideos.astro.cornell.edu}. The 
website offers several important functionalities beyond what 
machine-readable tables of measured quantities offer, such as, for
individual sources:

\begin{itemize}
\item The extraction positions and positions angles for 
the individual SL and LL nod spectra are overlaid on an optical survey map.
\item The nod1 and nod2 spectra are plotted together to allow assessment 
of the significance of spectral structure found in the final
nod-averaged spectrum.
\item Continuum, line and feature fits can be visually inspected by
choosing the SED fit of the appropriate wavelength range.
\item Amorphous and crystalline silicate band fits are plotted
side-by-side.
\item All fitted emission lines can be compared in one plot.
\item The excitation diagram of warm molecular hydrogen, and postage
stamp plots of the individual H$_2$ line fits are shown together.
\item Synthetic photometric points are overlaid on the SED.
\item The spectral decomposition into PAH and blackbody components
from either PAHFIT or QUESTFIT is visualized.
\end{itemize}

The portal also provides a way to rank and examine sources by any 
of the measured quantities, or to select a galaxy sample based on 
constraints built on inequalities of observables conjoined by 
logical operators. E.g. {\tt irs15flux} $>$ 2 $*$ {\tt mips24flux}, 
{\tt ne3\_15flux} $>$ {\tt ne2flux}, or {\tt redshift} $>$ 0.3.
The observables and spectra can also be downloaded as 
machine readable tables.

Within the portal we also link to auxilary pages that explain the 
characteristics of CHUNKFIT or how we came to adopt our spectral
resolution table.

%%%%%%%%%%%%%%%%%%%%%%%%%%%%%%%%%%%%%%%%%%%%%%%%%%%%%%%%%%%%%%%%%%
%%%%%%%%%%%%%%%%%%%%%%%%%%%%%%%%%%%%%%%%%%%%%%%%%%%%%%%%%%%%%%%%%%
%\setcounter{table}{1}
\begin{table*}[t!]
\renewcommand{\thetable}{\arabic{table}}
\centering
\caption{IDEOS observables: emission and absorption features} \label{tab:features}
\begin{tabular}{llll}
\tablewidth{0pt}
\hline
\hline
Column & Observable & Unit & Description\\
\hline
%\decimals: if set the decimals are displayed as entered
1     & Name   && IDEOS identifier \\
2     & Galaxy && NED galaxy identifier \\
3     & RA     & degrees & Right Ascension J2000.0 \\
4     & Dec    & degrees & Declination J2000.0 \\
5     & z      &         & Redshift (Paper \sc{i}) \\
6     & MirClass & & Mid-IR class designation \\
7--9  & Pah62Flux & W m$^{-2}$ & 6.2\,$\mu$m PAH flux measured using spline method\\
10--12& Pah62EQW & $\mu$m & Ice-corrected equivalent width of the 6.2\,$\mu$m PAH feature measured using spline method\\
13--15& Pah11Flux & W m$^{-2}$ & 11.2\,$\mu$m PAH flux (sum of PAH111 and PAH112) measured using spline method\\
16--18&Pah11EQW & $\mu$m & Equivalent width of the 11.2\,$\mu$m PAH feature measured using spline method\\
19--21&Pah127Flux & W m$^{-2}$ & 12.7\,$\mu$m PAH flux measured using spline method\\
22--24&Pah127EQW & $\mu$m & Equivalent width of the 12.7\,$\mu$m PAH feature measured using spline method\\
25--27&DrudePAH62FLUX & W m$^{-2}$ & 6.2\,$\mu$m PAH flux measured using drude method\\
28--30&DrudePAH62EQW & $\mu$m & Ice-corrected equivalent width of the 6.2\,$\mu$m PAH feature measured using drude method\\
31--33&DrudePAH77Flux & W m$^{-2}$ & 7.7\,$\mu$m PAH flux measured using drude method\\
34--36&DrudePAH77EQW & $\mu$m & Equivalent width of the 7.7\,$\mu$m PAH feature measured using drude method\\
37--39&DrudePAH11Flux & W m$^{-2}$ & 11.2\,$\mu$m PAH flux measured using drude method\\
40--42&DrudePAH11EQW & $\mu$m & Euivalent width of the 11.2\,$\mu$m PAH feature measured using drude method\\
43--45&DrudeTOTPahFlux & W m$^{-2}$ & Integrated flux of all 5--19\,$\mu$m PAH feature measured using drude method\\
46--48&DrudeMirFlux & W m$^{-2}$ & Integrated 5--25\,$\mu$m continuum flux using drude method\\
49    &DrudeOrigin  && Decomposition method: PAHFIT or QUESTFIT\\
50--52&SilStrength  && Strength of the 9.8\,$\mu$m amorphous silicate feature\\
53    &SilFitType   && Interpolation method used to create local continuum: powerlaw or spline\\
54    &SilAnchorType&& Origin of anchors for continuum interpolation\\

55--57&Cryst19Strength & & Strength of the 19\,$\mu$m crystalline silicate feature\\
58--60&Cryst23Strength & & Strength of the 23\,$\mu$m crystalline silicate feature\\
61--63&Cryst28Strength & & Strength of the 28\,$\mu$m crystalline silicate feature\\
64--66&Cryst33Strength & & Strength of the 33\,$\mu$m crystalline silicate feature\\
67--68&TauIce & & Optical depth of the 6\,$\mu$m ice feature \\
69--70&Tau685 & & Optical depth of the 6.85\,$\mu$m aliphatic feature \\
\hline
\end{tabular}
\tablecomments{The data for Tables 6--8 is published in its entirety in the
machine-readable format. The table shown here only provides a description  
of the included quantities.}
\end{table*}
%%%%%%%%%%%%%%%%%%%%%%%%%%%%%%%%%%%%%%%%%%%%%%%%%%%%%%%%%%%%%%%%%%
%%%%%%%%%%%%%%%%%%%%%%%%%%%%%%%%%%%%%%%%%%%%%%%%%%%%%%%%%%%%%%%%%%

%%%%%%%%%%%%%%%%%%%%%%%%%%%%%%%%%%%%%%%%%%%%%%%%%%%%%%%%%%%%%%%%%%
%%%%%%%%%%%%%%%%%%%%%%%%%%%%%%%%%%%%%%%%%%%%%%%%%%%%%%%%%%%%%%%%%%
\begin{table*}[t!]
\renewcommand{\thetable}{\arabic{table}}
\centering
\caption{IDEOS observables: emission lines} \label{tab:lines}
\begin{tabular}{llll}
\tablewidth{0pt}
\hline
\hline
Column & Observable & Unit & Description\\
\hline
71--73&Ar2Flux & W m$^{-2}$ & 6.99\,$\mu$m [Ar {\sc ii}] line flux\\
74--76&Ar3Flux & W m$^{-2}$ & 8.99\,$\mu$m [Ar {\sc iii}] line flux\\
77--79&Cl2Flux & W m$^{-2}$ & 14.37\,$\mu$m [Cl {\sc ii}] line flux\\
80--82&Fe2Flux & W m$^{-2}$ & 25.99\,$\mu$m [Fe {\sc ii}] line flux\\
83--85&H2S0Flux & W m$^{-2}$ & 28.22\,$\mu$m H$_2$ 0--0 S(0) line\\
86--88&H2S1Flux & W m$^{-2}$ & 17.03\,$\mu$m H$_2$ 0--0 S(1) line\\
89--91&H2S2Flux & W m$^{-2}$ & 12.28\,$\mu$m H$_2$ 0--0 S(2) line\\
92--94&H2S3Flux & W m$^{-2}$ & 9.66\,$\mu$m H$_2$ 0--0 S(3) line\\
95--97&H2S5Flux & W m$^{-2}$ & 6.91\,$\mu$m H$_2$ 0--0 S(5) line\\
98--100&H2S7Flux & W m$^{-2}$ & 5.51\,$\mu$m H$_2$ 0--0 S(7) line\\
101-103&Ne2Flux & W m$^{-2}$ & 12.81\,$\mu$m [Ne {\sc ii}] line flux\\
104--106&Ne3\_15Flux & W m$^{-2}$ & 15.56\,$\mu$m [Ne {\sc iii}] line flux\\
107--109&Ne3\_36Flux & W m$^{-2}$ & 36.01\,$\mu$m [Ne {\sc iii}] line flux\\
110--112&Ne5\_14Flux & W m$^{-2}$ & 14.32\,$\mu$m [Ne {\sc v}] line flux\\
113--115&Ne5\_24Flux & W m$^{-2}$ & 24.32\,$\mu$m [Ne {\sc v}] line flux\\
116--118&O4Flux & W m$^{-2}$ & 25.89\,$\mu$m [O {\sc iv}] line flux\\
119--121&S3\_18Flux & W m$^{-2}$ & 18.71\,$\mu$m [S {\sc iii}] line flux\\
122--124&S3\_33Flux & W m$^{-2}$ & 33.48\,$\mu$m [S {\sc iii}] line flux\\
125--127&S4Flux & W m$^{-2}$ & 10.51\,$\mu$m [S {\sc iv}] line flux\\
128--130&Si2\_Flux & W m$^{-2}$ & 34.82\,$\mu$m [Si {\sc ii}] line flux\\
\hline
131--133&Ar3Ar2Ratio & & 8.99\,$\mu$m [Ar {\sc iii}] / 6.99\,$\mu$m [Ar {\sc ii}]\\
134--136&Ne3\_15Ne2Ratio & & 15.56\,$\mu$m [Ne {\sc iii}] / 12.81\,$\mu$m [Ne {\sc ii}]\\
137--139&Ne5\_14Ne2Ratio & & 14.32\,$\mu$m [Ne {\sc v}] / 12.81\,$\mu$m [Ne {\sc ii}]\\
140--142&Ne5\_24Ne2Ratio & & 24.32\,$\mu$m [Ne {\sc v}] / 12.81\,$\mu$m [Ne {\sc ii}]\\
143--145&O4Ne2Ratio & & 25.89\,$\mu$m [O {\sc iv}] / 12.81\,$\mu$m [Ne {\sc ii}]\\
146--148&O4S3\_33Ratio & & 25.89\,$\mu$m [O {\sc iv}] / 33.48\,$\mu$m [S {\sc iii}]\\
149--151&O4S3\_18Ratio & & 25.89\,$\mu$m [O {\sc iv}] / 18.71\,$\mu$m [S {\sc iii}]\\
152--154&S3\_33S3\_18Ratio & & 33.48\,$\mu$m [S {\sc iii}] / 18.71\,$\mu$m [S {\sc iii}]\\
155--157&S4S3\_18Ratio & & 10.51\,$\mu$m [S {\sc iv}] / 18.71\,$\mu$m [S {\sc iii}]\\
158--160&Si2S3\_33Ratio & & 34.82\,$\mu$m [Si {\sc ii}] / 33.48\,$\mu$m [S {\sc iii}]\\
\hline
\end{tabular}
\tablecomments{The data for Tables 6--8 is published in its entirety in the
machine-readable format. The table shown here only provides a description  
of the included quantities.}
\end{table*}
%%%%%%%%%%%%%%%%%%%%%%%%%%%%%%%%%%%%%%%%%%%%%%%%%%%%%%%%%%%%%%%%%%
%%%%%%%%%%%%%%%%%%%%%%%%%%%%%%%%%%%%%%%%%%%%%%%%%%%%%%%%%%%%%%%%%%

%%%%%%%%%%%%%%%%%%%%%%%%%%%%%%%%%%%%%%%%%%%%%%%%%%%%%%%%%%%%%%%%%%
\begin{table*}[t]
\centering
\caption{IDEOS observables: synthetic photometry and rest
frame continuum flux densities} \label{tab:photometry}
\begin{tabular}{llll}
%\tablewidth{0pt}
\hline
\hline
Column & Observable & Unit & Description\\
\hline
161--162&IRAC8Flux & mJy & Synthetic IRAC 8\,$\mu$m photometry \\
163--164&IRS15Flux  & mJy & IRS blue peak-up photometry \\
165--166&IRS22Flux  & mJy & IRS red peak-up photometry \\
167--168&MIPS24Flux & mJy & MIPS 24\,$\mu$m photometry \\
169--170&MIRI56Flux & mJy & Synthetic MIRI 5.6\,$\mu$m photometry \\
171--172&MIRI77Flux & mJy & Synthetic MIRI 7.7\,$\mu$m photometry \\
173--174&MIRI10Flux & mJy & Synthetic MIRI 10\,$\mu$m photometry \\
175--176&MIRI11Flux & mJy & Synthetic MIRI 11.3\,$\mu$m photometry \\
177--178&MIRI13Flux & mJy & Synthetic MIRI 12.8\,$\mu$m photometry \\
179--180&MIRI15Flux & mJy & Synthetic MIRI 15\,$\mu$m photometry \\
181--182&MIRI18Flux & mJy & Synthetic MIRI 18\,$\mu$m photometry \\
183--184&MIRI21Flux & mJy & Synthetic MIRI 21\,$\mu$m photometry \\
185--186&MIRI25Flux & mJy & Synthetic MIRI 25.5\,$\mu$m photometry \\
187--188&WISE12Flux & mJy & WISE band 3 photometry \\
189--190&WISE22Flux & mJy & WISE band 4 photometry \\
\hline
191--193&CONT37Flux & mJy & Rest frame continuum flux density at 3.7\,$\mu$m \\
194--196&CONT42Flux & mJy & Rest frame continuum flux density at 4.2\,$\mu$m \\
197--199&CONT55Flux & mJy & Rest frame continuum flux density at 5.5\,$\mu$m \\
200--202&CONT97Flux & mJy & Rest frame continuum flux density at 9.7\,$\mu$m \\
203--205&CONT15Flux & mJy & Rest frame continuum flux density at 15\,$\mu$m \\
206--208&CONT24Flux & mJy & Rest frame continuum flux density at 24\,$\mu$m \\
209--211&CONT30Flux & mJy & Rest frame continuum flux density at 30\,$\mu$m \\
\hline
212--214&C15C55Ratio &    & Rest frame continuum slope 15\,$\mu$m/5.5\,$\mu$m \\
215--217&C24C55Ratio &    & Rest frame continuum slope 24\,$\mu$m/5.5\,$\mu$m \\
218--220&C30C55Ratio &    & Rest frame continuum slope 30\,$\mu$m/5.5\,$\mu$m \\
\hline
221     &SNR66       &    & Continuum S/N at 6.6\,$\mu$m \\
222     &SNR9        &    & Continuum S/N at 9.0\,$\mu$m \\
223     &SNR1125     &    & Continuum S/N at 11.25\,$\mu$m \\
224     &SNR1325     &    & Continuum S/N at 13.25\,$\mu$m \\
225     &SNR14       &    & Continuum S/N at 14\,$\mu$m \\
226     &SNR17       &    & Continuum S/N at 17\,$\mu$m \\
227     &SNR24       &    & Continuum S/N at 24\,$\mu$m \\
228     &SNR30       &    & Continuum S/N at 30\,$\mu$m \\
229     &SL2ScalingFactor&& Factor by which the SL2 spectral segment has been scaled\\
230     &SL1ScalingFactor&& Factor by which the SL1 spectral segment has been scaled\\
231     &LL2ScalingFactor&& Factor by which the LL2 spectral segment has been scaled\\
232     &LL1ScalingFactor&& Factor by which the LL1 spectral segment has been scaled\\
\hline
\end{tabular}
\tablecomments{The data for Tables 6--8 is published in its entirety in the
machine-readable format. The table shown here only provides a description  
of the included quantities.}
\end{table*}
%%%%%%%%%%%%%%%%%%%%%%%%%%%%%%%%%%%%%%%%%%%%%%%%%%%%%%%%%%%%%%%%%%

\section{Discussion} \label{sec:discussion}

An unexpected by-product of the spectral fitting of the IDEOS 
spectra has been the discovery of crystalline silicate 
emission and absorption features in the 23--34\,$\mu$m range of
almost all good S/N spectra with coverage of these features.
The large majority of these spectra show these forsterite features 
in emission: 417 times in emission and 9 times in absorption for 
the 33\,$\mu$m feature; 97 times in emission and 76 times in 
absorption for the 28\,$\mu$m feature; 320 times in emission 
and 235 times in absorption for the 23\,$\mu$m feature. Only 
the shorter wavelength 16 and 19\,$\mu$m forsterite features 
are fitted exclusively
in absorption in spectra that satisfy S$_{\rm sil}$$<$-0.5,
where they are detected 149 times.

We have found that the strength of the five fitted crystalline 
features is correlated with the strength of the 9.8\,$\mu$m amorphous 
silicate feature (Fig.\,\ref{fig:crystsil_silstrength}) and 
that the longer wavelength crystalline silicate bands require 
a deeper 9.8\,$\mu$m silicate absorption feature for them to be 
seen in absorption than their shorter wavelength counterparts 
(Figs.\,\ref{fig:model_spectra}\,\&\,\ref{fig:silicate_spectrum}).
These findings are reminiscent of the structure seen
in spectra of evolving AGB stars. The silicate bands in these 
sources are found to be in emission or absorption (see
Fig.\,\ref{fig:ulirg_stellar_18um_optdepth_profiles})
depending on the thickness of the dust shell 
\citep{sylvester99,dijkstra03,devries14,golriz14}.

For galaxies in our sample for which all silicate features
are seen in emission (105 sources)
in theory another kind of dust geometry may 
apply: that of a class {\sc i} or {\sc ii} protoplanetary disc 
\citep{meeus01,watson09,furlan11}. 
In these flared discs the dust temperature increases away from the
mid-plane, as the surface layer of the disc is directly exposed to 
the radiation of the protostar. As a result all silicate features 
are seen in emission. Most of the IDEOS sources that share these
spectral properties are AGN, whose dust geometries may have many 
similarities to those of protoplanetary disks.

In analogy to the evolving AGB stars discussed above, 
IDEOS galaxies with at least one forsterite band in absorption
(323 sources) should host a centrally heated dust structure, 
which we identify with the galaxy's nucleus. As illustrated in 
Fig.\,\ref{fig:model_spectra}, the 23 and 33\,$\mu$m 
forsterite bands can be used to classify
galaxies as hosting a deeply obscured, semi obscured
and unobscured nucleus. With the 28\,$\mu$m forsterite
feature as a proxy for the 33\,$\mu$m forsterite feature 
in sources for which the 33\,$\mu$m forsterite band is 
redshifted out of the IRS wavelength range, we are able 
to identify a total of 40 deeply obscured galactic nuclei.
As can be seen in Fig.\,\ref{fig:forkdiagram_c30c55_silstrength_crystdiff}, 
not all of these sources (color-coded as blue) are found 
in the upper left part of the Fork Diagram (class 3A), 
which is home to the most deeply obscured galactic nuclei. 
As demonstrated in Sect.\,\ref{sec:tracers_nucobs}, 
circumnuclear star formation shifts these sources along the
diagonal branch of the Fork Diagram towards class 1C. Class
3A is hence the locus of 'naked' buried nuclei: those that
are not 'dressed up' with circumnuclear star formation.
We predict that millimetre HCN-vib studies looking for Compact
Obscured Nuclei \citep[CONs;][]{falstad21} will
expose most of our 40 deeply obscured galaxies as CONs.
We are especially confident about this since all CONs and 
non-CONs in the sample of \cite{falstad21} align with our 
crystalline silicate-based classifications for the sources
that we have in common.

The case of Mrk\,231 offers another lesson. Our crystalline
diagnostics classify this ULIRG as containing a 
semi-obscured nucleus (the 33\,$\mu$m feature is in emission, 
the 23\,$\mu$m feature in absorption). A keyhole view through 
the obscuring dust shell \citep{marshall18} of this BAL QSO 
results in a significant 
reduction of the depth of the 9.8\,$\mu$m silicate feature, 
resulting in a significant downward shift in the Fork Diagram. 
The silicate strength is hence not a good proxy for the level 
of nuclear obscuration if:
1. There are light leaks (e.g. Mrk\,231). 
2. Circumnuclear star formation strongly contributes to the 
integrated SED (e.g. NGC\,4945).
3. The dust distribution is clumpy \citep{nenkova02}.
Sources well below the diagonal branch of the Fork Diagram 
may thus be affected by these AGN orientation effects or by
light leaks.

For sources with a centrally heated dust geometry one can attempt
to infer the crystallinity of the ISM, provided that both the 
amorphous and the crystalline silicate band used to probe the
column density are in the optically thick regime.
The shorter the wavelengths of the features used, the more easily
this precondition is met.
\cite{kemper04} exploited the proximity of the 11.2\,$\mu$m 
forsterite band to the 9.8\,$\mu$m amorphous silicate band to 
infer an upper limit of C$_{\rm ISM}$(11.2\,$\mu$m)=0.011 to the 
crystallinity of the Galactic Center line of sight.
Contamination by 11.2\,$\mu$m PAH emission forced \cite{spoon06} 
to use a longer wavelength forsterite band, at 16.3\,$\mu$m, 
for their crystallinity determinations in a sample of 12 ULIRGs.
They found 
C$_{\rm ISM}$(16\,$\mu$m)\footnote{Using our definition of crystallinity: Eq.\,2}
to range from 0.07 to 0.13.
For the centrally heated  sources in the IDEOS sample we have 
used the more commonly studied 19\,$\mu$m forsterite band
and were able to infer crystallinities for 254 sources. We
found an average crystallinity C$_{\rm ISM}$(19\,$\mu$m)=0.027 
with no apparent dependence on the silicate strength
(see Fig.\,\ref{fig:crystallinity}). These
crystallinities are far lower than previously reported by
\cite{spoon06} for their ULIRG sample and can be entirely explained
by our different assumption for the mass absorption coefficient 
and our different choice of forsterite band for the measurement
(see Appendix\,\ref{sec:appendix-c}).
X-ray absorption spectroscopy probing the dense ISM towards bright 
X-ray binaries in the Galactic Center region
\citep{rogantini20,zeegers19} has revealed 
average crystallinities to be markedly higher (on the order
of 10\%) than the upper limit (1.1\%) derived from the non-detection
of the 11.2\,$\mu$m forsterite band \cite{kemper04} towards
Sgr\,A$^*$. Our IR-derived crystallinities may thus be 
equally underestimating the true forsterite abundance
in galactic nuclei.

An intriguing result of our study is that it appears that 
crystalline silicate features are present in all galaxy spectra
obtained at sufficient S/N. Depending on the dust configuration
and the S/N at 16--34\,$\mu$m the detections may be limited to 
just the strong 33\,$\mu$m forsterite feature in emission, 
a 23\,$\mu$m feature in absorption, or a bucket-shaped 
18\,$\mu$m silicate absorption feature carved out by 
forsterite absorption at 16 and 19\,$\mu$m.
Fact is that our analysis shows that only one source, 3C\,390.3, 
positively shows no crystalline silicate bands to the strict 
limits we set for non-detection of the forsterite bands.
Even though we cannot exclude the existence of more sources
like 3C\,390.3 hiding among lower S/N spectra in our sample,
the results found based on 960 galaxies with coverage of all 
five crystalline silicate bands makes it plausible to assume
that crystalline silicates are a common component of the ISM
of all galactic nuclei in the Local Universe.
 
We can only speculate on why we find hardly any exceptions. 
Small number statistics may be one reason. 3C\,390.3 is 
part of a sample of only 158
very high\footnote{This sample has SNR$\geq$100 at 24 and 30\,$\mu$m 
and coverage of all 23--33\,$\mu$m forsterite bands.} 
S/N spectra, with 
802 other galaxies making up the rest of our study 
sample\footnote{The median SNR of our study sample is 48 and 61
at 24 and 30\,$\mu$m, respectively.}. On the other hand, of 
the 960 galaxies with coverage of all crystalline bands we 
detect forsterite in 538 and only in 4/960 we do not. Likely 
the large majority of the remaining 418 lower S/N 
sources will be like the 538 that do have detections.
Another possible reason is more astrophysical in nature.
The dust content of metal-rich galaxies is believed 
to be dominated by ISM dust production, which will form 
amorphous silicates \citep{asano13}. The stardust 
contribution (from evolved stars and supernovae) is minor. 
In addition, both the 
input of stardust and its destruction by supernovae depend 
on the star formation rate (SFR), which may result in a net 
weak dependence of the stardust abundance on the SFR, 
especially if a starburst occurs in a dusty galaxy. 
This may result in the observed apparent lack of 
correlation between galaxy type and detection rate of
crystalline silicates. 

In this context it would be interesting to study the dust 
content in low-metallicity galaxies that have an ongoing 
starburst. In such systems the freshly made stardust may dominate 
the total dust content, and would provide a clean measure 
of the stardust destruction rate by supernovae.

With the launch of JWST just a few days away,
a new era of mid-infrared spectroscopy will soon
arrive, providing the opportunity to disentangle the nuclear and
circumnuclear spectral properties for galaxies beyond the next
few Mpc. While this should remove some of the current guesswork, 
the long wavelength cut-off at 28\,$\mu$m will also limit the 
number of available diagnostic features, the crystalline 
silicates in particular.

\section{Conclusions}

The homogeneous analysis of the homogeneously extracted, stitched, 
and curated Spitzer-IRS low-resolution spectra in IDEOS has produced a 
wealth of observables for more than 3300 galaxies and galactic nuclei.
Among the 74 observables 
in IDEOS are PAH fluxes and their equivalent widths, the strength 
of the 9.8\,$\mu$m silicate feature, emission line fluxes, solid-state 
features, rest frame continuum fluxes, synthetic JWST, Spitzer and WISE 
photometry, and a mid-infrared spectral galaxy classification.
Our web portal allows each of our fits to be inspected.
Our spectral fits allowed us to identify 388 active galactic 
nuclei (AGN) based on the detection of emission lines of 
[Ne {\sc v}] at 14.3 or 24.3\,$\mu$m.

Emission line diagnostic diagrams based on the ratio of a high-ionization 
tracer like [O {\sc iv}] or [Ne {\sc v}] to a low-ionization tracer 
like [Ne {\sc ii}] or [S {\sc iii}] on one axis plotted 
against the ratio of a medium-ionization tracer like [Ne {\sc iii}] 
or [Ar {\sc iii}] to a low-ionization tracer like [Ne {\sc ii}] 
or [Ar {\sc ii}] on the other axis 
are effective tools to separate out low-metallicity galaxies 
from galaxies on the AGN/starburst scale. 
The former lack $>$54eV photons needed to create O$^{3+}$ and Ne$^{4+}$, 
but are able to drive ratios of medium-to-low ionization tracers 
like [Ne {\sc iii}]/[Ne {\sc ii}] and [Ar {\sc iii}]/[Ar {\sc ii}] 
to higher values than AGN-dominated galaxies do.

We have detected weak emission and absorption features of crystalline 
silicates at 16, 19, 23, 28 and 33\,$\mu$m in the spectra of 786
IDEOS galaxies.
Quantitative analysis of the detected crystalline and amorphous 
silicate bands in this large galaxy sample reveals several trends.
First, the 23, 28 and 33\,$\mu$m crystalline silicate bands can de
detected both in emission and in absorption. The 16 and 19\,$\mu$m 
bands are fitted only in absorption.
Second, the strengths of the various crystalline silicate bands 
show a positive correlation with the amorphous silicate strength.
Third, crystalline silicate bands switch from absorption to emission
at a more negative silicate strength as the wavelength of the 
crystalline band is longer.
These observed characteristics are consistent with
an origin for the amorphous and crystalline silicate features in 
a centrally heated dust shell -- a dusty or buried nucleus.
Based on the detection of one or more crystalline silicate bands
in absorption, we find at least 323 galaxies in the IDEOS sample
that host a centrally heated dust structure.
For a galaxy showing the 9.8\,$\mu$m silicate feature in emission 
our observations may also be consistent with an origin 
of the crystalline silicate detections in an AGN torus geometry.

Exploiting the nature of centrally heated dust structures,
we have used the sign of the 23 and 33\,$\mu$m crystalline 
silicate strengths -- emission or absorption -- to classify 
galaxies as deeply obscured, semi obscured or unobscured. 
Based on this diagnostic, we identify buried nuclei in a 
total of 40 galaxies.  %Total of 40 is based on 23&33 and 23&28um samples
Interestingly, the 5--20\,$\mu$m 
spectra of these galaxies range from fully 
absorption-dominated to fully PAH-dominated, which is 
likely the result of varying contributions of circumnuclear 
star formation to their galaxy-integrated spectra. In the
Fork Diagram these same galaxies span the full length
of the diagonal branch, with the most extreme member 
shifted all the way to the locus of dusty starburst 
galaxies by drowning the spectral signatures of its 
buried nucleus in PAH emission. Above 20\,$\mu$m
the circumnuclear spectrum is featureless, thus preserving
the crystalline silicate signatures of the buried nucleus
in the galaxy-integrated spectrum.

We have used the strength of the 9.8\,$\mu$m amorphous and the 
19\,$\mu$m crystalline silicate absorption bands to infer the
crystallinity of the ISM in obscured galactic nuclei 
(S$_{\rm sil}$$<$-1). For the 114 sources with a detection 
of the 19\,$\mu$m band 
crystallinities range from 0.7\% to 6\% with an average level of 
2.7\%, with no dependence on the level of obscuration.
Based on our detection statistics across all five forsterite
bands in the 16--34\,$\mu$m range, we conclude that crystalline 
silicates are a common component of the ISM of local (z$<$0.068)
galactic nuclei.

\acknowledgments

We would like to thank the anonymous referee for constructive 
comments which helped to improve this paper. We further would like 
to thank John Miles for help with setting up the 
precursor to CHUNKFIT, Jack Gallimore for sharing his adaptation 
of PAHFIT, Lee Armus, Vassilis Charmandaris, Daniel Dale, Andreas 
Efstathiou, Bill Forrest, Fr\'ed\'eric Galliano, Ismael Garcia-Bernete, 
Dimitra Rigopoulou, and J.-D. Smith for discussions, 
B. Sargent for providing Galactic comparison spectra, Jason Marshall 
for providing a galaxy disk template spectrum, Pedro Beirao, Karl
Gordon, Els Peeters and Miguel Pereira-Santaella for galaxy disk
spectra, and Margaret Meixner for suggesting to add JWST-MIRI 
bands to our suite of synthetic photometric fluxes.
HS would also want to acknowledge his friend Jack Waas for designing 
the logo of the IDEOS project for the web portal. 
The IDEOS project is funded by NASA under awards NNX13AE69G and 
NNX16AF26G.
This research has made use of the NASA/IPAC Extragalactic 
Database (NED), which is operated by the Jet Propulsion Laboratory, 
California Institute of Technology, under contract with the National 
Aeronautics and Space Administration.

\facilities{Spitzer(IRS)}

\appendix

\section{5.5--8\,$\mu$m absorption profile template} \label{sec:appendix-a}

\begin{figure}[]
\begin{center}
\includegraphics[scale=1]{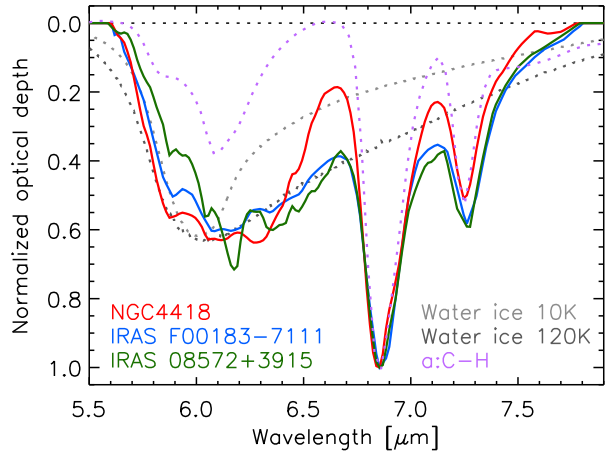}
\end{center}
\caption{Comparison of the normalized 5.5--8\,$\mu$m optical depth 
profiles of three galactic nuclei.  Overlaid with dotted lines are
the profiles of pure water ice and a a:C-H hydrogenated amorphous
carbon analog.
\label{fig:5-8um_bands}}
\end{figure}

We have searched for a template to fit the absorption features
commonly seen in the 5.5--8\,$\mu$m spectra of strongly obscured 
galactic nuclei. 
To qualify, the spectra needed to be free of PAH contamination in the
5.5--8\,$\mu$m range, have a silicate strength  S$_{\rm sil}<$-2.5, 
and be detected at high S/N. The resulting sample consists of just
three galaxies, NGC\,4418 \citep{spoon01}, IRAS\,F00183--7111
\citep{spoon04}, and IRAS\,08572+3915 \citep{spoon06}. We infered their
optical depth spectra by performing a spline fit as described in
Sect.\,\ref{sec:underlyingcont} and illustrated in 
Fig.\,\ref{fig:silfit_spline_n4418}. The optical depth spectra for 
the three galaxies are shown in Fig.\,\ref{fig:5-8um_bands} together
with two laboratory spectra of pure water 
ice\footnote{Sackler Laboratory Ice Database: http://icedb.strw.leidenuniv.nl/}, 
and one of an a:C--H analog \citep{dartois05}. 

The three galaxy spectra clearly show resemblance to features seen in 
the laboratory profiles: the optical depth peaks at 6.85 and 7.25\,$\mu$m 
agree in peak position and width with the structure in the spectrum of
the a:C--H hydrogenated amorphous carbon analog \citep{dartois07},
while the position and width of the broad 6\,$\mu$m feature agrees
with the profile of 120\,K water ice. 
Neither compound, nor a combination of
the two, will, however, reproduce the empirical profiles to a degree
that they can be used as templates in the fitting of the IDEOS
spectra. The true composition will likely be far more complex and
diverse than can be infered from these spectra.

%%%%%%%%%%%%%%%%%%%%%%%%%%%%%%%%%%%%%%%%%%%%%%%%%%%%%%%%%%%%%%%%%%
%%%%%%%%%%%%%%%%%%%%%%%%%%%%%%%%%%%%%%%%%%%%%%%%%%%%%%%%%%%%%%%%%%

%%%%%%%%%%%%%%%%%%%%%%%%%%%%%%%%%%%%%%%%%%%%%%%%%%%%%%%%%%%%%%%%%%
\begin{figure}[]
\begin{center}
\includegraphics[scale=1]{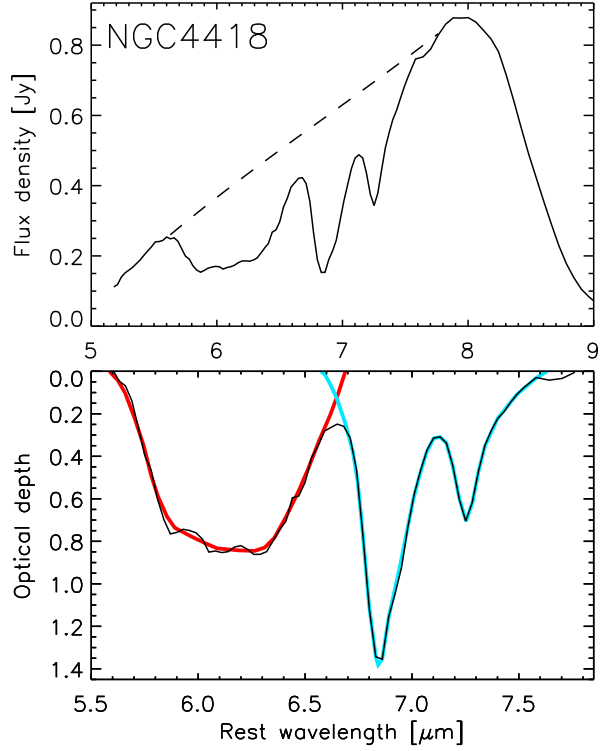}
\end{center}
\caption{Absorption features in the 5--8\,$\mu$m spectrum of
NGC\,4418. 
Upper panel: the Spitzer-IRS low-resolution spectrum of
NGC\,4418 \citep{spoon07} with the local 5.6--7.8\,$\mu$m
absorption-corrected continuum shown as a dashed line.
Lower panel: Optical depth spectrum resulting from adopting the
local 5.6--7.8\,$\mu$m spline continuum shown in the upper panel. 
Overlaid in red is the 6\,$\mu$m water ice profile and overlaid 
in blue is the aliphatic CH deformation mode feature to be used 
for fitting IDEOS spectra.
\label{fig:ngc4418}}
\end{figure}
%%%%%%%%%%%%%%%%%%%%%%%%%%%%%%%%%%%%%%%%%%%%%%%%%%%%%%%%%%%%%%%%%%

We therefore resorted to using an empirical optical depth profile as
our template, and separated the contribution of the CH deformation 
modes at 6.85 \& 7.25\,$\mu$m from the rest of the optical depth 
profile (6\,$\mu$m profile hereafter) to maximise the ability to fit 
these features in other IDEOS spectra. 
We chose the spectrum of NGC\,4418 as our template\footnote{Note that 
the spectral decomposition codes CAFE 
\citep{marshall07} and QUESTFIT \citep{veilleux09} prefer the optical 
depth profiles of IRAS\,F00183--7111 and IRAS\,08572+3915,
respectively, for their spectral fitting.}, 
over the spectra of IRAS\,F00183--7111 and 
IRAS\,08572+3915, because its more symmetric 6\,$\mu$m profile 
provides better fits to most spectra than the highly asymmetric 
6\,$\mu$m profiles of IRAS\,F00183--7111 and IRAS\,08572+3915
(Fig.\,\ref{fig:5-8um_bands}). 

Following \cite{dartois07}, we fit the 6.85 and 7.25\,$\mu$m aliphatic
CH deformation absorption bands with gaussian profiles and subtract
these from the optical depth spectrum. The remaining optical depth
profile in the 5.6--6.7\,$\mu$m range is then attributed to 6\,$\mu$m
water ice and is smoothed so that it can be used as a water ice template for
fitting IDEOS spectra. This profile is overlaid in red in the lower
panel of Fig.\,\ref{fig:ngc4418}. The blue profile in the same panel
is the combined aliphatic CH deformation mode profile.

\section{PAH profile templates} \label{sec:appendix-b}

The PAH emission features at 6.22, 11.25 and 12.65\,$\mu$m (PAH62,
PAH112, and PAH127) have asymmetric profiles, requiring either a 
combination of gaussian probability functions, or a single 
asymmetric probability function to accurately describe the 
observed shapes.

We chose the type IV probability function of \cite{pearson1895}, 
used to model pull distributions with non-gaussian tails, for our purposes.
The functional form for this Pearson-IV profile, P(x), is as follows:

\begin{equation}
P(x)= k \ [ 1 + (\frac{x-\lambda}{a})^2 \ ]^{-m} \ {\rm exp} [ \ -\nu
\ {\rm tan}^{-1}(\frac{x-\lambda}{a}) \ ]
\end{equation}

where the normalization coefficient k is defined by \cite{heinrich04} 
for m$>$1/2 as:

\begin{equation}\label{eq:two}
k = \frac{1}{\sqrt{\pi} \ a} \ \frac{\Gamma(m)}{\Gamma(m-1/2)} \
\left| \frac{\Gamma(m \ + \ i\nu/2)}{\Gamma(m)} \right|^2
\end{equation}

where $\Gamma$ is defined as 

\begin{equation}
\Gamma(x) = \int_0^{\infty} t^{x-1} \ e^{-t} \ dt
\end{equation}

The last term in Eq.\,\ref{eq:two} can be computed using a piece of
computer code, {\tt GAMMAR2}, provided in Sect.\,5.1 of \cite{heinrich04}.

\begin{figure}[t]
\begin{center}
\includegraphics[scale=0.6]{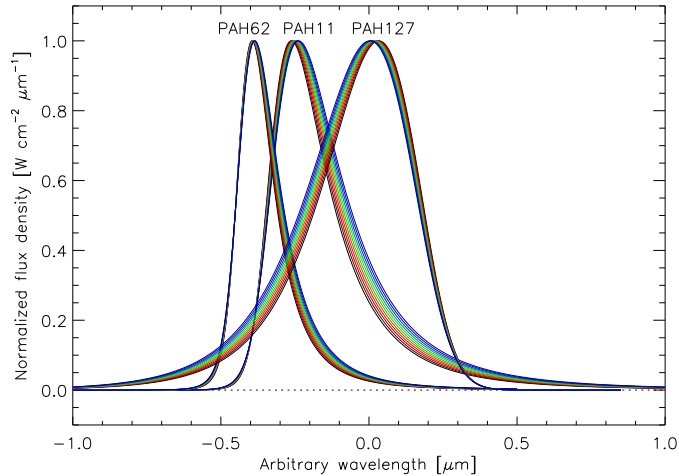}
\end{center}
\caption{Normalized pearson-IV profiles used to model the PAH62, PAH112
and PAH127 features in the SED fitting of the 5.39--7.25\,$\mu$m and
9.8--13.5\,$\mu$m ranges. Different colors are used to show the effect
of varying the A parameter within the allowed ranges.
\label{fig:pearson4}}
\end{figure}

Table\,\ref{tab:pearson4} shows for each of the three PAH profiles 
the allowed parameter ranges for the four parameters in the Pearson-IV 
function. Only $a$ and $\lambda$ are allowed to vary. The former to 
accommodate the variation in width of the PAH feature resulting from
the factor $\sim$2 variation in spectral resolution. The latter to 
allow compensation for the shift of the peak as a result of a change 
in $a$. See Fig.\,\ref{fig:pearson4}.

%%%%%%%%%%%%%%%%%%%%%%%%%%%%%%%%%%%%%%%%%%%%%%%%%%%%%%%%%%%%%%%%%%
\begin{table}{t}
\centering
\caption{Pearson IV parameters used to define shapes of PAH profiles} \label{tab:pearson4}
\begin{tabular}{lrrr}
\hline
\hline
 & PAH62 & PAH112 & PAH127 \\
\hline
$\lambda$ &  6.14--6.152 &  11.07--11.14 & 13.02--13.04\\
a         &  0.095--0.11 &  0.13--0.155  & 0.345--0.375\\
m         &  2.0         &  2.0          & 5.0  \\
$\nu$     & -2.5         & -4.0          & 10.0 \\
\hline
\end{tabular}
\end{table}
%%%%%%%%%%%%%%%%%%%%%%%%%%%%%%%%%%%%%%%%%%%%%%%%%%%%%%%%%%%%%%%%%%

%%%%%%%%%%%%%%%%%%%%%%%%%%%%%%%%%%%%%%%%%%%%%%%%%%%%%%%%%%%%%%%%%%
%%%%%%%%%%%%%%%%%%%%%%%%%%%%%%%%%%%%%%%%%%%%%%%%%%%%%%%%%%%%%%%%%%

\section{Crystalline silicate templates} \label{sec:appendix-c}

\begin{figure}[t]
\begin{center}
\includegraphics[scale=0.6]{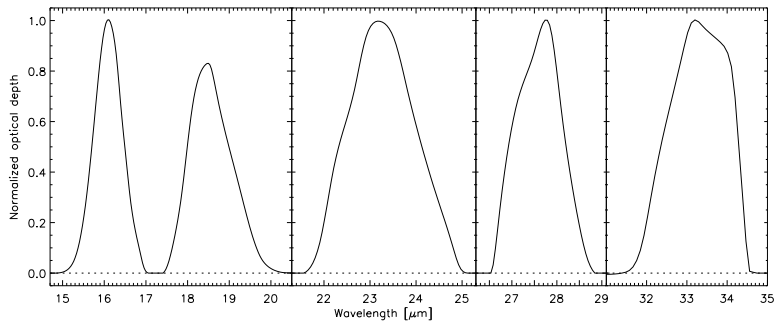}
\end{center}
\caption{Normalized optical depth profiles of crystalline
silicates found in spectra of IDEOS galaxies. The panel 
of the left shows the profiles of the 16.1 and 18.5\,$\mu$m 
forsterite bands, which are fitted together. The other forsterite
bands, centered at 23.3, 27.6 and 33.2\,$\mu$m, are fitted 
individually. 
\label{fig:crystsil}}
\end{figure}

Following \cite{spoon06}, we attribute the 16.1, 18.5, 23.3, 27.6, 
and 33.2\,$\mu$m absorption and emission bands seen in the IDEOS galaxy sample
to forsterite crystalline silicates and fit these bands with the empirical 
profiles described below and as shown in Fig.\,\ref{fig:crystsil}.

%%%%%%%%%%%%%%%%%%%%%%%%%%%%%%%%%%%%%%%%%%%%%%%%%%%%%%%%%%%%%%%%%%
%%%%%%%%%%%%%%%%%%%%%%%%%%%%%%%%%%%%%%%%%%%%%%%%%%%%%%%%%%%%%%%%%%

\begin{figure}[t]
\begin{center}
\includegraphics[scale=1]{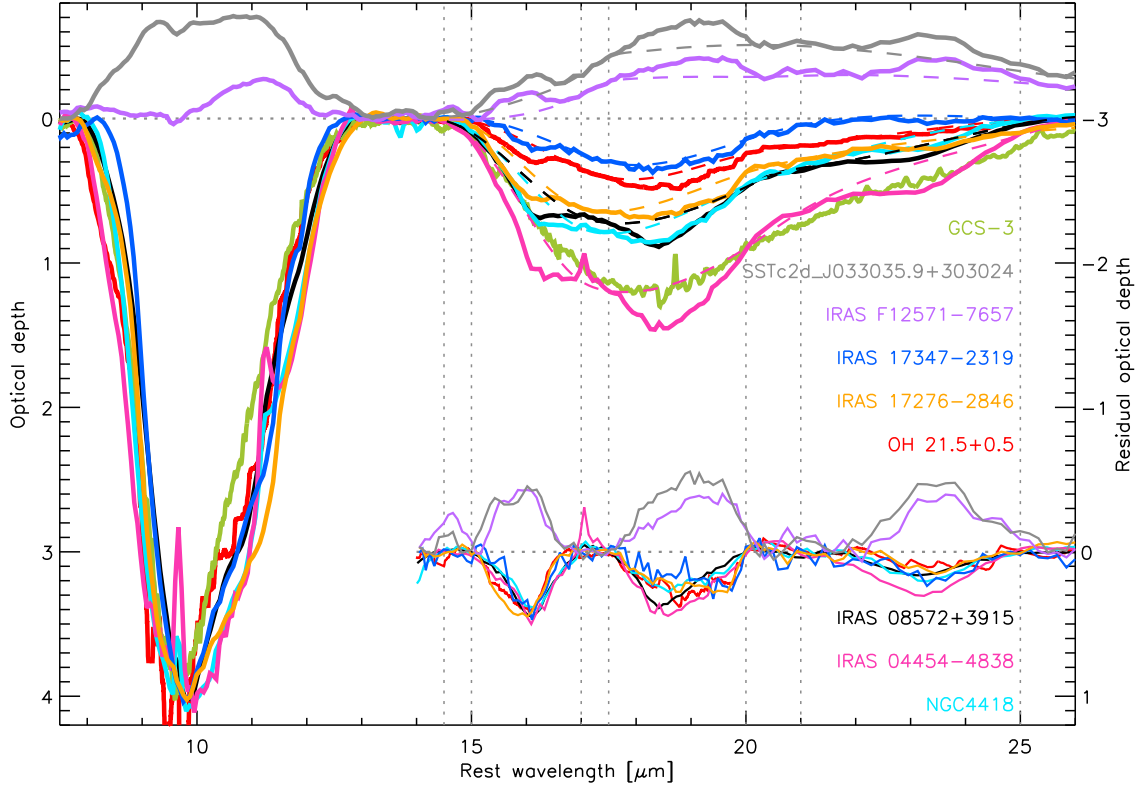}
\end{center}
\caption{Silicate optical depth profiles of three deeply obscured
galactic nuclei (NGC\,4418, IRAS\,08572+3915, and IRAS\,04454-4838), 
five stellar sources (SSTc2d\,J033035.9+303024, IRAS\,F12571--7657,
IRAS\,17347--2319, IRAS\,17276--2846, and OH\,21.5+0.5
\citep{sloan03}), and the Galactic Center Source GCS-3 \citep{chiar06}.
The silicate absorption spectra have been scaled to a 9.8\,$\mu$m 
optical depth of 4. The silicate emission sources have not.
The inset shows the residual optical depth features in the
14.5--26\,$\mu$m range after subtraction of the spline-fitted
amorphous silicate component \citep{spoon06}. GCS-3 is not included as it does not
show crystalline signatures. For clarity, the residuals have been 
scaled to a common 16.1\,$\mu$m absolute value.
\label{fig:ulirg_stellar_18um_optdepth_profiles}}
\end{figure}

In Galactic sources the 16.1\,$\mu$m forsterite band can be easily 
identified in the spectra of select young and old stars
(Fig.\,\ref{fig:ulirg_stellar_18um_optdepth_profiles}), 
both in emission (e.g. T\,Tauri stars IRAS\,F12571--7657 and 
SSTc2d\,J033035.9+303024; see the spectra in \cite{olofsson09}) 
and in absorption (e.g. OH/IR stars IRAS\,17347--2319, IRAS\,17276--2846, 
and OH\,21.5+0.5; see the spectra in \cite{jones12}). 
The band profile looks the same as in deeply obscured 
galactic nuclei (NGC\,4418, IRAS\,08572+3915, and IRAS\,04454--4838),
as can be seen in the inset of
Fig.\,\ref{fig:ulirg_stellar_18um_optdepth_profiles},
after removal of the spline-fitted amorphous silicate component
\citep{spoon06}.
We therefore pick the 16.1\,$\mu$m profile of 
the highest S/N galactic nucleus, IRAS\,08572+3915, to serve as 
the profile to be used in the spectral fitting described in 
Sect.\,\ref{sec:partial1421}. 

For some sources in Fig.\,\ref{fig:ulirg_stellar_18um_optdepth_profiles}
the strength of the 16.1\,$\mu$m band is almost
equal to the strength of the broader forsterite band peaking at 
18.5--19\,$\mu$m. The profile of the latter shows some variation among
the galactic nuclei, with some sources peaking near 18.5\,$\mu$m (e.g.
IRAS\,08572+3915 and IRAS\,04454--4838) and others near 19.0\,$\mu$m 
(e.g. NGC\,4418). The latter is where the profiles of the stellar 
sources peak.
Since the 16.1 and 18.5\,$\mu$m forsterite features are fitted 
jointly with a fixed profile ratio, we choose the high S/N spectrum
of IRAS\,08572+3915 to also provide the template for the 18.5\,$\mu$m
forsterite profile.

%%%%%%%%%%%%%%%%%%%%%%%%%%%%%%%%%%%%%%%%%%%%%%%%%%%%%%%%%%%%%%%%%%
%%%%%%%%%%%%%%%%%%%%%%%%%%%%%%%%%%%%%%%%%%%%%%%%%%%%%%%%%%%%%%%%%%

The peak-to-peak profile ratio for the 16.1 and 18.5\,$\mu$m bands in
IRAS\,08572+3915 is 1.0:0.83 (Fig.\,\ref{fig:crystsil}). This is far 
higher than calculated opacity curves for forsterite under various 
assumptions of grain size, grain shape, and optical constants predict.
Take for example the forsterite opacity curves shown in Fig.\,2 of 
\cite{devries15}, computed using the DHS grain shape 
distributions of \cite{min05} using optical constants of \cite{suto06};
or the forsterite opacity curve shown in Fig.\,2 of \cite{spoon06},
which uses the optical constants of \cite{fabian01} and the assumption
of a CDS grain shape distribution \citep{bohren83}.
Invariably the 16.1\,$\mu$m forsterite band is much weaker than 
the 19\,$\mu$m band.

%%%%%%%%%%%%%%%%%%%%%%%%%%%%%%%%%%%%%%%%%%%%%%%%%%%%%%%%%%%%%%%%%%
\begin{figure}[t]
\begin{center}
\includegraphics[scale=0.6]{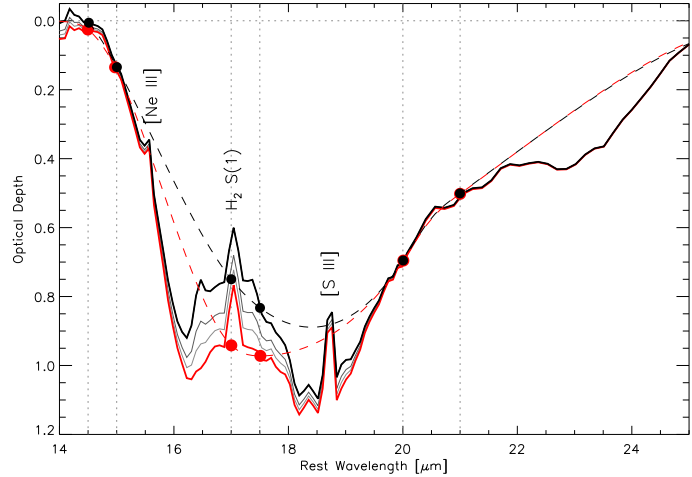}
\end{center}
\caption{Optical depth profile for ESO\,374-IG\,032\,NED02 in 
the 18\,$\mu$m silicate absorption band ({\it black}). Dashed 
lines trace the amorphous 18\,$\mu$m profile resulting from a 
spline fit to pivots at 14.5, 15.0, 17.0, 17.5, 20.0, and 21.0\,$\mu$m.
Residual optical depth features centered at 16.1, 18.5, and 
23.3\,$\mu$m are attributed to forsterite absorption. 
The profiles shown in red and in shades of gray are the result 
of subtracting the noise-free pure PAH template \#3 of \cite{smith07}
in increasing amounts. \label{fig:17um_complex_contamination}}
\end{figure}
%%%%%%%%%%%%%%%%%%%%%%%%%%%%%%%%%%%%%%%%%%%%%%%%%%%%%%%%%%%%%%%%%%

This leads us to explore an alternative origin for the 16.1 and 
18.5\,$\mu$m absorption bands in galaxies, despite strong empirical 
evidence provided by the stellar spectra shown in 
Fig.\,\ref{fig:ulirg_stellar_18um_optdepth_profiles}. Could
the appearance of 16.1 and 18.5\,$\mu$m {\it absorption} features 
in the spectra of buried galactic nuclei be ascribed to 
the presence of PAH {\it emission} from the 17\,$\mu$m 
PAH complex \citep{smith07} instead?
Fig.\,\ref{fig:17um_complex_contamination}
shows the 14--25\,$\mu$m spectrum of ESO\,374-IG\,032\,NED02, a deeply 
obscured galactic nucleus with clear signs of PAH emission in the
6--13\,$\mu$m range and at 16.4 and 17.4\,$\mu$m.
Using the noise-free PAH template \#3 of \cite{smith07}, we subtracted
as much of the 17\,$\mu$m PAH complex emission from the galaxy
spectrum as is consistent with the amount of PAH emission found in 
the 6--13\,$\mu$m range. 
As can be seen in Fig.\,\ref{fig:17um_complex_contamination}, 
the 16.4 and 17.4\,$\mu$m PAH bands disappear as the amount of
subtracted PAH template emission increases, yet the 16.1 and
18.5\,$\mu$m absorption structures remain even in the red
spectrum, which represents moderate oversubtraction of 6--13\,$\mu$m 
PAH emission.
A full fit to the latter spectrum (the red dashed line in
Fig.\,\ref{fig:17um_complex_contamination}) would measure almost the same 
forsterite strength as a fit to the original spectrum would. 
We thus conclude that the 16.1 and 18.5\,$\mu$m forsterite absorption
bands cannot be mimicked by emission from the 17\,$\mu$m PAH complex.

The profiles used to fit the 23.2\,$\mu$m and the 27.6\,$\mu$m 
forsterite bands in the 21.5--36\,$\mu$m spectral range 
(Sect.\,\ref{sec:partial1936}) have been adapted from the average
crystalline silicate optical depth spectrum of 12 ULIRGs
obtained by \cite{spoon06}. This average profile has higher S/N in 
the 21--30\,$\mu$m range than the profile of IRAS\,08572+3915.
The resulting profiles, characteristic of forsterite, are shown 
in the middle panels of Fig.\,\ref{fig:crystsil}.

The profile of the 33.2\,$\mu$m crystalline silicate band is best
determined from high S/N spectra of galaxies in mid-IR classes
1A--C and 2A, which show the feature in emission. The galaxies we 
selected have the following IDEOS IDs:
10510592\_0, 10511872\_0, 10870272\_0, 10871808\_0, 11290112\_1, 
12432128\_3, 14839552\_0, 14840320\_0, 14842112\_0, and 14869504\_0.
To obtain the feature profile we first averaged the continuum-normalized
spectra, and then removed the profile of the 33.48\,$\mu$m 
[S {\sc iii}] line. The resulting forsterite profile is shown in the 
right panel of Fig.\,\ref{fig:crystsil}.

\end{document}